\documentclass[preprint]{elsarticle}
\usepackage{amssymb}
\usepackage{amsmath}
\usepackage{color}
\usepackage{graphics}
\usepackage{epsfig,epstopdf}
\usepackage{graphicx}
\usepackage{epsfig}
\usepackage{float}
\usepackage{color}
\usepackage[normalem]{ulem}
\newcommand\red[1]{#1}
\newcommand\forget[1]{}

\newcommand{\be}{\begin{equation}}
\newcommand{\ee}{\end{equation}}
\newcommand{\nd}{\end{equation}}
\newcommand{\bea}{\begin{eqnarray}}
\newcommand{\eea}{\end{eqnarray}}

\renewcommand\Re{{\rm Re}\,}

\newcommand\Mo{{\rm Mo}\,}

\newcommand\Ca{{\rm Ca}}

\newcommand\eps\epsilon
\newcommand\Or{{\cal O}}
\newcommand\N{{\bf n}}

\journal{Journal of Computational Physics}

\begin{document}

\begin{frontmatter}
\title{Transition in a numerical model of contact line dynamics and forced dewetting}
\author[afkhami]{S.~Afkhami\footnote{Corresponding author email address: shahriar.afkhami@njit.edu}
}
\address[afkhami]{Department of Mathematical Sciences, New Jersey Institute of Technology, Newark, NJ, USA}
\author[guin]{J. Buongiorno}
\author[guin]{A. Guion}
\address[guin]{Nuclear Science and Engineering Department, Massachusetts Institute of Technology, Cambridge, MA, USA}
\author[zaleski]{S. Popinet}
\address[zaleski]{Sorbonne Universit\'{e}, CNRS, Institut Jean le Rond d'Alembert, UMR 7190, F-75005, Paris, France}
\author[zaleski]{Y. Saade}
\author[scardovelli]{R. Scardovelli}
\address[scardovelli]{DIN -- Laboratorio di Montecuccolino, Universit\`{a} di Bologna, 40136 Bologna, Italy}
\author[zaleski]{S. Zaleski\footnote{Corresponding author email address: zaleski@dalembert.upmc.fr}}

\begin{abstract}
We investigate the transition to a Landau--Levich--Derjaguin film in forced dewetting
using a quadtree adaptive solution to the Navier--Stokes equations with surface tension. 
We use a discretization of the capillary forces near the receding contact line that yields an equilibrium
for a specified contact angle $\theta_\Delta$ called the numerical contact angle. 
Despite the well-known contact line singularity, dynamic simulations can proceed without
any explicit additional numerical procedure.
We investigate angles from $15^\circ$ to $110^\circ$ 
and capillary numbers from $0.00085$ to $0.2$ {where the mesh size $\Delta$
is varied in the range of $0.0035$ to $0.06$ of the capillary length $l_c$}.
To interpret the results, we use
Cox's theory which involves a microscopic distance $r_m$ and
a microscopic angle $\theta_e$.
In the numerical case, the equivalent of $\theta_e$ is the angle $\theta_\Delta$ 
and we find
that Cox's theory also applies.
We introduce the  scaling factor or gauge function $\phi$ so that 
$r_m = \Delta/\phi$ and estimate this gauge function by comparing our numerics
to Cox's theory. 
The comparison provides a direct assessment of the agreement of
the numerics with Cox's theory and reveals a critical feature of the numerical
treatment of contact line dynamics: agreement is poor at small angles while it is better
at large angles.
This scaling factor is shown to depend only on   
$\theta_\Delta$ and the viscosity ratio $q$. In the case of small $\theta_e$, 
we use the prediction by Eggers [Phys. Rev. Lett., vol. 93, pp 094502, 2004]
of the critical capillary number for the 
Landau--Levich--Derjaguin forced dewetting transition.
We generalize this prediction to large  $\theta_e$ and arbitrary $q$ and express
the critical capillary number as a function of  $\theta_e$ and $r_m$. 
This implies also a prediction of the critical capillary number
for the numerical case as a function of  $\theta_\Delta$ and $\phi$.
The theory involves a logarithmically small parameter $\epsilon = 1/\ln (l_c/r_m)$ and is thus of 
moderate accuracy. 
The numerical results are however in approximate agreement in the general case, 
while good agreement is reached in the
small $\theta_\Delta$ and $q$ case.
An analogy can be drawn between the numerical contact angle condition and 
a regularization of the Navier--Stokes equation by a partial Navier-slip model. 
The analogy leads to 
a value for the numerical length scale $r_m$ proportional to the slip length. Thus
the microscopic length found in the simulations is a kind of numerical
slip length in the vicinity of the contact line.  
The knowledge of this microscopic length scale and the associated gauge function can be used
to realize grid-independent simulations that could be matched to microscopic physics 
in the region of validity of Cox's theory. This version of the paper includes the corrections indicated in 
\cite{corrigendum}. 
\end{abstract}


\end{frontmatter}
\section{Introduction}


Wetting of solids by liquids, in which a liquid displaces another
fluid on a solid substrate, is an ubiquitous phenomenon with applications
ranging from coating \cite{troian_jap2_00} and tear films on the
cornea \cite{Li2014} to micro-layer formation in wall boiling
\cite{Kim2011,Guin2013} and CO2 sequestration
\cite{Kimbrel2015}. However, despite the abundance of applications,
the precise mechanism of wetting is only partially understood. From
the numerical modeling point of view, difficulties arise due to the
highly multiscale nature of the problem (length scales extending from
the macroscopic to the molecular sizes). Another major challenge in
numerical simulations is the so-called contact line singularity that
arises when a continuum description of moving contact lines is used in
combination with a no-slip boundary condition at the liquid-solid
interface.  
Because of this singularity, the continuum description is untenable
below a certain scale. Thus a transition to a different, nonsingular
physics must occur as the scale is reduced. The most obvious 
such transition is the appearance of molecular effects at nanometer
scales. However a variety of other ``microscopic'' contact line
physics, some of which 
would ``kick-in'' at scales much larger than the nanometer, have
been considered in the literature. It is difficult to
be exhaustive but these involve precursor film models
\cite{MahadyvdW15}, diffuse-interface models
\cite{Ding2007,YueFeng2011,seppecher96,pismen00b} and the related issue of
evaporation \cite{Pomeau11}, interface formation models
\cite{Shikhmurzaev,Sprittles2013}, and surface roughness \cite{hock76}.
The reader may find references to other mechanisms in review papers
\cite{Blake2006,bonn_rmp2009,Snoeijer13,Sui14}.  Slip of
the contact line is of particular interest as a possible physical
mechanism to allow motion of the contact line on the microscopic scale,
mostly because it conveniently does not require to change the Navier--Stokes 
equations, see
e.g.~\cite{HuhScriv,Greenspan,Dussan79,cox1986,RENARDY2001,Spelt2005,Devauchelle07,
Afkhami_jcp09,Manservisi:2009gs}.
However, numerical simulations involving slip-length modeling are unfeasible in 
most physical problems since the true slip should be
related to molecular interactions between the liquid and the solid
substrate \cite{Blake2006}, which based on experimental measurements
lies in the nanometer range \cite{Joseph2005,Lauga2007}.
Thus a regularization of the numerics based on the slip length leads to  
computationally inconvenient large ratios of scales. 
On the other hand, if no regularization is performed or 
if the slip length is dependent on grid spacing, 
the moving contact line solutions become themselves dependent
on grid spacing (see e.g.~\cite{Afkhami_jcp09,Moriarty92,Weinstein08}). 

This paper pursues several goals. One is to describe the method used in the numerical framework,
Gerris \cite{popinetGerris} to 
impose the contact angle and the related ``banded'' advection of the Volume-Of-Fluid method. 
(Although the contact angle in a version of Gerris has been described 
in \cite{AB2008,AB2009}, the exact method used in the mainline distribution of the Gerris code
has not been described before.) 
Another  goal is to attempt to extend numerically the theory of
Eggers and his coworkers \cite{Eggers2004b,Eggers2005,Chan12} for the dewetting transition to the case of finite
microscopic angles.  One consequence of this analysis applied to the 
numerical case is a precise description of the behavior of any numerical method
in the vicinity of the contact line. Another consequence is to aid models
such as those in \cite{Afkhami_jcp09,Legendre2015} or \cite{moataz}
that are used to perform grid-independent models. 

In this paper we do not attempt to implement a sophisticated
{dynamic} contact angle model, but instead take an existing, 
simple numerical method already documented in \cite{AB2008,AB2009} for static cases and apply it
``as is''  for dynamic simulations. 
This is a kind of ``implicit modeling'' approach
similar to the implicit subgrid scale modeling frequent in Large Eddy Simulations 
of turbulence. This approach has the merit of simplicity, and it then
remains to assess how this ``numerical boundary condition'' affects the flow. 

\begin{figure}[]
\begin{center}
\includegraphics[scale=0.35,trim=0 30mm 120mm 30mm, clip=true]{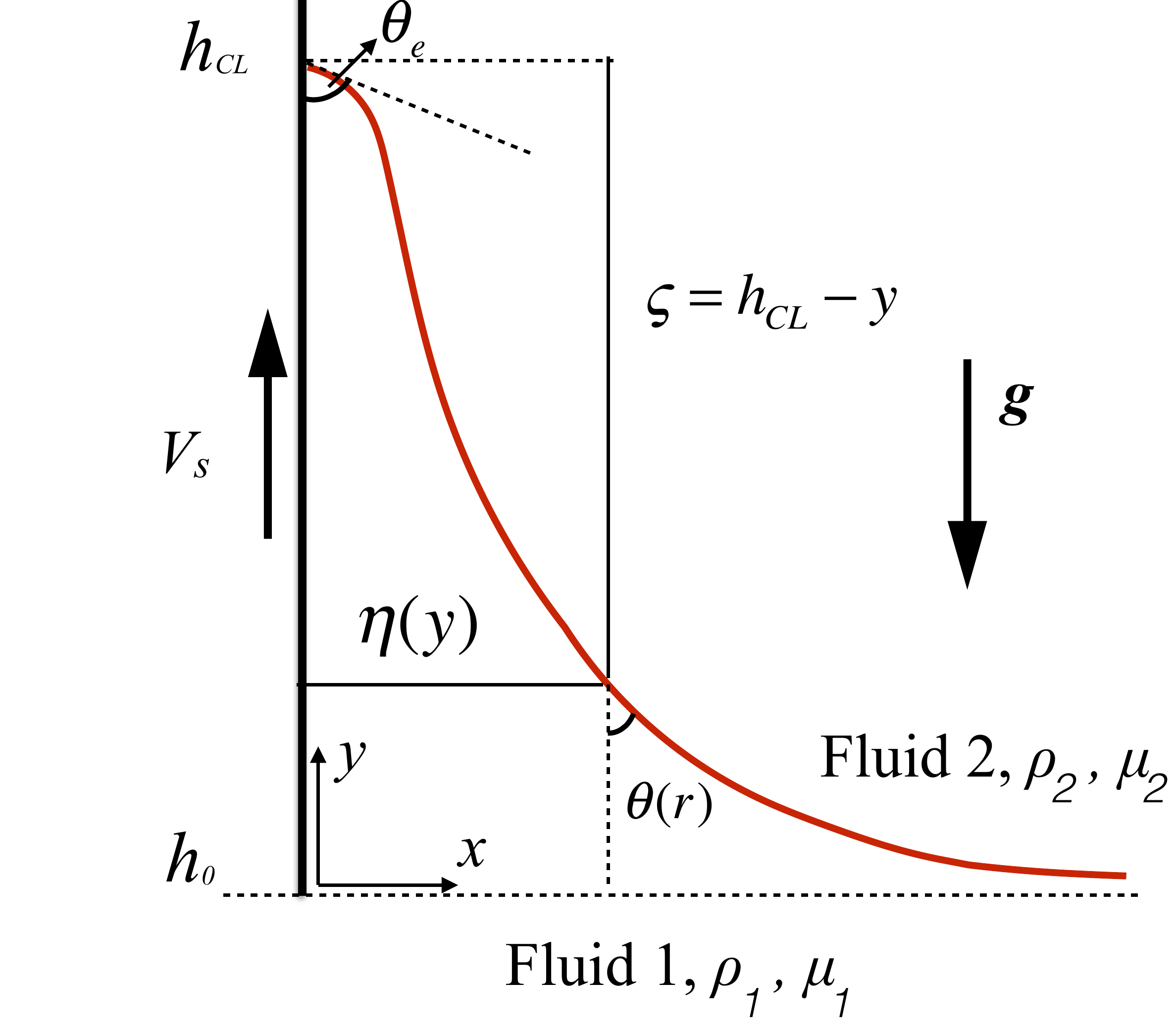}
\end{center}
\caption{Schematic illustrating the contact angles $\theta_{e}$ and $\theta(r)$, corresponding to
length scales $\lambda$ and $r$, respectively.
$h_{CL}$ represents the contact line height and $h_0$ is the unperturbed initial interface profile.
$\zeta = h_{CL}-y$ is the vertical distance of the interface at $y$ from the contact line.
$\eta(y)$ is the horizontal distance of the interface from the solid substrate.
}
\label{fig:1}
\end{figure}

We study a specific, complex physical problem: the dewetting transition. 
In a number of applications, the interface is forced to move along a solid
in a manner that can result either in a receding contact line or
in the formation of a thin film on the solid.
One example of such a flow is the withdrawing-tape experiment 
whose geometry is illustrated on Fig.~\ref{fig:1}.
A solid substrate is withdrawn on the left from a viscous liquid pool
of quiescent liquid. 
The interface may either sustain a stationary state meniscus,
if below a critical capillary number, ${\rm Ca}_{cr}$, or continue to move up 
the substrate until depositing a thin film to arbitrary heights.
The latter is called a Landau--Levich--Derjaguin (LLD) film \cite{LandauLevich1942,Derjaguin1943}.
This transition can be understood in terms of  the imbalance between the surface tension, 
gravity and viscous 
forces that leads to the vanishing of the contact line.
The analysis of the transition process on a partially wetting substrate is however
complicated due to the singularity of the  moving contact line.
On one hand, Eggers and his coworkers, in a series of papers \cite{Eggers2004b,Eggers2005,Chan12}, provided 
a hydrodynamic prediction, based on the lubrication approximation theory,
of the critical capillary number $\mbox{Ca}_{cr}$.
Cox \cite{cox1986} and Voinov \cite{Voinov1976} described, 
on the other hand, how the singularity drove a peculiar
curved form of the fluid wedge at small $\Ca$.
We use {these theories to predict the numerically observed 
transition. We shall show that 1) the microscopic length $r_m$ is entirely described by a gauge
function depending only on the equilibrium contact angle imposed numerically and 2) that the
knowledge of this numerical gauge function can be used 
to mimic the effect of actual subgrid scale 
microscopic physics. This second point is related to the 
notion of grid-independent simulations in \cite{Afkhami_jcp09,Legendre2015}.}

The rest of this paper is organized as follows. 
In Sec.~\ref{sec:setup}, we describe the physical problem and explain our
choice of parameters. 
In Sec.~\ref{sec:num}, we describe the computational setup and the numerical method
for the implementation of the contact angle.
In Sec.~\ref{sec:results}, we report the results of our simulations including the 
method for determining the critical capillary number ${\rm Ca}_{cr}$.
In Sec.~\ref{sec:theory}, we introduce the main new theoretical concept of this paper, 
the gauge function $\phi$. We discuss the general theoretical features of the transition
and study it in two cases, the small angle, small viscosity ratio case, free-surface case, where 
 Eggers's results can be applied directly and the more general case where the use of matched asymptotics and 
Airy functions is replaced by heuristic arguments.
In Sec.~\ref{sec:dynamic}, we discuss the connection with theories where
the microscopic angle varies with the capillary number as in the 
Molecular Kinetic Theory \cite{Blake1969}. 
In Sec.~\ref{sec:neo-AZB}, we apply the concepts developed above to describe
an improved procedure for obtaining grid-independent computations of problems with dynamic contact lines. 
Finally, we draw our conclusions in Sec.~\ref{sec:con}.

\section{Problem setup: forced dewetting}
\label{sec:setup}
We consider a solid plate being withdrawn from a liquid reservoir with
a constant velocity $V_s > 0$. The computational domain is $0\le x,y \le L$, 
with fluid~1 occupying $y<h_0$ and fluid~2 occupying $y>h_0$ at $t=0$ (see Fig.~\ref{fig:1}).
The viscosity and density of fluid $i=1,2$ are $\mu_i$ and $\rho_i$, respectively.
The capillary number is  defined as
$$\mbox{Ca}=\mu_1 V_s/\sigma,$$
where $\mu_i$ is the viscosity of fluid $i$ and
$\sigma$ the surface tension. 
We set $L\approx9\,l_c$, where $l_c$ is the capillary length
$l_c = \sqrt{\sigma/[(\rho_1 - \rho_2) g]}$
with $g$ the gravitational acceleration. 
The Reynolds number is then defined based on 
the capillary length as
$\mbox{Re} = \rho_1 V_s l_c/\mu_1$.
Thus $\Re = \Ca \,\Mo^{-1/2}$ where the Morton number is 
$$
\Mo = \frac{(\rho_1 - \rho_2)\, g \mu^4_1}{\rho_1^2 \sigma^3}.
$$
We define a ``gravity-wave-damping'' number as
$$
N_G =  \frac{\rho_1^2 V_s^3}{\mu_1 (\rho_1- \rho_2) \,g}.
$$
Indeed, the wavelength of gravity waves traveling
at the same speed as the withdrawing tape is $L_{gw} = \rho_1 V_s^2 / [(\rho_1- \rho_2) g]$
and it can be connected to $N_G$ through
$N_G  = \rho_1 L_{gw} V_s / \mu_1$. Thus $N_G$ is also the Reynolds number
based on the wavelength of gravity waves. It is related to the capillary length Reynolds number by
\be
\Re = N_G^{1/2} \Ca^{-1/2}. \label{ng_re}
\nd
We use several setups for the simulations, presented in Tab.~\ref{tab1}. In Setups A and B,
for relatively more efficient computations, we set the ratios of physical properties to 
moderate values with the viscosity ratio $\mu_1/\mu_2=1$ and
the density ratio $\rho_1/\rho_2=5$. 
The other parameters are chosen in the following way. 
In Setup A, the number $N_G$ is arbitrarily set to 
$N_G=25/64$.
(It is the value corresponding to the arbitrary choices of  $\rho_1=5, \rho_2=1, g=16,\mu_1=1, V_s=1$.)
Thus from Eq.~(\ref{ng_re}) the Reynolds number based on $l_c$ varies as
\be
\Re = \frac58 \Ca^{-1/2}. \label{re_ca}
\nd
As a result the Reynolds number  based on $l_c$ increases  as the capillary number decreases.
Varying the Reynolds number between $0$ (Stokes approximation) and $\Re=3$ has no effect on
the results, however increasing $\Re$ beyond this value introduces significant
inertial effects and interface oscillations that modify the conclusions of our
investigations. At small $\Ca$, we therefore switch to Setup B, where
$N_G$ is free to vary and the Reynolds number based on $l_c$ is fixed to 
$\Re=1$. 
In a final set of simulations (Setup C), we keep
 $\rho_1/\rho_2=5$ and $\Re=1$ but let $\mu_1/\mu_2=50$.
This allows to bring the simulations somewhat closer to air/water conditions without
encountering the numerical problems arising with very large density ratios. 
\begin{table}
\begin{center}
\begin{tabular}{ccccc}
\hline
Setup  & $N_G={\rho_1^2 V_s^3}/({\mu_1 \, (\rho_1- \rho_2) \,g})$ &  $\Re={\rho_1 V_s l_c}/{\mu_1}$
& $1/q = \mu_1/\mu_2$ & $\rho_1/\rho_2$ \\
\hline 
A & ${25}/{64}$  &   -  &  1 & 5 \\
B &        -          &   1      &  1 & 5 \\
C &        -          &   1      &  50 & 5 \\
\hline
 \end{tabular}
 \end{center}
\caption{Summary of the simulation Setups. When no value is indicated, the corresponding
number is computed from the other numbers using Eq.~(\ref{ng_re}).}
  \label{tab1}
\end{table}

We begin the simulations by considering a flat 
interface between the two fluids initially at the
height $h_0\approx 3.1\,l_c$. 
A no-slip boundary condition is prescribed at the substrate ($x=0$).
Symmetry boundary conditions are imposed on the right ($x=L$),
top ($y=L$), and bottom ($y=0$) boundaries of the domain. 
We note that we have checked that the results are insensitive to  
the computational domain size.

In the neighborhood of the contact line,
no specific choice of parameters is required
except the equilibrium or static contact angle that is specified in the 
numerical model. It is also expected that the numerical model
leads to a solution varying continuously with the withdrawing tape velocity, 
so that the contact angle tends to the static contact angle as
$\Ca$ tends to zero.
How simulations with a contact angle are performed is described 
in the next section.

\section{Numerical model}\label{sec:num}

We use {Gerris} \cite{popinetGerris,Popinet03,Popinet2009} to numerically 
solve the Navier--Stokes, continuity and density equations,
\begin{equation}
\rho\left( \partial_t \textbf{u} + \textbf{u} \cdot \nabla \textbf{u} \right) 
= - \nabla p + \nabla\cdot\left[ \mu\left(\nabla
    \textbf{u} + \nabla\textbf{u}^\top\right)\right]
  + \sigma\kappa\delta_s\textbf{n} +  \rho \textbf{g},
\end{equation}
\begin{equation}
\nabla\cdot\textbf{u} = 0,
\end{equation}
\begin{equation}
\partial_t \rho + \textbf{u} \cdot \nabla \rho=0,
\label{eq:vof}
\end{equation}
respectively. Here, $\textbf{u}$ is the velocity field, $p$ the pressure, 
$\rho = \rho_1 \chi + \rho_2 (1 - \chi)$, 
$\mu = \mu_1 \chi + \mu_2 (1 - \chi)$,
$\kappa$ is the interface curvature,
$\textbf{n}$ the normal to the interface (pointing from fluid 1 to fluid 2), 
$\delta_s$ the delta function centered at the interface, 
$\rho\textbf{g} = - \rho g{\hat y}$ the body force due to gravity
and ${\hat y}$ the unit vector in the $y$-direction, and 
$\chi$ ($=1$ in fluid 1 and $0$ in fluid 2)  the characteristic function, 
where $\delta_s\textbf{n} = \nabla \chi$.
Note that Eq.~(\ref{eq:vof}) is equivalent to 
\begin{equation}
\partial_t \chi + \textbf{u} \cdot \nabla \chi =0,
\label{eq:vof2}
\end{equation}
which is solved using the Volume-Of-Fluid (VOF) interface capturing method
\cite{Popinet2009,zaleski99,AB2008,PopinetARFM}. The Continuous Surface Force (CSF) method 
is used for the surface tension force with curvature computed using the 
Height-Function method \cite{Popinet2009,PopinetARFM}. Viscous forces are implemented using 
a partially implicit method described in \cite{Lagree:2011bq}. 

\begin{figure}[]
\begin{center}
\includegraphics[width=4.5in]{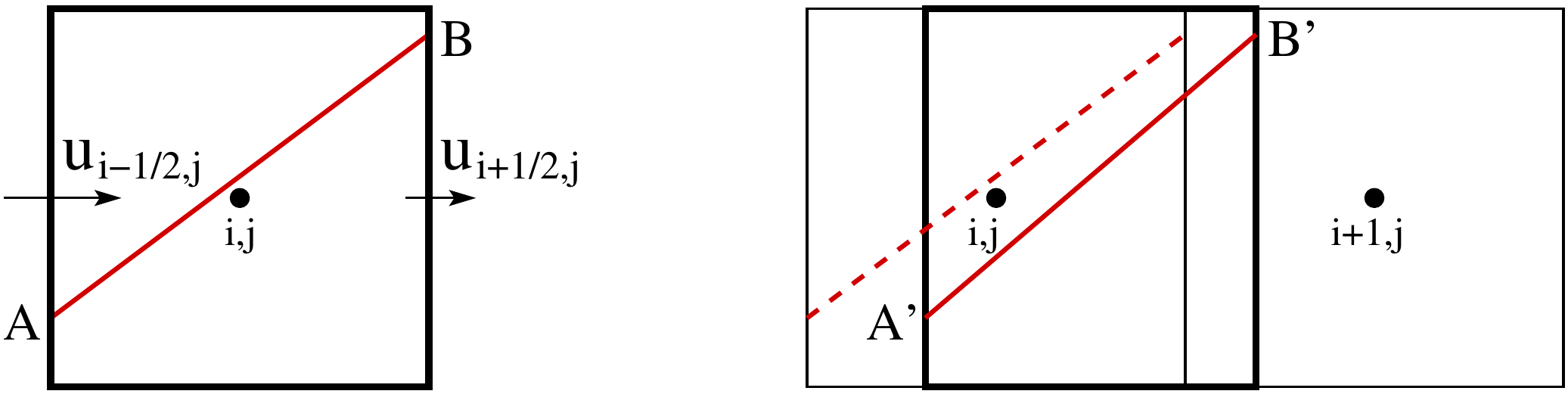} 
\end{center}
\caption[]{Advection of the interface along the $x$-direction: standard 
``Lagrangian Explicit'' or ``CIAM'' method with the cell centered at grid point 
$i,j$ being advected and expanded/compressed by the flow, 
in this case it is compressed since $\partial u / \partial x < 0$;
before advection (left) and after advection (right).}
\label{vofx}
\end{figure}

\begin{figure}[]
\includegraphics[width=4.5in]{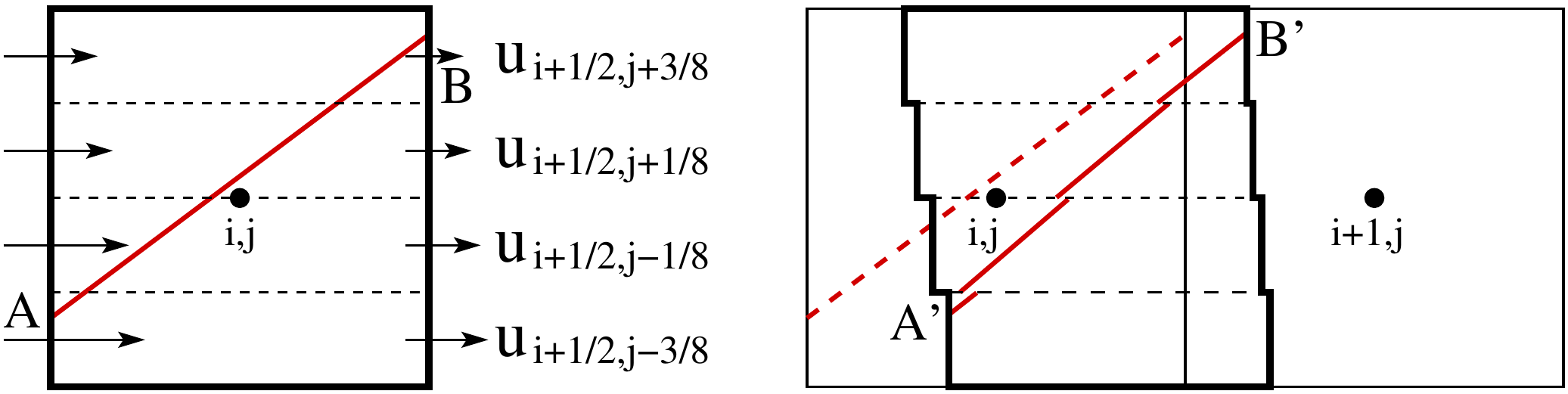}
\caption[]{Advection of the interface along the $x$-direction
with the cell subdivided in $m=4$ equal bands: 
with ``banding'' the effect of shearing, that is a 
$\partial u / \partial y$ derivative, is taken into account more accurately  
than in the standard method;
before advection (left) and after advection (right).}
\label{vofx2}
\end{figure}

{It is useful to describe the procedure used near the contact line in the main 
distribution of the Gerris code, which we used in the computations reported below.} 
First, without giving the full details that can be found in the references, 
let us outline the procedure used away from the contact line. There, the VOF 
discretization of Eq.~(\ref{eq:vof2}) consists in the definition of a variable 
$C_{i,j}$ on each grid point $i,j$ that is equal to the volume fraction of the 
reference fluid, fluid ``1'', in the cell. The VOF method proceeds in two steps, 
first the reconstruction of the interface followed by its advection. In the 
first part of the reconstruction step, the interface normal $\N=(n_x,n_y)$ in 
cell $i,j$ is determined from the values $C_{i,j}$ in neighboring cells, using
the ``Height-Function'' method described below (see e.g.~\cite{AB2008,Popinet2009}) or
the ``mixed Youngs-centered'' (MYC) method (see e.g.~\cite{Tryggvason11}), if the former method fails. 
In the second part of the reconstruction step, the position of a linear segment 
representing the interface in the cell is determined using elementary
geometry (see \cite{Tryggvason11}) from the knowledge of $\N$ and $C_{i,j}$.
Thus the equation of the segment is written
\be
n_x x + n_y y = \alpha, \label{aeq}
\nd
where the scalar $\alpha$ characterizes the position of the interface. 
The knowledge of $\N$ and $\alpha$ is then used in turn in the second, 
advection step, where the interface is displaced by the fluid velocity field.
On Fig.~\ref{vofx}, the standard ``Lagrangian-Explicit'' (see 
\cite{Tryggvason11}) advection step is represented. It is useful to describe 
the advection process in some detail. The collocated velocities  
$u_{i,j}, v_{i,j}$, defined on cell centers, are used to compute an auxiliary 
set of velocities $u_{i+1/2,j}$ and $v_{i,j+1/2}$ on different cell face 
centers using a projection method. 
The determination of the motion of the piece of interface shown on 
Fig.~\ref{vofx} is identical to the ``Lagrangian Explicit''  or ``CIAM'' 
method \cite{Popinet2009,Tryggvason11}. However a recent implementation in  
{Gerris} uses a ``banded'' advection. In this approach, the cell is subdivided in 
$m$ equal bands, the default being $m=4$ as on Fig.~\ref{vofx2}, and 
the advection is performed separately in each band. After that, the bands 
in the cells are aggregated to produce the final volume fraction. 
This ensures a better representation of shearing or rotating velocity fields, 
while volume conservation is enforced by the requirement that the average 
of the horizontal velocity in the bands, for example, on the right side of the cell 
in Fig.~\ref{vofx2} (left panel) to be equal to the face center velocity 
$u_{i+1/2,j}$. 

\begin{figure}[]
\begin{center}
\begin{tabular}{cc}
\includegraphics[width=2.25in]{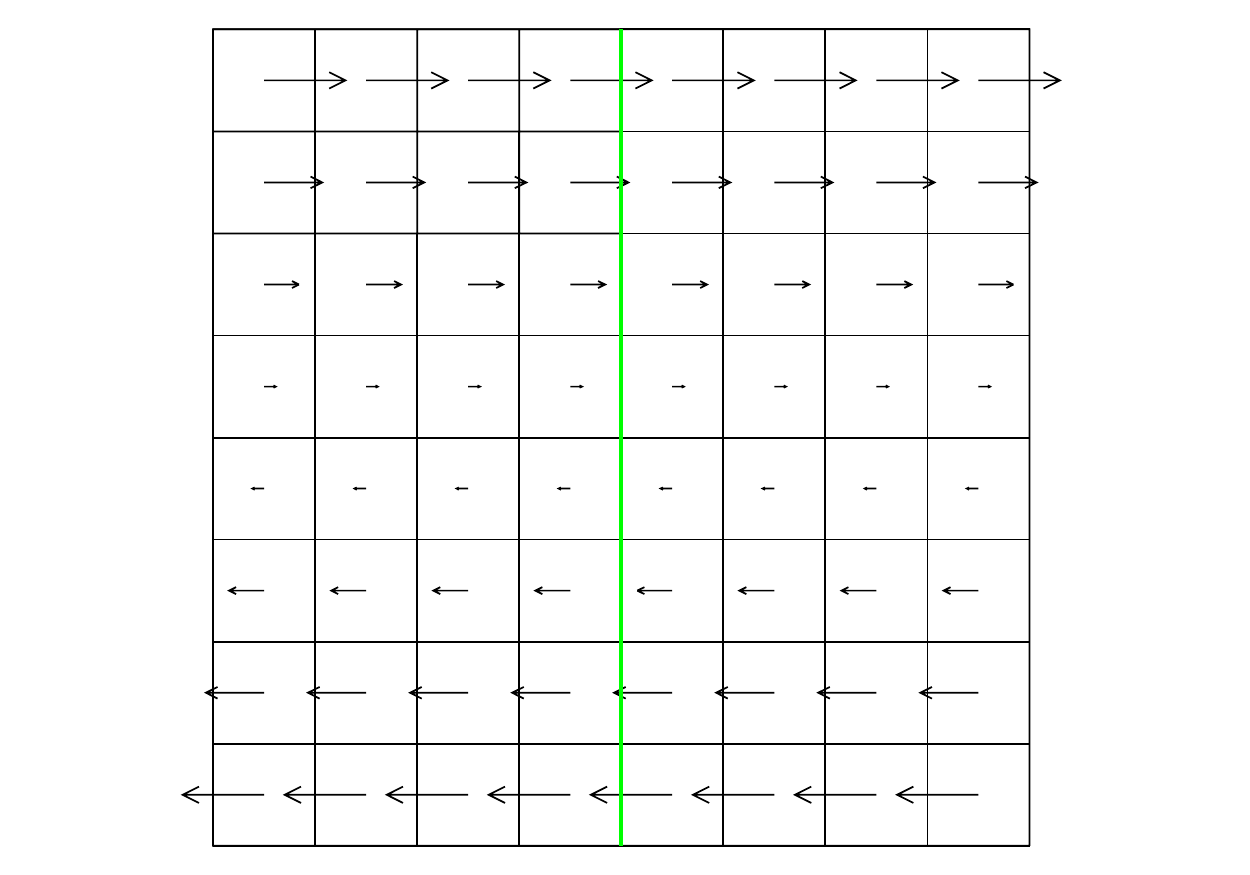} &
\includegraphics[width=2.25in]{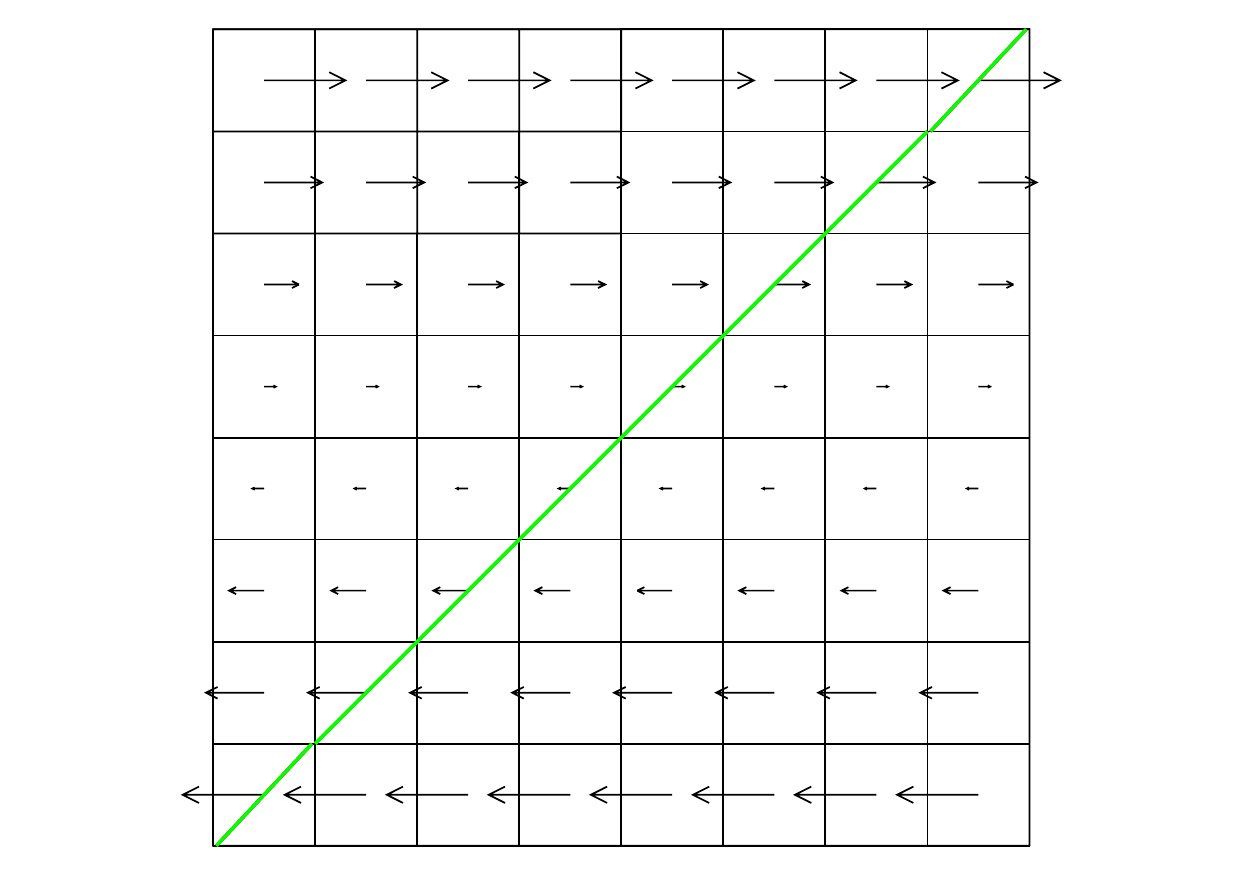}\\
(a)&(b)\\
\includegraphics[width=2.25in]{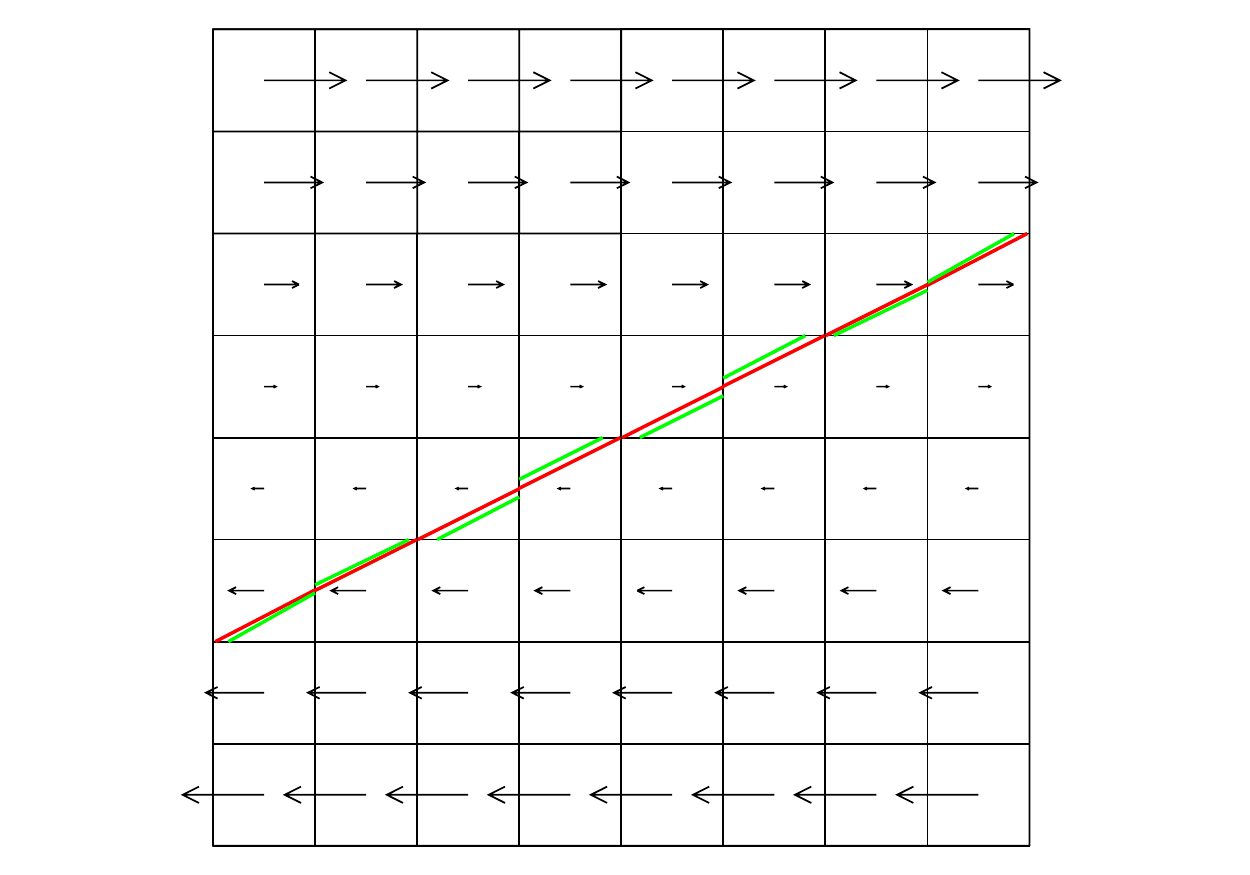} &
\includegraphics[width=2.25in]{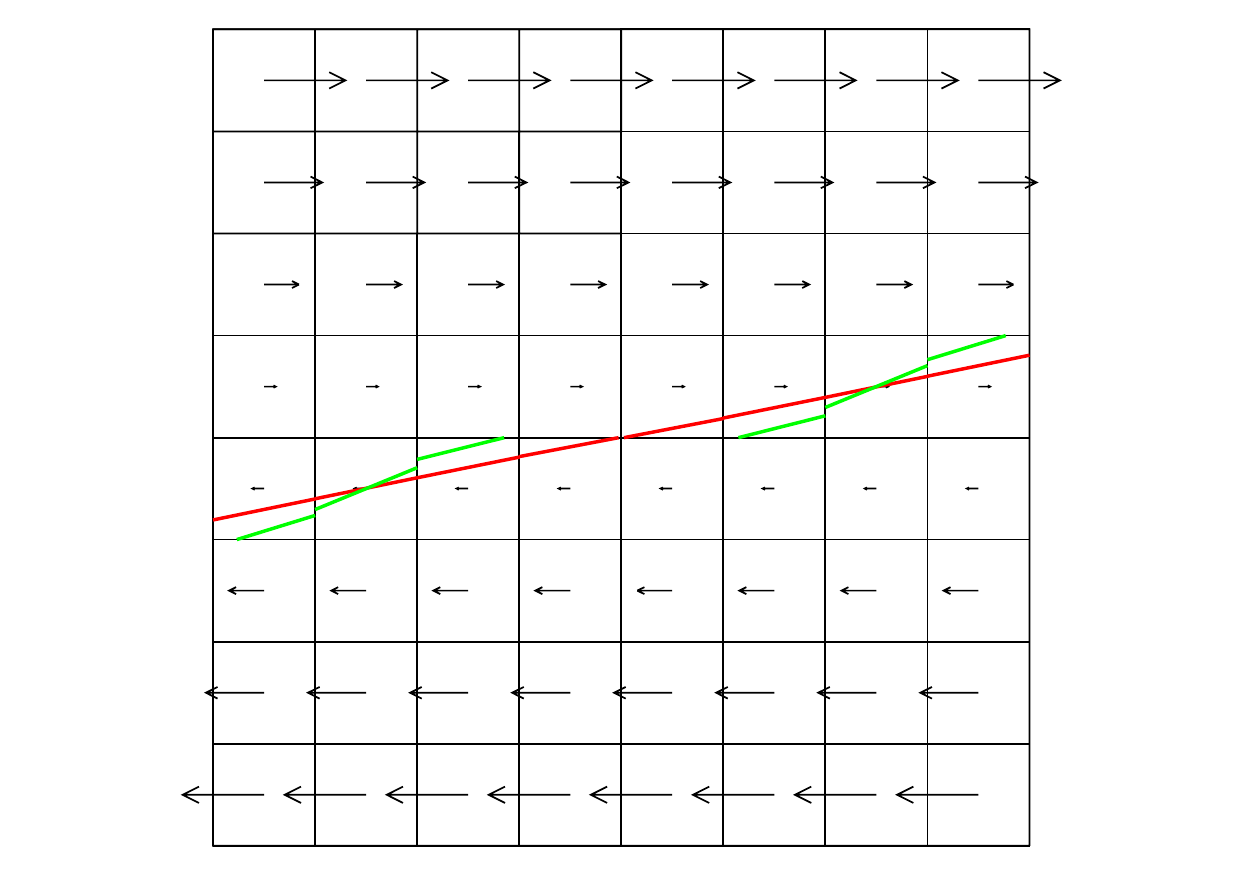}\\
(c)&(d) 
\end{tabular}
\end{center}
\caption[]{Evolution of the VOF interface in a shear flow at a shear rate of $1$. 
(a) $t=0$, (b) $t=1$, (c) $t=2$, and (d) $t=5$; for $m = 1$ ({\color{green}{\rule[0.5ex]{2mm}{.9pt}}}) and $m = 4$ ({\color{red}
{\rule[0.5ex]{2mm}{.9pt}}}). Note that in (d), the VOF reconstructions for $m=1$  
({\color{green}{\rule[0.5ex]{2mm}{.9pt}}})
in the center of the domain become identical to the black lines of the grid and hidden by them.}
\label{gerristest}
\end{figure}

\begin{figure}[]
\begin{center}
\includegraphics[width=3.5in]{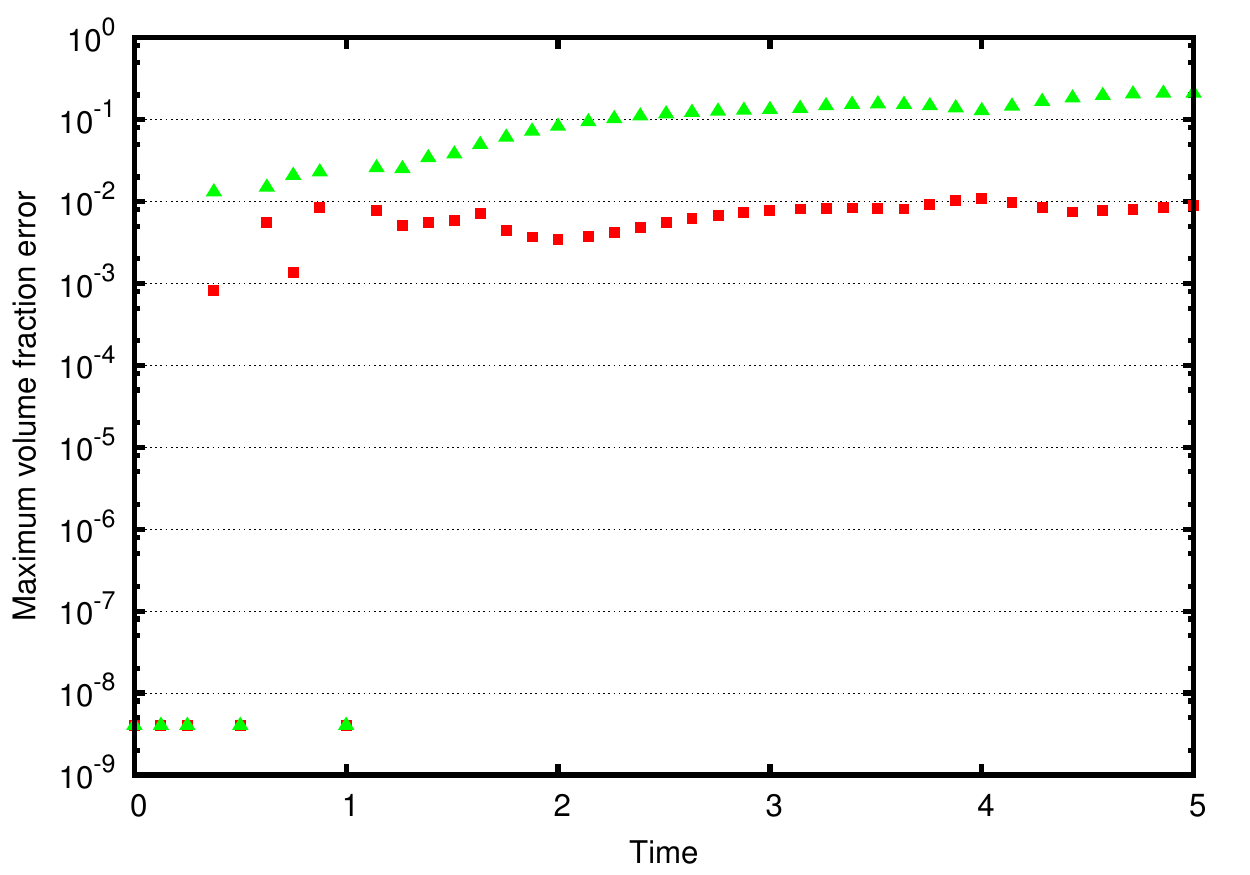}
\end{center}
\caption[]{Volume fraction error as a function of time; for $m=1$ ({\color{green}$\blacktriangle$}) and
                 $m=4$ ({\color{red}{\tiny $\blacksquare$}}).}
\label{gerristesterror}
\end{figure}

We next present a test case to clearly illustrate the improvement
when using the ``banded'' advection \cite{TestGerris}.  In this test case, a straight interface is advected
by a pure shear flow, at a shear rate of $1$, in a $1\times1$ computational domain.
The exact solution is simply a rotation of the interface around the 
center of the domain. We note that both the interface and the velocity field are described 
exactly by a second-order method. Fig.~\ref{gerristest} illustrates the rotation of the straight interface
under the shear flow. The green segments are the VOF reconstructed
interfaces obtained with $m=1$ and the red segments are when using 
a ``banded'' advection method with $m = 4$ bands.
Fig.~\ref{gerristesterror} illustrates the evolution of the error with time. 
For $t=1$, the interface is at $45$ degrees and the errors in fluxes cancel out exactly.

In order to compute capillary forces, we use the Height-Function method, 
in which the local height of the interface is computed from summing over 
a column of cells with 
$$
h_i = \sum_{j=j_1}^{j_2} C_{i,j}
$$
where $j_1$ is the index of the bottom cell in the column and $j_2$ the index 
of the top cell.  When the bottom cell is full ($C_{i,j_1} = 1$) and the top 
cell is empty ($C_{i,j_2} = 0$), and there is a single interface in the 
column, the height $h_i$ approaches the exact interface height to second order
\cite{Bornia11}. Using finite differences of the local Height-Function 
then provides the curvature, as well as the interface normals, used to compute
the surface tension force by the CSF method (see  \cite{Popinet2009,PopinetARFM}). 

The numerical method given in this paper has often been tested only
for regular grids. For example tests of the curvature estimation by
the height function method were performed only on circles or spheres
where the curvature is uniform. A uniform curvature is naturally
associated in these tests to a uniform grid. However, there are some
indirect tests, where near a singularity, a range of scales of curvature
appear, such as the pinching thread test in \cite{Popinet2009}. The
contact line flow is another such case where the singular flow near
the dynamic contact line forces a wide range of curvature. Tests of more complex
flows as performed in this paper are also useful to assess the accuracy of 
capillary force modeling on non-uniform quadtree grids.

\begin{figure}[]
\begin{center}
\begin{tabular}{cc}
\multicolumn{2}{l}{\includegraphics[width=5in]{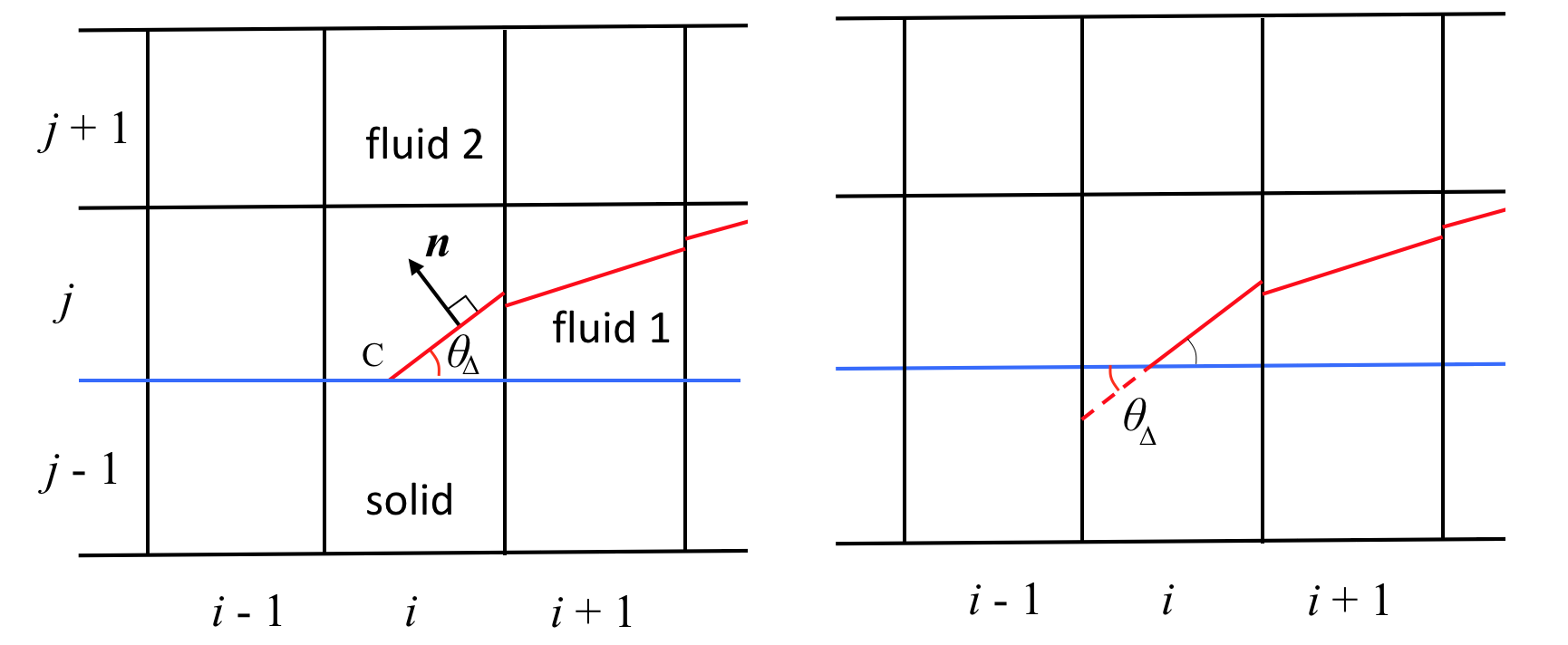} }\\
(a)&(b)\\
\multicolumn{2}{l}{\includegraphics[width=5in]{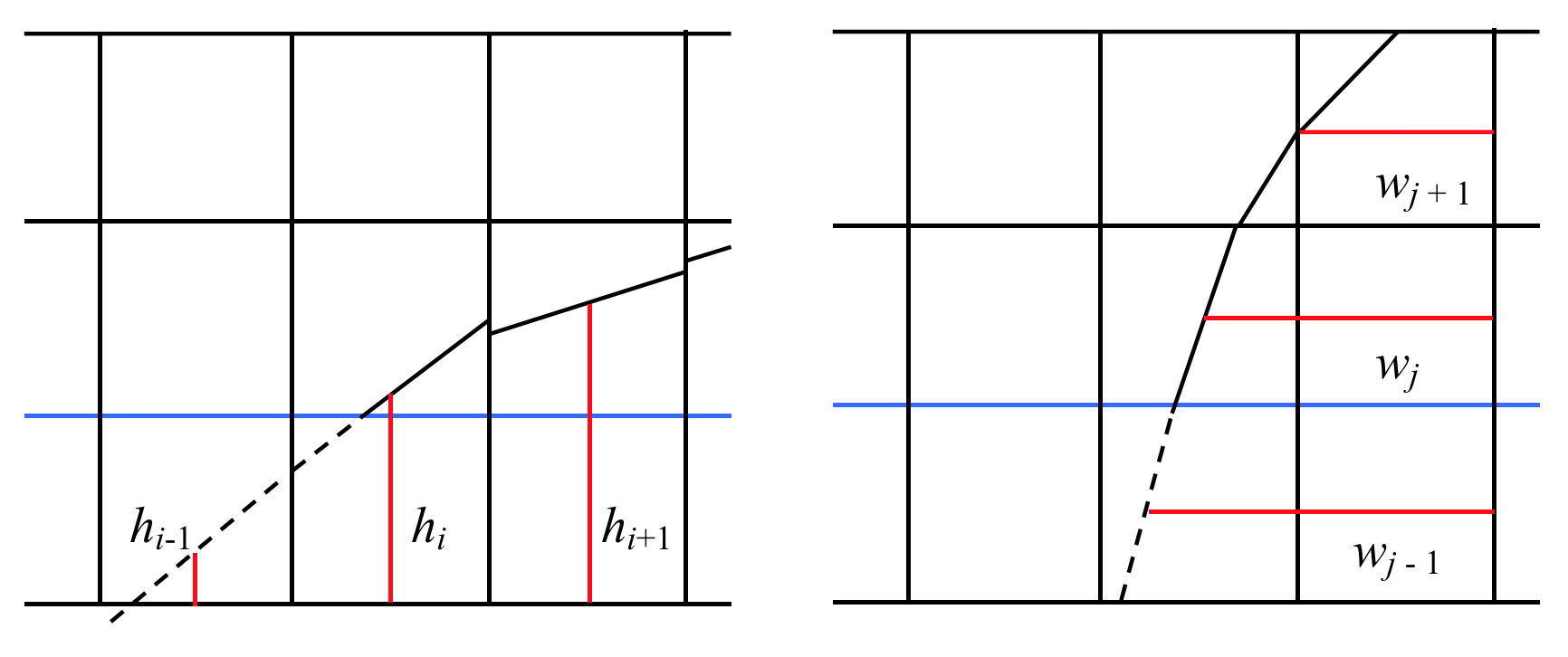} }\\
(c)&(d) 
\end{tabular}
\end{center}
\caption[]{Reconstruction of the interface and height functions in the 
vicinity of the contact line. (a) Reconstruction in cell $i,j$. (b) 
Linear extension to cell $i,j-1$. (c) Standard Height-Function method 
in columns $i$ and $i+1$ together with extrapolation of the height 
to cell $i-1$. (d) Same as (c) but with widths in horizontal segments 
$j-1$, $j$ and $j+1$.}
\label{vofcl}
\end{figure}

Near the contact line, we consider a cell $i,j$ containing the contact line C 
as shown on Fig.~\ref{vofcl}(a). The solid-fluid boundary 
is located exactly on the lower boundary of this cell. (Other locations for 
the solid boundary have not been explored by the authors.) For such a computational cell, 
the normal vector is recovered 
trivially from the specified contact angle $\theta_\Delta$ as
$n_x = - \sin(\theta_\Delta)$, $n_y = \cos(\theta_\Delta)$. The value 
of $\alpha$ in Eq.~(\ref{aeq}) is then obtained using elementary geometry. 
The interface can then be linearly extended into the solid cell $i,j-1$ 
as shown by the dashed line on Fig.~\ref{vofcl}(b). It is assumed there is 
no other contact line in the immediate neighborhood. The computation of 
the normal in cell $i+1,j$ is not immediately possible 
but this difficulty is easily circumvented by assigning ``ghost'' 
$C_{i,j}$ values to the cells in the first solid layer $j-1$.

Then as shown on  Fig.~\ref{vofcl}(c), the heights $h_i$ and $h_{i+1}$ can 
be reconstructed in columns $i$ and $i+1$. To construct the height $h_{i-1}$ 
in the column beyond the contact line a first-order extrapolation of the form 
$h_{i-1} = h_i - \tan (\theta_\Delta) \Delta$ is used, where $\Delta$ is the 
local grid size. 
\forget{Or higher-order extrapolations are used: first the approximate
position of the contact line $x_c$ is obtained from Eq.~(\ref{aeq}), 
second the height is fitted by a polynomial so that 
$h(x) = a x^2 + b x + c$, cells are located at $x = i \Delta$ and 
the coefficients $a,b$ and $c$ are computed from the constraints 
$h_i =  a x_i^2 + b x_i + c$, $h_{i+1} =  a x_{i+1}^2 + b x_{i+1} + c$ 
and $\tan(\theta) = h'(x_c)=2 a x_c + b$. Then $h_{i-1}$ can be obtained, 
although it is not really necessary since the heights are used to computed 
the curvature, which is in this case directly available from the polynomial 
fit.}

The heights over the three columns $i-1,i,i+1$ can be used to compute
the curvature in the cell containing the contact line C. This is a good 
approximation when both the interface slope and its curvature are small
enough. Alternatively, it is possible to fit a parabolic approximation
of the interface through the two heights $h_i$ and $h_{i+1}$ and the
contact point C of Fig.~\ref{vofcl}(a) computed from Eq.~(\ref{aeq}).
\forget{
In all the developments above, the Height-Function is only available over 
the three columns $i-1,i,i+1$ when the interface is of sufficiently small 
slope and the curvature is small enough.}

For vertical interfaces one uses ``widths'' $w_j$ instead, as shown on 
Fig.~\ref{vofcl}(d). Finally when neither heights nor widths are available, 
several other strategies are used to compute the curvature, as outlined by 
Popinet \cite{Popinet2009}, using either a mixture of heights and widths 
(the so-called mixed-heights method) or if the mixed-heights method fails,  
a polynomial fit to the mid points of the segments in each cell. 
These alternative strategies can be adapted to the vicinity of the contact 
line, provided extrapolations of the volume fraction $C_{i,j}$ and slope 
in cells $i-1,j-1$ and $i,j-1$ are used. It is important to remark that
the Height-Function method for curvature (without mixed heights) always 
provides a curvature for $\kappa \Delta$ small enough. The critical value 
of $\kappa\Delta$ was estimated independently by one of the authors 
through tests on a large number of random circles showing that the minimum 
value of $\kappa \Delta$ at which the Height-Function always works is about 
$0.06$. 

Once the interface positions and the curvature are computed, there is no 
special difficulty in computing the velocity field using the standard methods. 
No special provision is made for the discontinuity of velocities or the 
divergence of viscous stresses and pressures, which are computed as elsewhere 
in the domain using finite volumes and finite differences. 

We note however that the {Gerris} code uses the staggered (face) velocities to 
advect interface pieces, and therefore, in the contact line cell C on Fig.~\ref{vofcl},
the tangential (face) velocities used for advecting the interface in that cell will not be 
equal to the solid velocity, since they are defined half-a-cell above the solid boundary 
which is at $j-1/2$ on Fig.~\ref{vofcl}.
Intuitively, this allows a kind of effective slip. 
However the amount of slip is reduced by the banding method described above 
and shown in Fig.~\ref{vofx2}. 

It is interesting to note that at the scale $\Delta$ of the cell $i,j$ 
the interface and the velocity field are represented in a coarsely averaged 
manner that is very far from capturing the flow reversal expected inside 
the fluid 1 wedge (although on scales larger than $\Delta$ this flow is well 
captured). Whether it is possible or desirable to have a more sophisticated 
discretization approach near the contact line is beyond the scope of this paper. 

\section{Results: transition to film formation in forced dewetting}\label{sec:results}

\begin{figure}[]
\begin{center}
\includegraphics[width=2.75in]{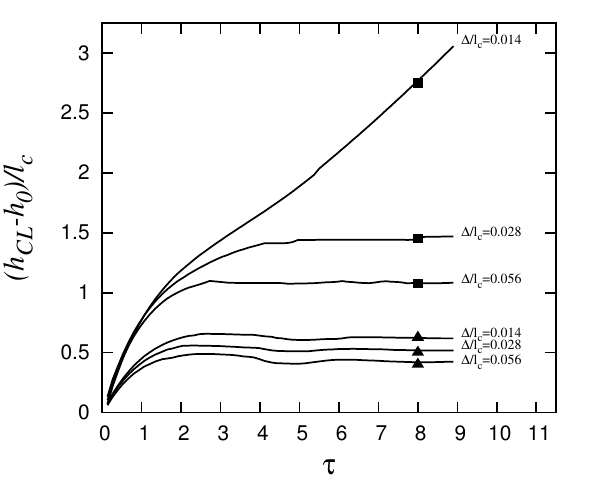}
\end{center}
\caption[]{Contact line height as a function of time
                 for $\theta_{\rm_\Delta}=90^\circ$ ($\blacktriangle$) 
                 and $\theta_{\rm_\Delta}=60^\circ$ ({\tiny $\blacksquare$}) at various $\Delta/l_c$.
                 For $\Delta/l_c=0.014$ and $\theta_{\rm_\Delta}=60^\circ$, the contact line
                 elevation continues increasing as $\tau$ increases. In this figure, $\Ca=0.03$.}
  \label{fig:3}
\end{figure}

We focus on the problem of a partially wetting substrate withdrawn from a 
liquid reservoir. We find two parameter ranges: first, the stationary regime, for which the
contact line motion along the substrate can evolve to a steady state,
as  $\tau\to\infty$, where the nondimensional time is $\tau = V_s t/l_c$; 
second, the unsteady regime for which a steady state solution cannot 
be found and the contact line height continues to increase, covering
the substrate by a film. All the results presented in this section are for Setup A,
unless stated otherwise. 
We begin by presenting various scenarios characterized by different nondimensional
grid sizes, $\Delta/l_c$, and the imposed contact angle, $\theta_{\rm_\Delta}$. 
Fig.~\ref{fig:3} shows the instantaneous contact line height, $h_{CL}$, 
from the reference height, $h_0$, nondimensionalized by $l_c$. 
Fig.~\ref{fig:3} shows that depending on $\Delta/l_c$, different equilibrium configurations can be
obtained. Also, it shows that when the contact angle is decreased, the contact line
is raised to a new equilibrium height for the large grid sizes while a stationary
meniscus cannot be achieved for the smallest grid size.    
{Here we take $0.007<\Delta/l_c<0.057$. We note that
for the capillary numbers that we consider (from $0.001$ to $0.1$),
it is very difficult to have a larger range of grid sizes since one needs to satisfy
$\Delta \ll \ell \ll l_c \ll L$, where $\ell$ is the thickness of the film at the transition, 
of the order $\ell = \Ca^{2/3} l_c$. We  however note that 
the {Gerris} code uses quadtree grids (or octree in 3D), that allow 
to refine the grid where necessary. This is a very useful feature for 
dynamical contact line problems as it allows to use a very small grid size 
$\Delta$ in the immediate neighborhood of 
the contact line and a larger grid size elsewhere.} 

\begin{figure}[]
\begin{center}
\begin{tabular}{cc}
\includegraphics[width=2.35in]{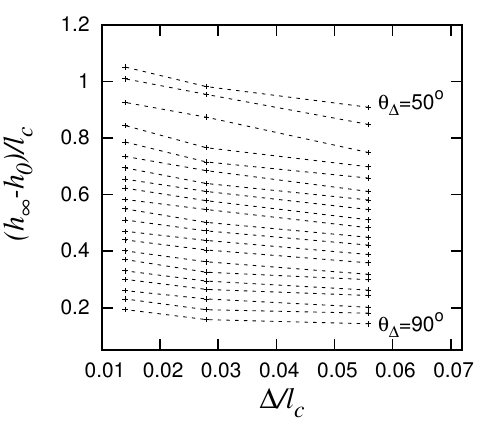} &\includegraphics[width=2.35in]{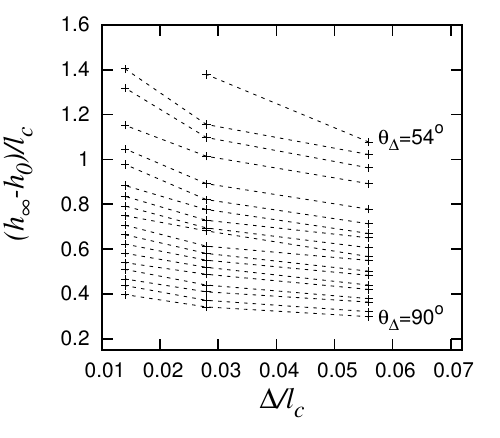}\\
(a)&(b)\\
\includegraphics[width=2.35in]{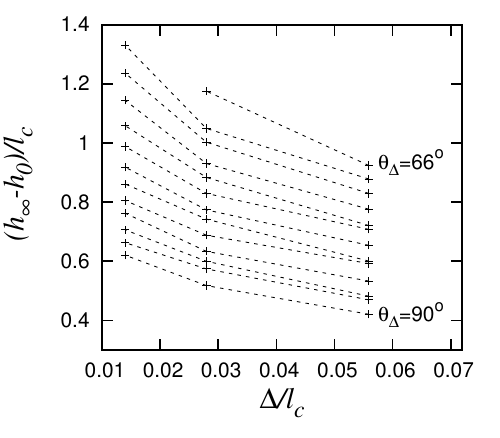} &\includegraphics[width=2.35in]{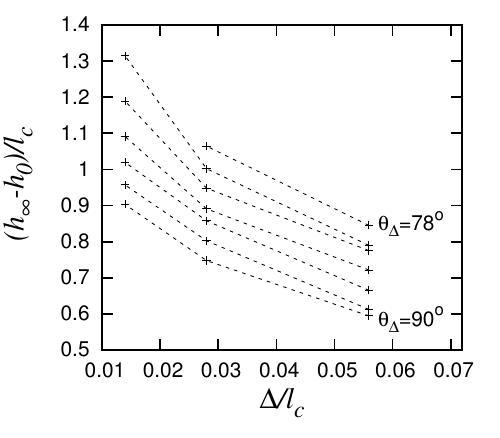}\\
(c)&(d)
\end{tabular}
\end{center}
\caption[]{Nondimensional stationary contact line height, $ (h_{\infty} - h_0)/l_c$,  as a
function of the nondimensional mesh size, $\Delta/l_c$, for 
(a) $\Ca=0.01$, (b) $\Ca=0.02$, (c) $\Ca=0.03$, and (d) $\Ca=0.04$,
for various contact angles, $\theta_{\rm\Delta}$; the contact angle difference between each set is $2^\circ$.
For ($\Ca$,$\theta_{\rm \Delta}$)=($0.02$,$54^\circ$), ($0.03$,$66^\circ$), 
and ($0.04$,$78^\circ$), no steady state menisci can be attained when $\Delta/l_c\le0.014$.}
  \label{fig:4}
\end{figure}

Next, we elaborate on the stationary state results in detail.
Figs.~\ref{fig:4}(a)-(d) show the nondimensional 
stationary contact line height $(h_{\infty} - h_0)/l_c$,  
where $h_{\infty}=h_{CL}(\tau\to\infty)$,
as a function of the nondimensional mesh size, $\Delta/l_c$, 
for various capillary numbers
and for various contact angles, $50^\circ\le\theta_{\rm_\Delta}\le 90^\circ$.
The results show the grid sizes, $\Delta/l_c$, 
where a stationary meniscus forms and no film is deposited on the substrate,
for the range of considered $\theta_{\rm_\Delta}$. 
\begin{figure}[]
\begin{tabular}{ccc}
\includegraphics[width=2.35in]{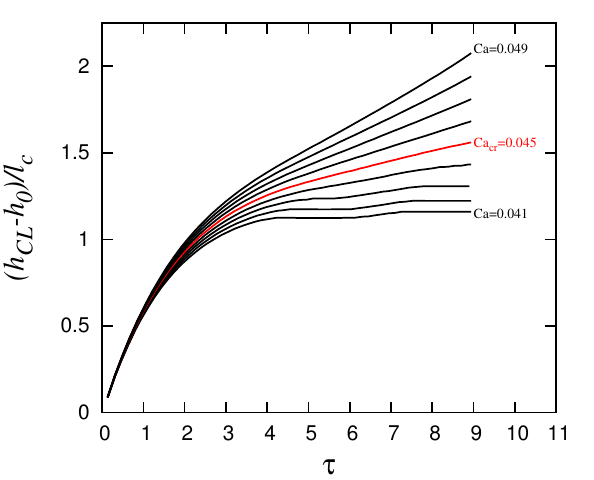} & 
\includegraphics[width=2.35in]{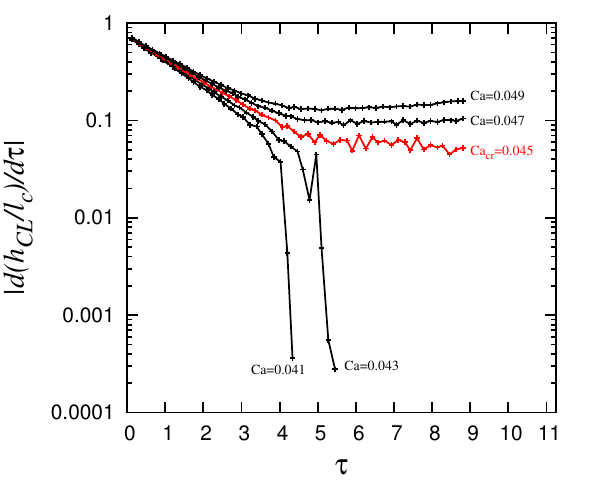}\\
(a)&(b)
\end{tabular}
\caption[]{(a) Contact line height as function of time. The increment 
                 in $\Ca$ is 0.01 and the red line indicates the critical capillary number ${\rm Ca}_{cr}=0.045$.
                 (b) Contact line velocity relative to the substrate velocity as a function of time   
                 At the critical capillary number ${\rm Ca}_{cr}=0.045$, depicted in red, the transition occurs.
                 $\theta_{\rm \Delta}=90^\circ$ and $\Delta/l_c=0.007$.
                 }
  \label{fig:5}
\end{figure}

More importantly, the results show that the computed height of the contact line 
is a function of $\Delta/l_c$ and that for small enough $\theta_{\rm \Delta}$
no steady state menisci can be attained when $\Delta/l_c\le0.014$.
This lower limit of $\theta_{\rm \Delta}$, for which steady state contact
lines can be achieved, gets larger as $\Ca$ is increased.  
The results clearly depend on the chosen value for the smaller mesh size
near the contact line, an effect that is expected and will be explained below. 
This dependence of the results with mesh refinement becomes more marked
as $\Ca$ is increased or $\theta_{\rm \Delta}$ is decreased.

Next we present the results, for which the contact line continues 
to move upward and a liquid film is then deposited
on the substrate.  We can understand the onset of film deposition, i.e.~the forced dewetting 
transition, as when the balance between the surface tension
and viscous forces close to the contact line region can no longer hold, resulting in wetting failure.
(We note that gravity is only involved, asymptotically, in the outermost region.)
At the transition, however, we will need to allow the computations to run for a very long
time in order to determine when the contact line motion along the wall 
cannot reach a stationary state.
For efficient and accurate determination of
the numerical values of the transition $\Ca_{cr}$, we propose a procedure as follows. 
Fig.~\ref{fig:5}(a) shows 
contact line heights as a function of time for various values of $\Ca$, when 
$\theta_{\rm \Delta}=90^\circ$ and $\Delta/l_c=0.007$.
As shown, for sufficiently small $\Ca$, a stationary meniscus can 
be reached while for large $\Ca$, the contact line height keeps increasing. 
We then use the information in Fig.~\ref{fig:5}(a) to obtain
Fig.~\ref{fig:5}(b).  We then pick the 
transition capillary number, for which the relative velocity of the 
contact line, $|d(h_{CL}/l_c)/d\tau|$, does not go to zero as a function of $\tau$.
This critical capillary number ${\rm Ca}_{cr}$ is depicted in red in 
Figs.~\ref{fig:5}(a) and (b). 
Using the procedure above, we can therefore determine ${\rm Ca}_{cr}$ 
with  a very good precision.
\begin{figure}[]
\begin{center}
\begin{tabular}{cc}
\includegraphics[width=2.35in]{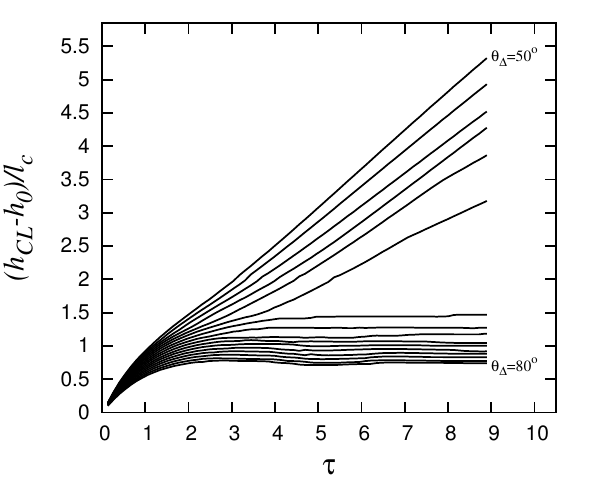}&
\includegraphics[width=2.35in]{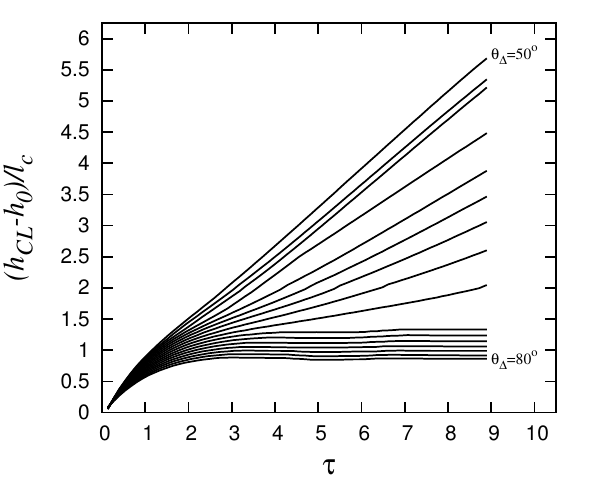}\\
(a)&(b)
\end{tabular}
\end{center}
\caption[]{Contact line height as function of time
                 for $50^\circ\le\theta_{\rm \Delta}\le80^\circ$ and (a) $\Delta/l_c=0.028$ and 
                 (b) $\Delta/l_c=0.014$; the increment in $\theta_{\rm \Delta}$ is $2^\circ$.  $\Ca=0.03$
                 }
  \label{fig:6}
\end{figure}
Figs.~\ref{fig:6}(a)-(b) show the contact line height, $ (h_{CL} - h_0)/l_c$,  as a
function of nondimensional time, $\tau$, for two mesh sizes, $\Delta/l_c$, when varying the
wall contact angle $\theta_{\rm \Delta}$, for a fixed $\Ca$. 
As shown, the transition from a stationary meniscus 
not only depends on $\theta_{\rm \Delta}$, but also on $\Delta/l_c$. 
As illustrated, the transition occurs at a larger
$\theta_{\rm \Delta}$ for smaller $\Delta/l_c$.
This observation begs a further exploration of how the critical capillary number 
depends on the contact angle and the grid size. We study these effects in what follows.

\begin{figure}[]
\begin{center}
\begin{tabular}{cc}
\includegraphics[width=3.in, angle=90, trim=30mm 65mm 0 70mm]{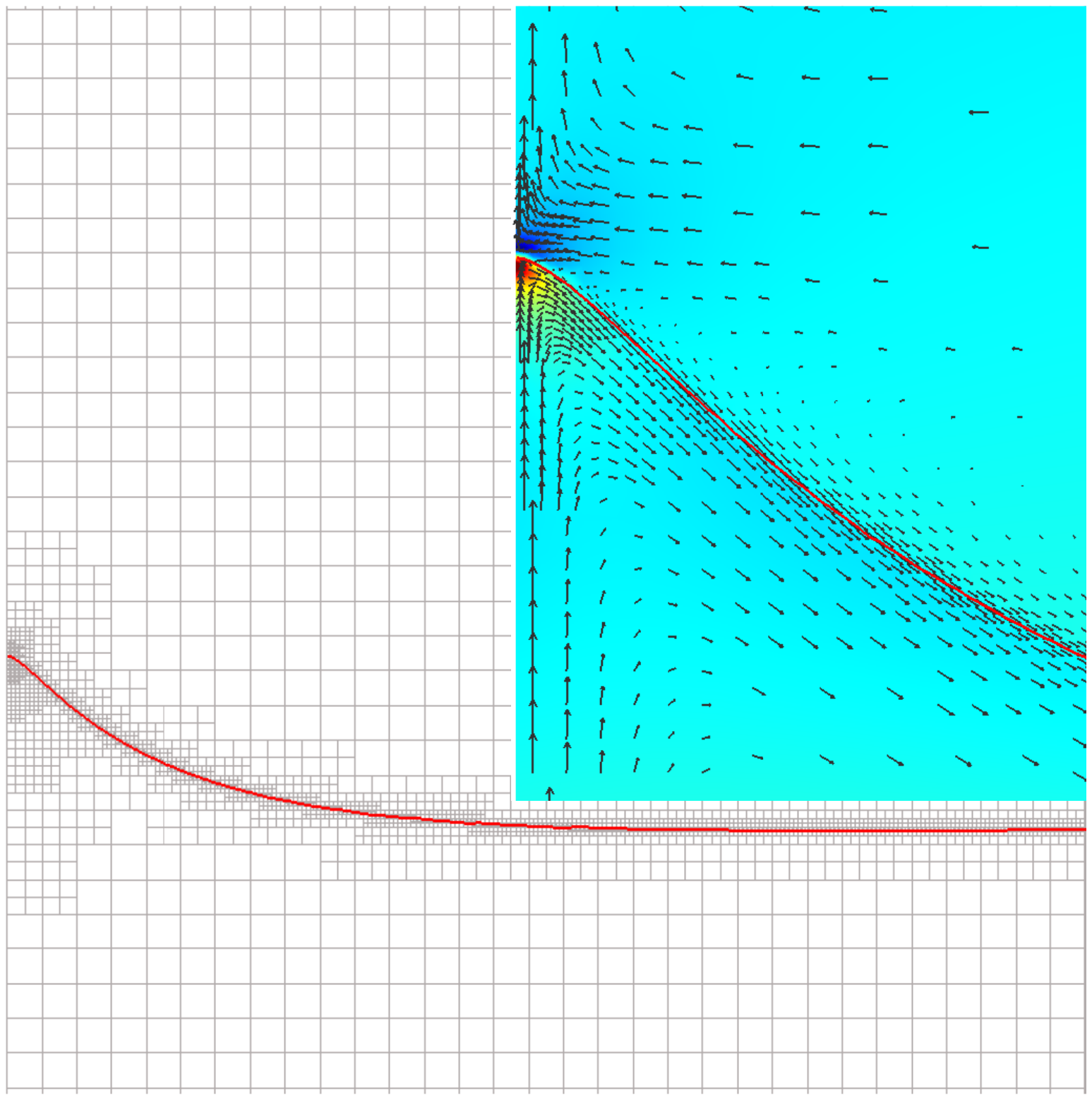}&
\includegraphics[width=2.2in]{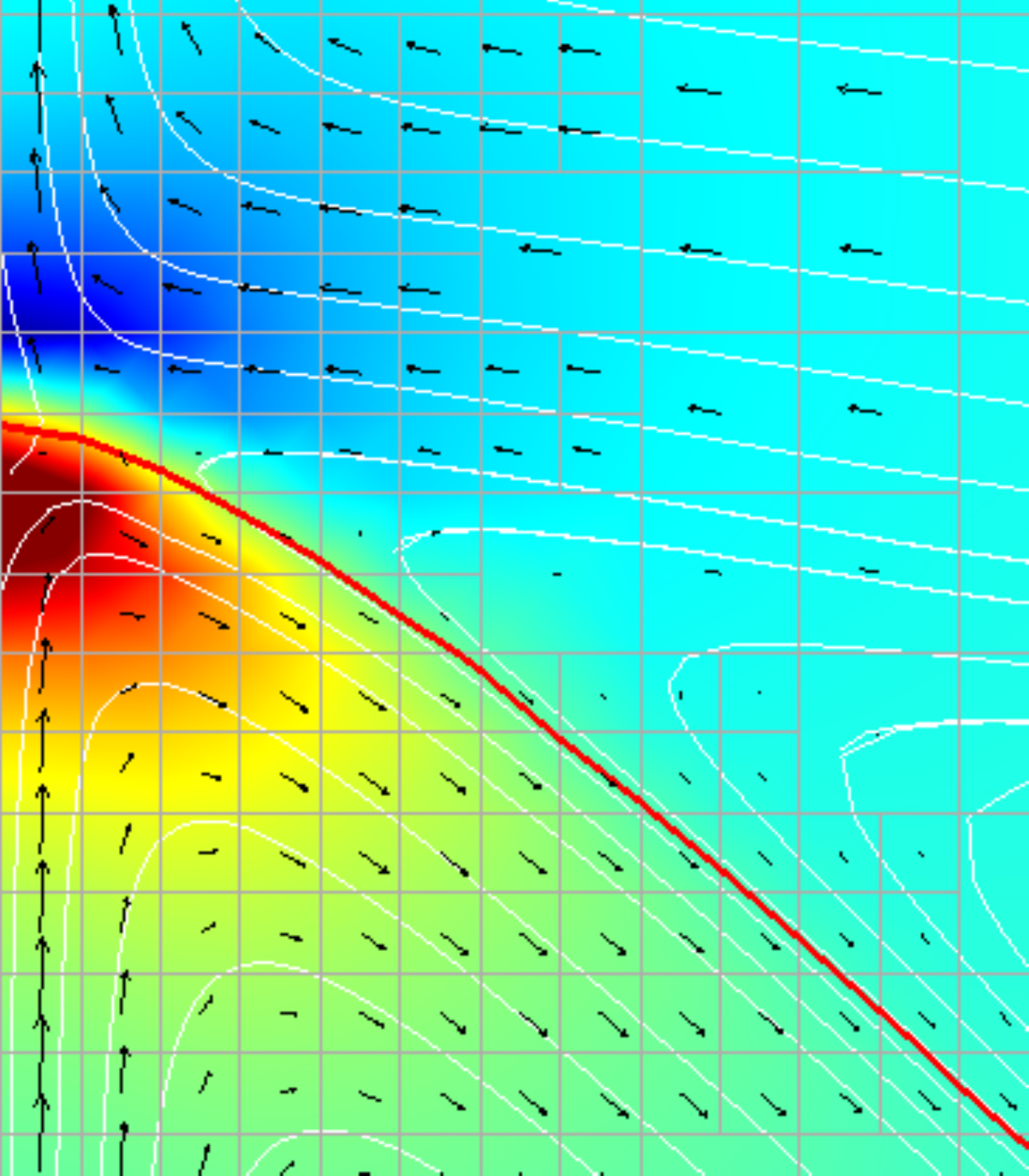}\\
(a)&(b)
\end{tabular}
\end{center}
\caption[]{(a) A stationary meniscus forms when $\Ca<\mbox{Ca}_{cr}$ as $\tau\to\infty$.
                The inset shows the magnified flow field and the pressure distribution. 
                $\Ca=0.043$, $\theta_{\rm \Delta}=90^\circ$, 
                and $\Delta/l_c=0.014$. (b) A magnified view of the contact line region and the computational mesh;
                The fine structure of  the  flow field and the pressure distribution in the contact line region
                are illustrated. The pressure colors show the maximum (dark red) and minimum (dark blue)
                of the pressure distribution.}
  \label{fig:7}
\end{figure}

To shed more light on the transition mechanism, we analyze the flow, for both
when a stationary meniscus forms and when a steady state contact line cannot
be attained. Fig.~\ref{fig:7}(a) provides an example of a stationary meniscus for
$\Ca=0.043$, $\theta_{\rm \Delta}=90^\circ$, and $\Delta/l_c=0.014$ 
(for this set of parameters, $\mbox{Ca}_{cr}=0.52$).
The inset shows the magnified flow field and the pressure distribution.
Fig.~\ref{fig:7}(b) shows a magnified view of the contact line region and the computational mesh.
The fine structure of  the  flow field and the pressure distribution in the contact line region
are illustrated.
As shown, large gradients of velocity and pressure necessitate
a high mesh resolution around the contact line region. 
As illustrated, the interface is highly curved close to the contact line, 
leading to an intensified
pressure gradient around that region, while the pressure gradient remains weak outside the
vicinity of the contact line,
leading to gentle bending of the interface away from the contact line.

\begin{figure}[]
\begin{center}
\begin{tabular}{c}
\includegraphics[width=2.05in, trim=30mm 60mm 30mm 40mm,angle=90]{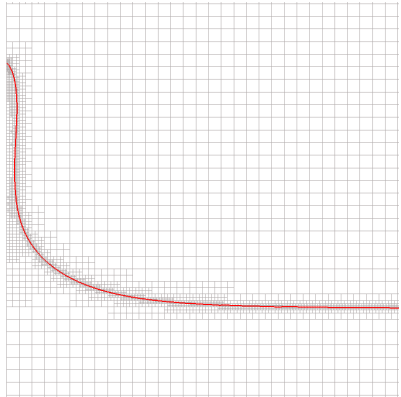}\\
(d)\\
\includegraphics[width=2.05in, trim=30mm 60mm 30mm 40mm,angle=90]{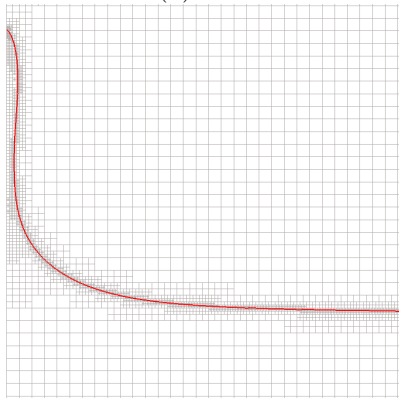}\\
(e)\\
\includegraphics[width=2.05in, trim=30mm 60mm 30mm 40mm,angle=90]{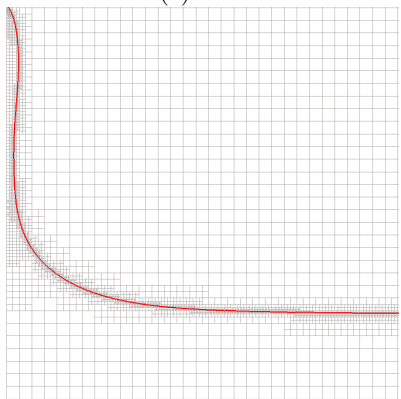}\\
(f)
\end{tabular}
\end{center}
\caption[]{Time evolution of the interface for $\Ca>\mbox{Ca}_{cr}$ at
                $\tau=$ 4 (a), 4.8 (b), 5.9 (c), 6.8 (d),  7.9 (e), 8.7 (f).
                The insets show the magnified flow field and the pressure distribution.
                (a)-(c) Right panels show a magnified view of the contact line region and the computational mesh;
                The fine structure of  the  flow field and the pressure distribution in the contact line region
                are illustrated. $\Ca=0.048$, $\theta_{\rm \Delta}=60^\circ$, and $\Delta/l_c=0.014$.
                The pressure colors show the maximum (dark red) and minimum (dark blue)
                of the pressure distribution.
                 }
  \label{fig:8}
\end{figure}

Fig.~\ref{fig:8} shows an example of when the contact line cannot attain a steady state, 
leading to the formation of a film deposited on the substrate, for
$\Ca=0.048$, $\theta_{\rm \Delta}=60^\circ$, and $\Delta/l_c=0.014$ 
(for this set of parameters, $\mbox{Ca}_{cr}=0.024$).
The figure shows a typical evolution of the interface and the transition to the film. Figs.~\ref{fig:8}(d)-(f) 
also show the sagging of the interface behind the contact line after the film formation.  
The insets of Figs.~\ref{fig:8}(a)-(c) show the magnified flow field and the pressure distribution. 
Figs.~\ref{fig:8}(a)-(c) also show a further
magnified view of the contact line region and the flow streamlines.
The fine structure of  the  flow field and the pressure distribution in the contact line region
are illustrated. The results reveal the strong pressure gradients close to the contact
line region; when capillary forces can no longer balance this strong 
pressure gradient, the consequence is the wetting failure. 
\begin{figure}[]
\vspace{-10mm}
\begin{center}
\begin{tabular}{ccc}
\includegraphics[width=2.5in, angle=90, trim=30mm 65mm 0 70mm]{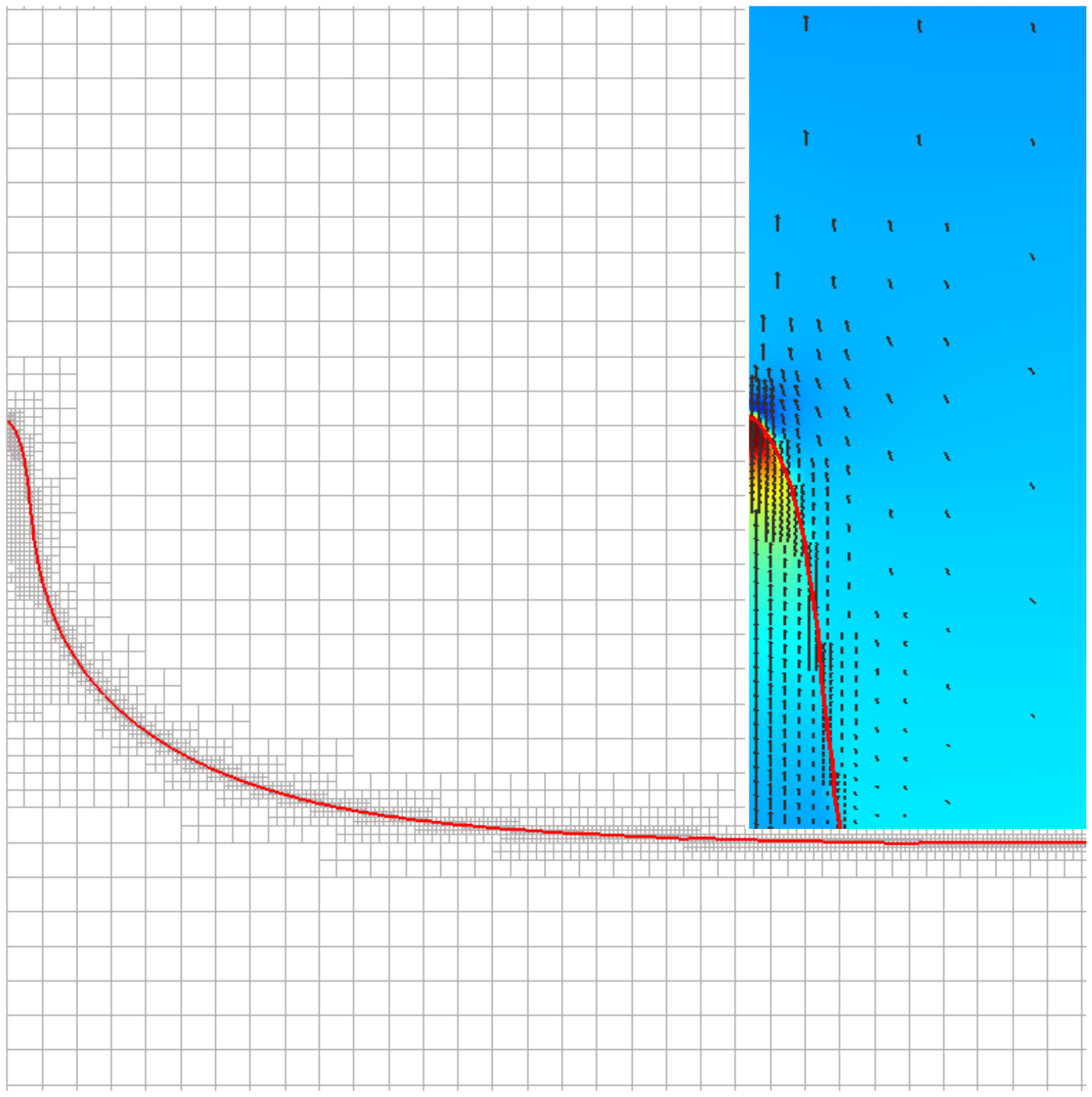}&
\includegraphics[width=1.65in]{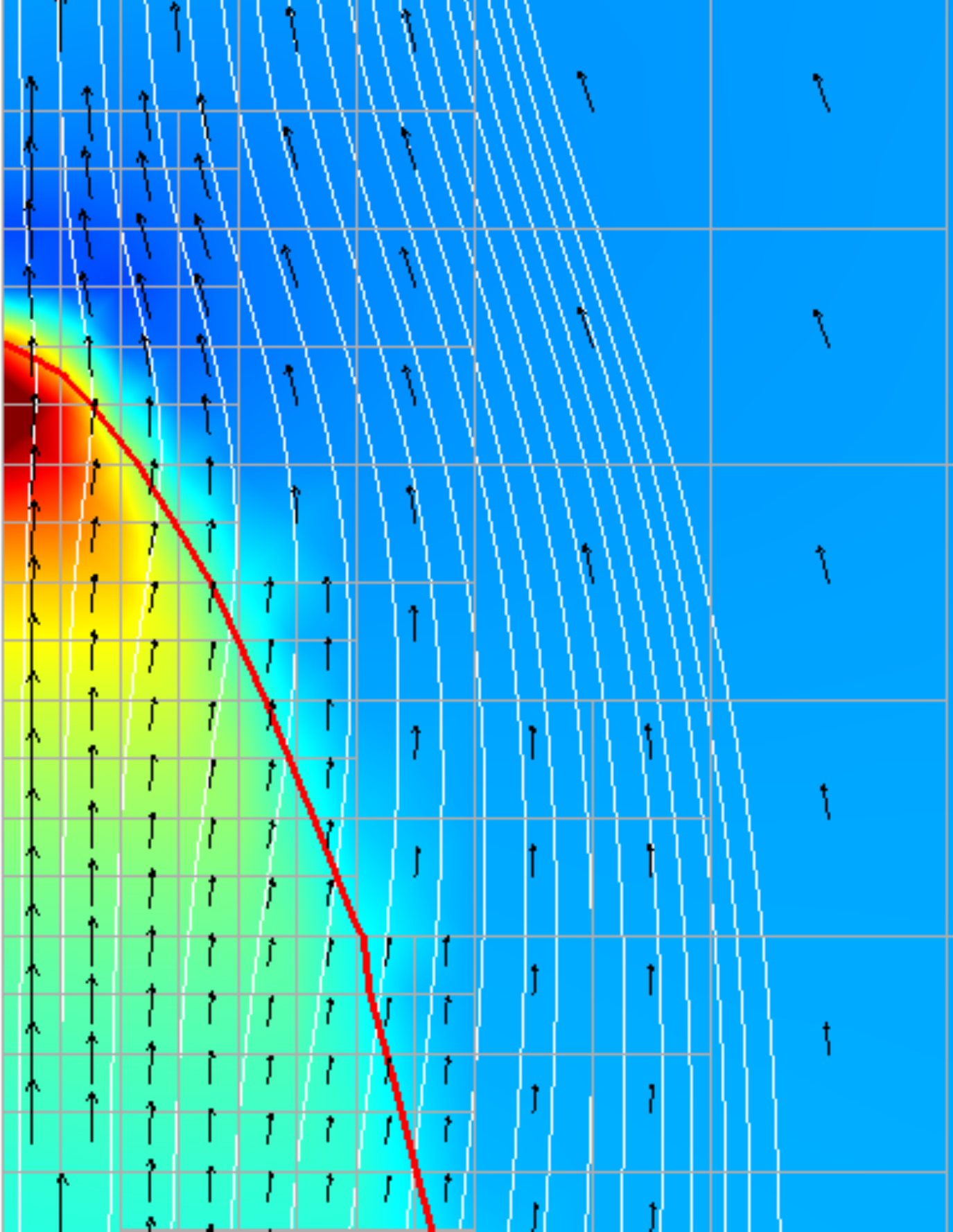}\\
(a)\\
\includegraphics[width=2.5in, angle=90, trim=30mm 65mm 0 70mm]{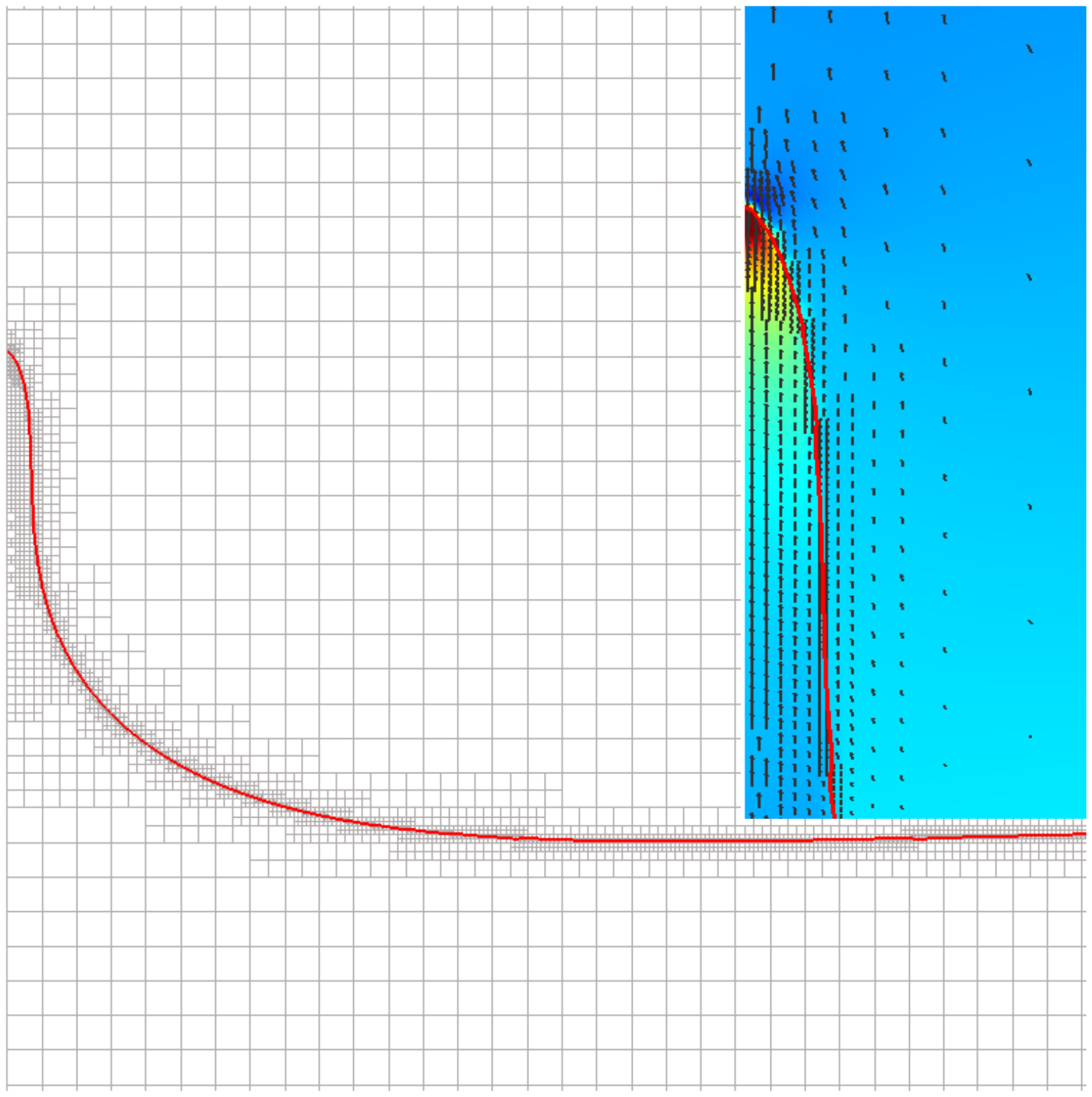}&
\includegraphics[width=1.65in]{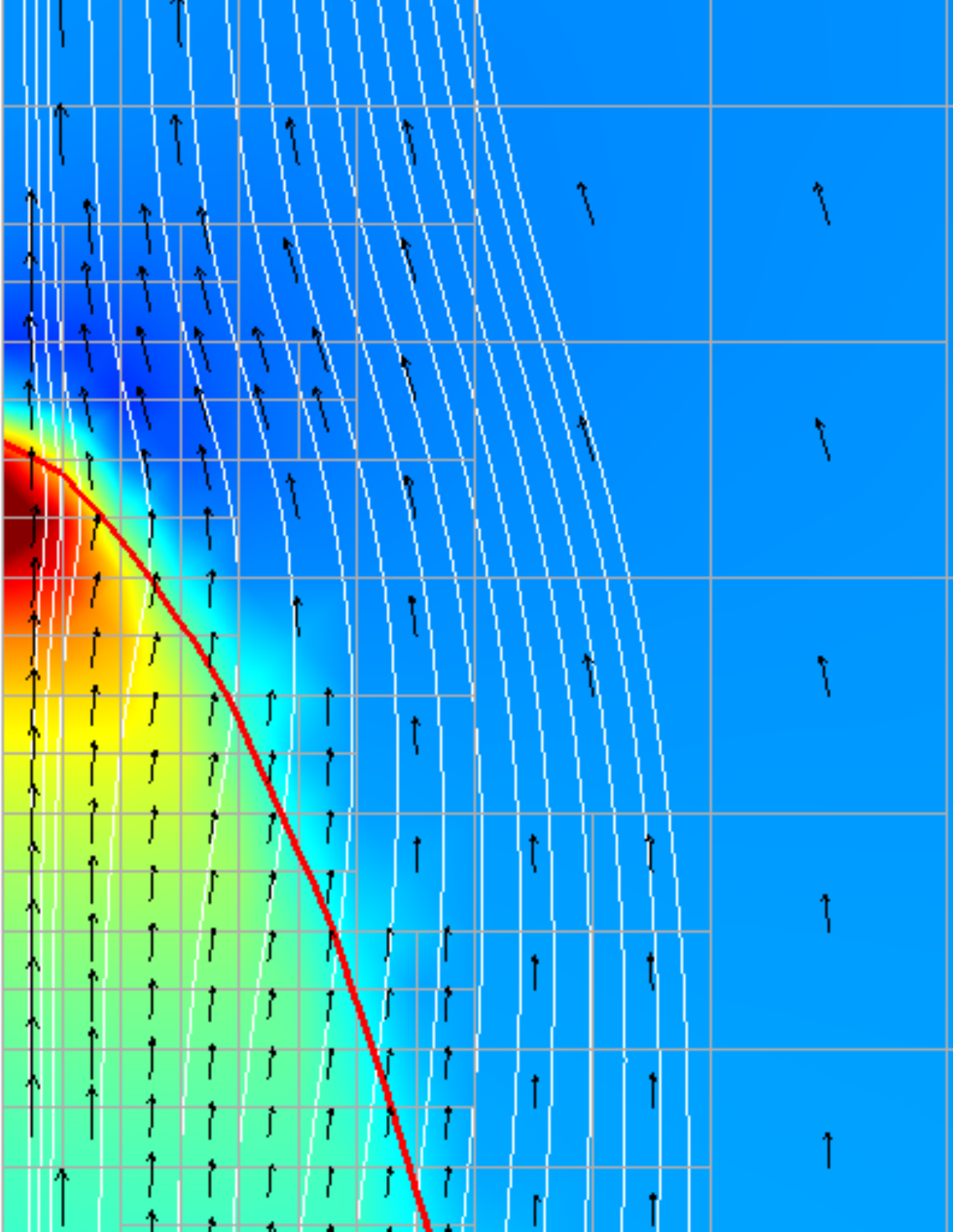}\\
(b)\\
\includegraphics[width=2.5in, angle=90, trim=30mm 65mm 0 70mm]{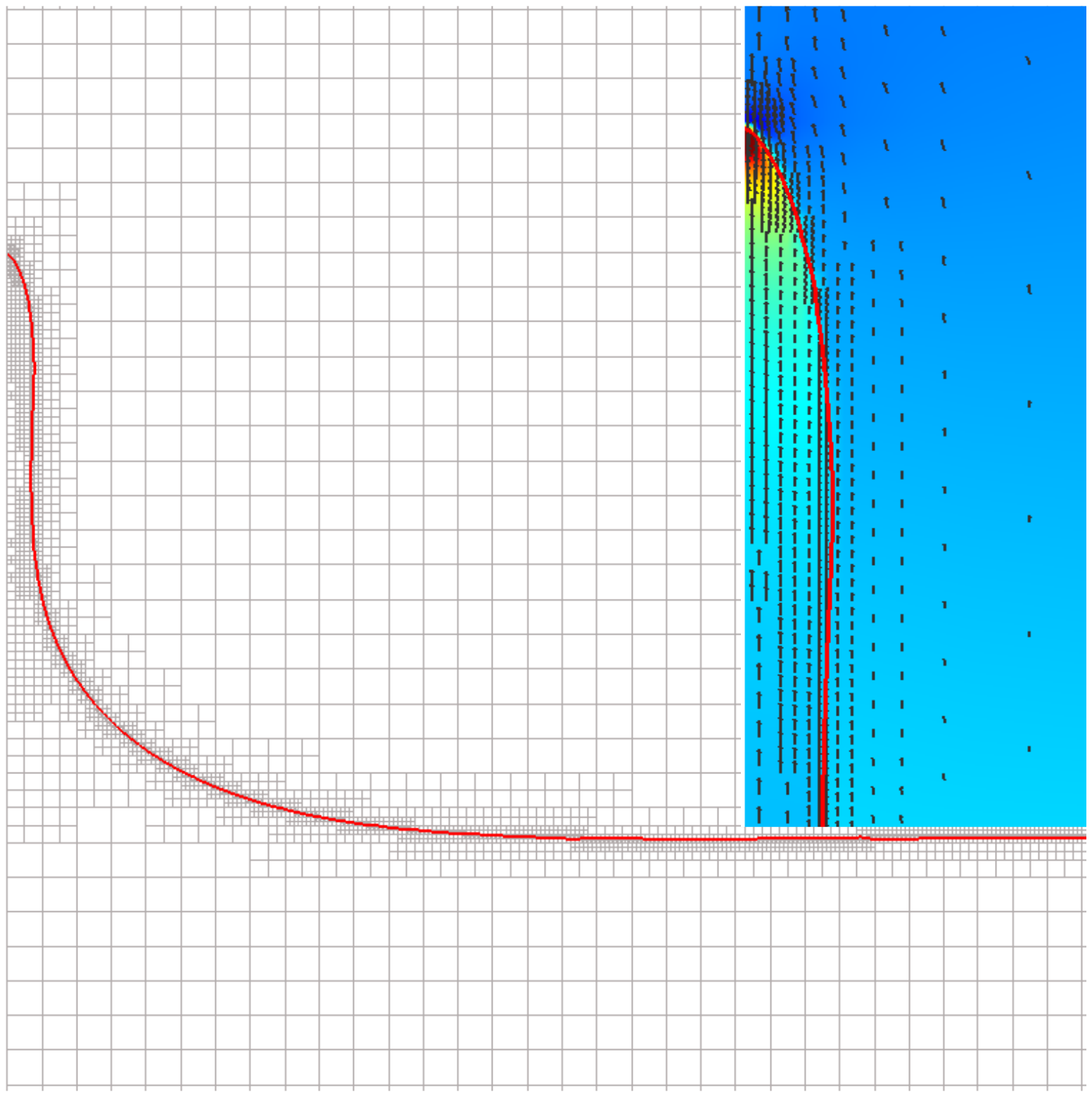}&
\includegraphics[width=1.65in]{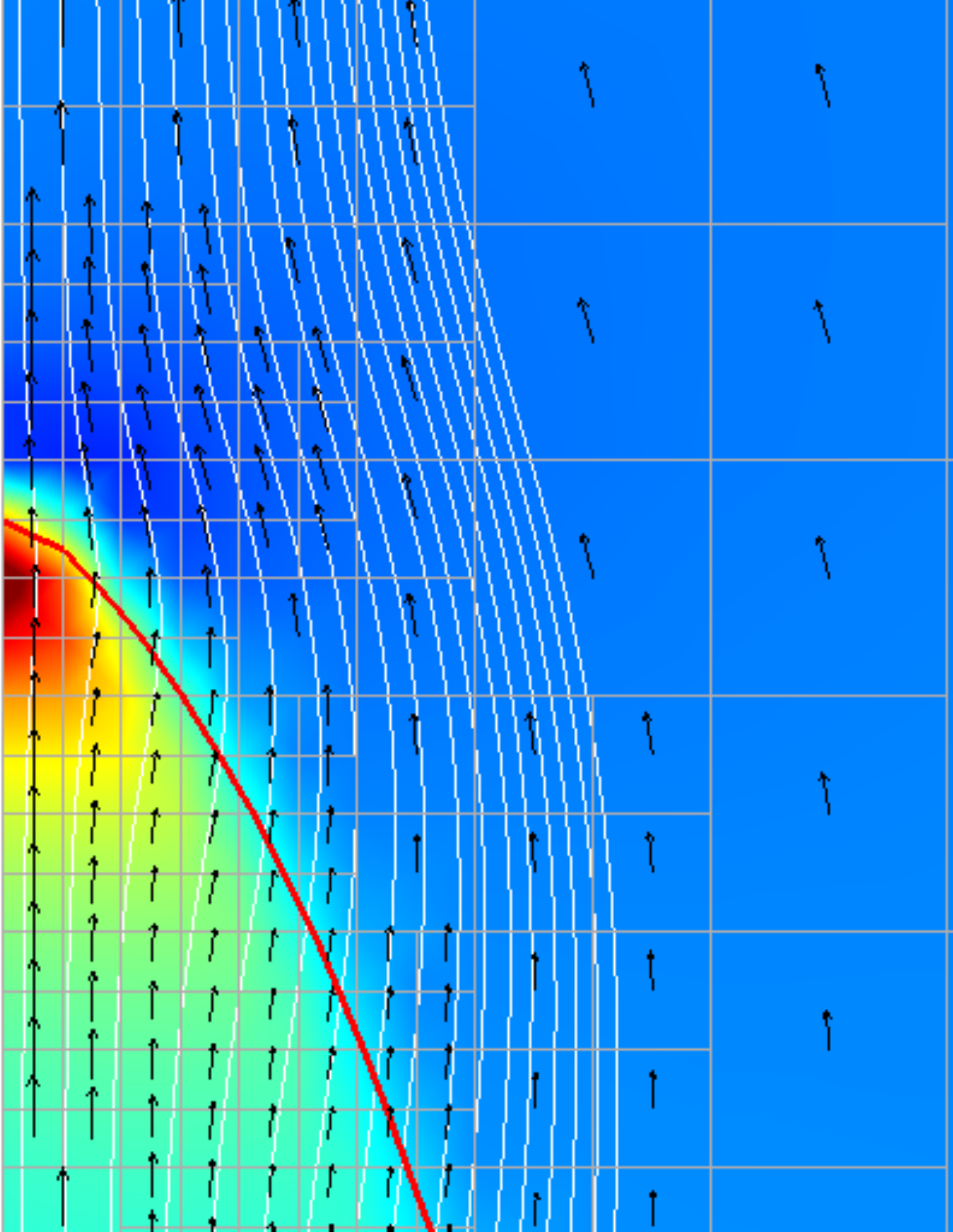}\\
(c)
\end{tabular}
\end{center}
\end{figure}
\begin{figure}[]
\begin{center}
\includegraphics[width=2.75in]{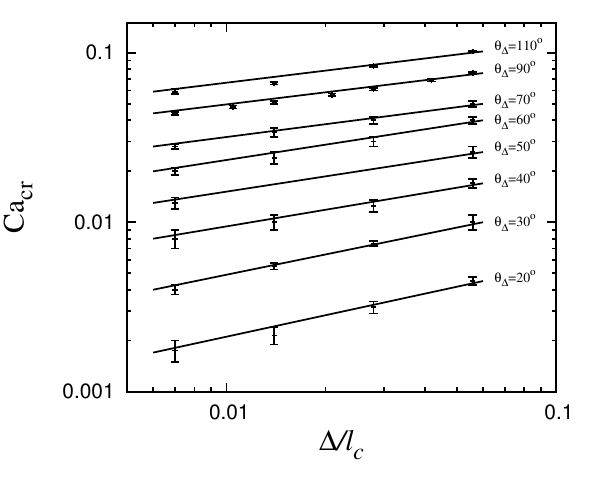}\\
(a)\\
\includegraphics[width=2.75in]{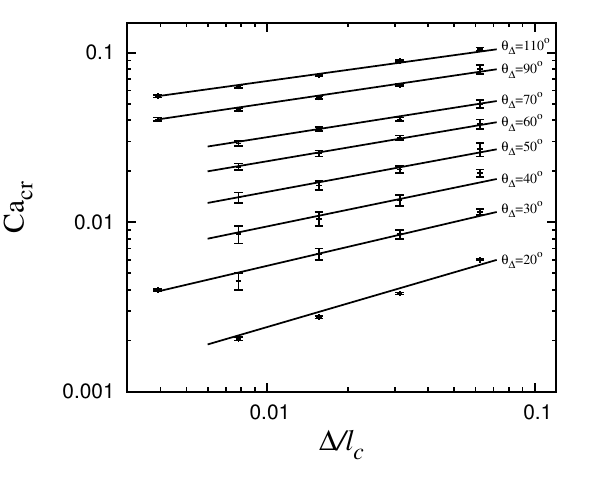}\\
(b)\\
\includegraphics[width=2.75in]{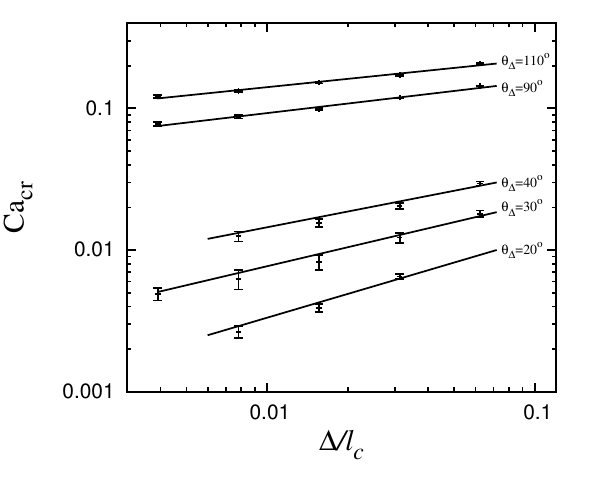}\\
(c)
\end{center}
\caption[]{${\rm Ca}_{cr}$ as a function of $\Delta/l_c$
for a range of $\theta_{\rm \Delta}$ for Setups A (a), B (b), and C (c).
Symbols present the numerical results and the solid lines are
drawn to guide the eye. The error bars represent the increment in the
capillary number in the procedure of finding ${\rm Ca}_{cr}$. 
}
\label{fig:9}
\end{figure}

We now turn our attention to computing the critical capillary number, ${\rm Ca}_{cr}$,
of the dewetting transition, as a function of $\Delta$, $\theta_\Delta$, and $q$.
We carry out a series of simulations when incrementing the capillary number, 
holding $\Delta$ and $\theta_\Delta$ fixed, until we find ${\rm Ca}_{cr}$.
We repeat the procedure for Setups A, B, and C in Tab.~\ref{tab1}.
We note that the smallest mesh size we choose here is $\Delta/l_c \approx 0.004$; 
such an unprecedented fine grid is only made possible owing to the adaptive
mesh refinement capability of  the {Gerris} code, that allows focusing the 
grid refinement at around the interface and in particular the contact line region
while the grid size is free to adapt elsewhere.
Fig.~\ref{fig:9} presents ${\rm Ca}_{cr}$ as a function of $\Delta/l_c$
for a range of $\theta_{\rm \Delta}$ for Setups A, B, and C. The symbols are direct simulation results and 
the solid lines are drawn to guide the eye through a set of points corresponding to
the same  $\theta_{\rm \Delta}$. The error bars represent the increment in the
capillary number in the procedure of finding ${\rm Ca}_{cr}$.
These results illustrate the logarithmic dependence of the dewetting transition 
on the mesh size for all the ranges of the contact angles that are
considered here for various Setups. Moreover, as expected, for large capillary
numbers, Setup A and B behave more or less the same. In addition, we show
the effect of the viscosity ratio on computed ${\rm Ca}_{cr}$; in general, lowering
the viscosity of fluid 2 results in an increase in ${\rm Ca}_{cr}$ compared to when 
the viscosity of both fluids is matched, with this effect being more prominent at
large contact angles. As shown, a decrease in contact angle promotes the dewetting
transition.     
In the next section,  we will
develop an improved understanding of the onset of this dewetting transition.  We
study the effects of the contact angle and the grid size on
$\mbox{Ca}_{cr}$ and give the scaling of it with $\Delta/l_c$.

\section{Hydrodynamic theories of the dynamic contact line and the dewetting transition}
\label{sec:theory}

\subsection{Generalities about the asymptotic description}
\label{sec-gen-theo}

To interpret our numerical results, we will use a 
theoretical framework that extends the work in \cite{Eggers2004b,Chan12}. The theory in \cite{Eggers2004b,Chan12}
is valid for small capillary number and equilibrium
contact angles, large viscosity and density ratios,
and several specific slip length models. Since our numerical approach is not based on a slip
length model, we cannot apply this theory directly. However, as we shall see, it is
straightforward to replace the slip length analysis, by an asymptotic matching, with the small length
scale behavior, using the theory of Cox~\cite{cox1986}.
 In the general physical case, it is also 
assumed that any microscopic physics with a sufficiently small length scale can be 
represented in Cox's theory. 

In the small $\Ca$ limit, the problem is analyzed by asymptotic matching of three or four different regions:

1) A region I near the contact line, where microscopic effects dominate in 
the physical reality and numerical effects dominate in our simulations. The size of that region 
is the microscopic scale $\lambda$.
We make no assumptions about region I 
except that it is limited in length scales to a microscopic scale $\lambda$
that remains much smaller than all other relevant length scales.

2) A region II at length scales large enough that continuum mechanics and the no-slip condition
on the wall hold, but close enough to the contact line that a special logarithmically-scaling solution
holds, as discussed by Cox in \cite{cox1986}. This region starts at lengths $\ell \gg \lambda$ and ends at lengths 
$\ell \ll \Ca^{1/3} l_c$. In that region, capillary forces are balanced by viscous drag forces.
The interface thus bends, 
and as shown below, it has a curvature proportional to $\Ca/r$ where $r$ is the distance to the contact line. 

3) An outer region III, where viscous effects are negligible and surface tension
balances gravity. This region scales like the capillary length $l_c$. 
The solution in region III is the famous static meniscus solution
\cite{landlif}. {We define the ``apparent contact angle'' $\theta_{\rm a}$  as the angle seen
in meniscus variables that is on scales such that $y\sim l_c$. The meniscus solution quadrature gives 
the curvature 
\be
\kappa_{\rm III} = l_c^{-1} \sqrt{2 - 2 \sin \theta_{\rm a}} \label{kappaIII}
\nd

The general properties of the solution can be seen as follows at small 
capillary numbers. 
At the overlap of region I and region II the angle is the microscopic angle noted $\theta_e$.
It is often assumed that it is the equilibrium contact angle, but this is true only if
the microscopic physics in region I are close to equilibrium. It is quite possible to assume that
$\theta_e$ differs from the equilibrium angle in a manner determined by experiments, or by microscopic theories
of the physics
beyond the scope of this paper. 
For small capillary numbers, the bending in region II is sufficiently small that 
the angle changes little from region I to region III and thus $\theta_{\rm a} = \theta_e$. This condition
closes the problem and gives in particular the height of the contact line 
$h_{cl} = l_c \sqrt{2 - 2 \sin \theta_{e}}$.
As the capillary number increases, the curvature and the bending increase in region II. 
The apparent contact angle decreases below $\theta_e$ while the height of the contact line increases. 

It can thus be hypothesized that the critical condition for dewetting
can be expressed as the condition of vanishing apparent contact angle. 
For a vertical plate, this means that the inner solution in region II  matches into
the outer solution in the meniscus region III with a vertical slope 
or yet that at the inflection point, the slope vanishes.
Thus a region IV, overlapping regions II and III is found in which the slope is small. 
This hypothesis is implicit in the works of  Derjaguin \cite{Derjaguin1943} and Landau and Levich \cite{LandauLevich1942}
since their assumption of a lubrication theory implies a near-vertical slope in the matching region, the existence
of region IV and a zero apparent contact angle. 
Moreover Landau and Levich \cite{LandauLevich1942} have shown that in region IV and at small 
capillary number, the film thickness behaves as $h \sim \Ca^{2/3} l_c$, the famed 
Landau-Levich film thickness. 

This hypothesis of transition at zero apparent contact angle
has been considered by Eggers in the small angle case \cite{Eggers2004b,Chan12}, 
but has not yet been investigated for large angles or when the free-surface model has to be replaced 
by an interfacial model at non-vanishing viscosity ratio $q$. 

The matched asymptotics allowing to connect the regions defined above are valid upon the hypothesis that
a parameter $\eps$ is small enough. This parameter is not simply $\Ca$ as the Landau-Levich film scaling
would seem to imply, but $\eps \sim \Ca/G(\theta_e)$ where $G(\theta_e)$ is the function introduced in \cite{cox1986}
and defined below. It can also
be shown that at the transition, an equivalent estimate of the small parameter is  $\eps \sim 1/\ln(l_c/r_m)$. 
This value of $\eps$ is not exceedingly small, which has consequences which will be spelled out below. 

}

\subsection{The theory of Cox}

We thus focus on the analysis of region II. 
In this region, the 
wedge solution of Huh \& Scriven is assumed \cite{HuhScriv}. It is then
possible to use it as done by Cox \cite{cox1986} to obtain the variation of
pressure in the wedge and thus by Laplace's law the variation of 
curvature. After integration, one may obtain the variation of
slope. 
Cox's solution in his ``intermediate region'' identified with our region II, 
from his Eqs.~(7.13) and (7.18) is 
\be
G[\theta(r)] = G(\theta_e) - \Ca \ln (r/\lambda_c)  - \Ca \frac{ Q_i }{f(\theta_e,q)}
+ o(\Ca) \label{eq:cox_eq_2_m},
\nd
where $\lambda_c$ is a characteristic scale for microscopic effects, 
$\theta_e$ is the equilibrium angle in agreement with our
assumption above for the $\Ca\to0$ limit, $f$ and $G$ are defined
below, and $q=\mu_2/\mu_1$ is the viscosity ratio. $Q_i$ is an integration constant
that is obtained by matching with region I and thus depends on region
I characteristics. 
Since we actually write the first order of an expansion in small $\Ca$, 
higher-order terms exist whose form is however unknown to the authors. 

\newcommand\romander[2]{\frac{{\rm d} #1}{{\rm d}#2}}

Finally the functions $f$ and $G$ are given by 
\begin{equation}
G(\theta) = \int_{0}^{\theta} \frac{d\theta'}{f(\theta',q)},
\label{eq:g}
\end{equation}
and
\begin{equation*}
f(\theta,q) = \frac{2\sin
\theta\{q^2(\theta^2-\sin^2\theta)+2q[\theta(\pi-\theta)+\sin^2\theta]+[(\pi-\theta)^2-\sin^2\theta]\}}
{q(\theta^2-\sin^2\theta)[(\pi-\theta)+\cos\theta\sin\theta]+[(\pi-\theta)^2-\sin^2\theta](\theta-\cos\theta\sin\theta)}.
\end{equation*}
\noindent 
In some specific asymptotic limits, special forms of $G(\theta)$ can be used. 
For $\theta \ll 1$ and $q=0$ as in thin liquid film in air situations, 
it can be shown that 
\be 
G(\theta) \approx \theta_{}^3/9 \label{thinfilms}.
\nd
Also, Sheng and Zhou \cite{Sheng92} show that 
$G(\theta) - G(\theta_{e}) \approx (\cos{\theta_{e}} - \cos{\theta})/5.63$
when $q=1$ and $|\cos{\theta}|<0.6$. 
In this work, we account for $G(\theta)$ using Eq.~(\ref{eq:g}) directly.

A more convenient notation is to introduce a ``microscopic length scale'' 
\be
r_m=\lambda_c \exp[f(\theta_e,q)/Q_i] \label{rm_Qi},,
\nd
as in Voinov \cite{Voinov1976,Voinov00} 
so that in the general case
\be
G(\theta) = G(\theta_e) - \Ca \ln (r/r_m), \label{Voinovform}
\nd
The microscopic length scale $r_m$ encompasses the properties
of the microscopic region. 
Thus the effect of the corresponding small scales can be summarized with two parameters
$\theta_e$ and $r_m$. 

Eq.~(\ref{eq:cox_eq_2_m}) is very attractive because it provides
a universal description giving the dependence of $\theta(r)$ on $\Ca$,
to the leading order, without any specification of the microscopic physics  or
any necessity to calculate the details of the macroscopic flow in the
outer region. In this paper, we also claim that it is valid when the concept of
``microscopic physics'' is
replaced by ``numerical scheme''; the numerical scheme introducing deviations from the 
continuum equations at scales of order $\lambda_c \sim \Delta$. 

The microscopic length $r_m$ can be obtained in a number of special cases, or by dimensional analysis. 
Eggers \cite{Eggers2004b} has assumed several slip length models. For the Navier-slip model, and using our notations above,
\be
\left. v\right\vert_{x=0} - V_s  = \lambda \left.\frac{\partial v}{\partial x}\right\vert_{x=0}, \label{nslip}
\nd
where $\lambda$ is the slip length. Using lubrication theory valid for small slopes or angles, 
one obtains the equation \cite{duffy1997third}
\be
\eta^{'''} = \frac{3 \Ca}{\eta^2 + 3\lambda \eta}, \label{lub-slip}
\nd
where we introduced 
a variable $\zeta$ that measures distance away from the contact line along the 
direction of the solid surface. We also note that
$\eta(\zeta)$ is the local thickness of the film and  we use the notation
$\eta' = d\eta/d\zeta$, $\eta'' = d^2\eta/d\zeta^2$, etc..  
A solution of Eq.~(\ref{lub-slip}) can be obtained  \cite{Eggers:2005uq} in the form
\be
\eta'(\zeta) \sim \theta_e - \frac{3\Ca}{\theta^2_e} \ln \left(\frac{ \zeta}{r_m}\right),
\label{eh}
\nd
where 
\be
r_m = \frac{3 \lambda}{{\rm e} \,\theta_e} + \Or(\Ca). \label{leq}
\nd
(Notice that equation (33) of \cite{Eggers:2005uq} misses the above factor of $\mathrm e$ 
because of an incorrect
derivation from equation (32) of \cite{Eggers:2005uq}).
This can be compared to the solution obtained by Cox in \cite{cox1986}. 
From  Eq.~(\ref{thinfilms}) and Eq.~(\ref{Voinovform}), using the fact 
that for small angles $r \sim \zeta = h_{CL} - y$ and $\theta \sim \eta'(\zeta)$, one obtains
\be
\eta'(\zeta)^3 \sim \theta_e^3 - 9 \Ca \ln (\zeta/r_m). \label{coxhp3}
\nd
For $\Ca \ll \theta^3_e$, the above yields again Eq.~(\ref{eh}).
Thus the length scale $r_m$ can be written as
\be
r_m = \lambda/\phi, \label{phi-lub} 
\nd
where $\phi$ is a gauge function depending on $\theta_e$ equal, at first order, to 
\be
\phi(\theta_e) =  \frac{{\rm e} \,\theta_e} {3}. \label{eh2}
\nd
Different slip-length models give different gauge functions $\phi$. Moreover, alternate models of the
microscopic physics beyond slip length, such as Van der Waals forces or phase field, 
would still result in an expression
of the form Eq.~(\ref{Voinovform}), with $\phi$ having a dependence on $\theta_e$ and other
microscopic model parameters to be determined. 


In the numerical case, the equivalent expression for the microscopic length scale in region II is 

\be
r_m = \Delta/\phi_{\rm num}, \label{phi-num}
\nd
where the gauge function $\phi_{\rm num}$ may depend on $q$, $\theta_e$ and the details of
the numerical implementation such as the number of shear bands $n$, any specific numerical slip etc. 
In what follows, for simplicity, we shall drop the ``num'' subscript and leave
the dependence of $\phi$ on the viscosity ratio $q$ implicit. Comparing expressions
(\ref{phi-lub}) and (\ref{phi-num}) leads to the relation 
\be
\lambda = \phi \Delta/\phi_{\rm num}, \label{effslip} 
\nd
which shows that the numerical model can be interpreted as having an effective slip.
Using the definition of the numerical gauge function Eq.~(\ref{phi-num}), Cox's solution
(Eq.~(\ref{eq:cox_eq_2_m}))  can be rewritten
in the form
\be
G(\theta_e) - G[\theta(r)] =  \Ca \ln (r/ \Delta) + \Ca \ln \phi. \label{fit-phi}
\nd
This form is verified numerically in the next Section.

\subsection{Numerical verification of Cox's theory and matched asymptotics}
\label{sec:comp}

We next present the numerical results of computed curvature and interface shapes,
for all the three Setups A, B, and C, and compare the results  
with the theoretical predictions presented in previous section. 
We consider $15^\circ\le\theta_{\rm \Delta}\le110^\circ$ and 
$0.004\le\Delta/l_c\le0.016$. In what follows, we show that
in general, the curvature is negative and higher
close to the contact line and it decreases away from the contact to zero,
where it then turns and becomes positive to assume the curvature of the meniscus and eventually
approaches the zero curvature of the flat film far away from the wall. This
means that the curvature effects are more prominent close to the contact line,
and that there always exists an inflection point. 
We show that interface bending becomes significant as we increase the 
contact angle and/or decrease the mesh size.  
The divergence of the curvature near the contact line is as expected
and discussed in the previous section.
We also present the interface 
apparent slope along with the curvature to enlighten the connection of the 
interface apparent zero slope with the presence of an inflection point. 
Indeed, we notice that the angle is near zero at inflection point when the curvature vanishes,
as assumed in Sec.~\ref{sec-gen-theo}.  
The interface
slope results also confirm that we approach the imposed numerical contact angle,  
$\theta_{\rm \Delta}$, as we approach the contact line. We nondimensionalize 
the curvature by $l_c$ and present the results as a function of the
vertical height from the contact line, i.e.~$\zeta/l_c = (h_{CL}-y)/l_c$,
(see Fig.~\ref{fig:1} for the illustration of the distance $\zeta$).
  
\begin{figure}[]
\begin{center}
\begin{tabular}{cc}
\includegraphics[width=2.25in]{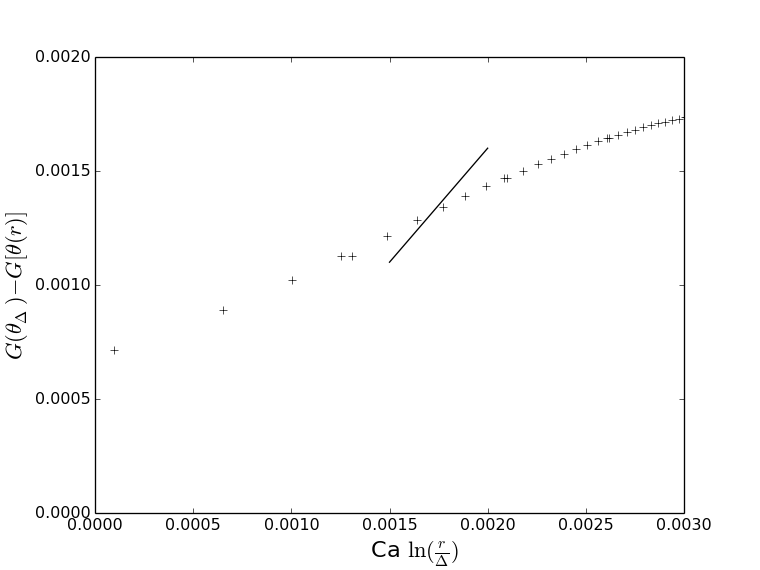} &
\includegraphics[width=2.25in]{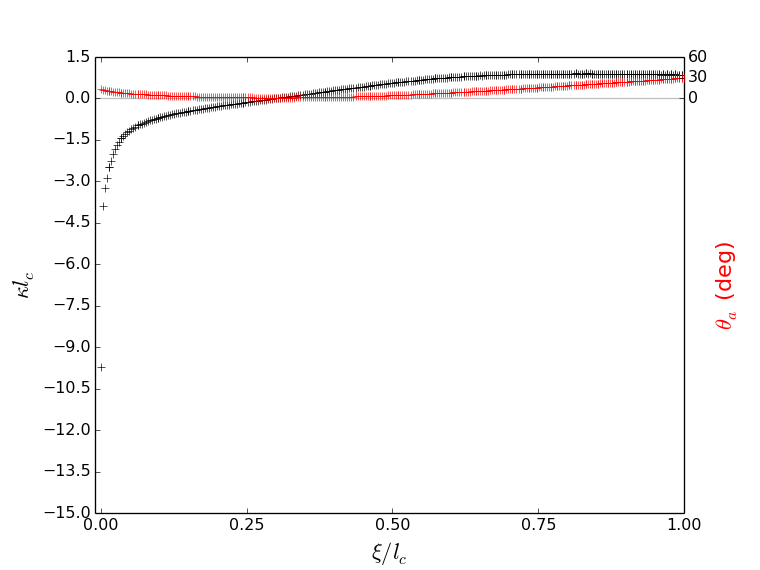}\\
(a)&(b)\\
\includegraphics[width=2.25in]{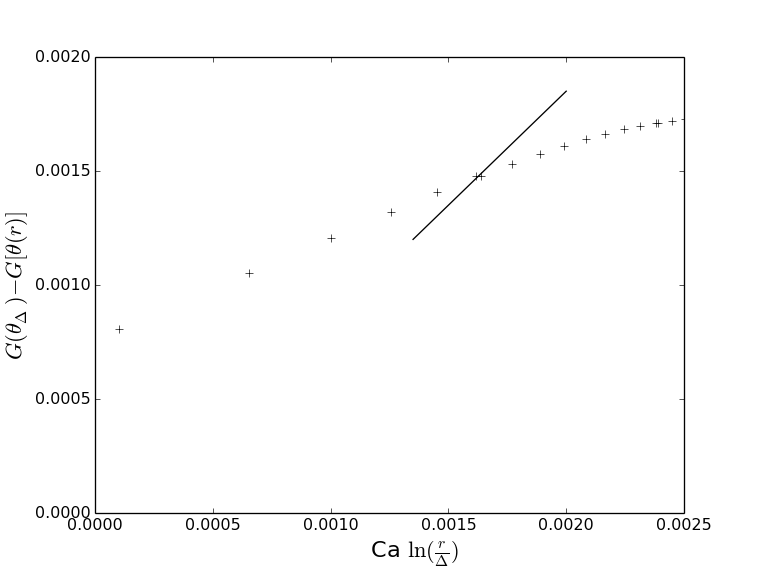} &
\includegraphics[width=2.25in]{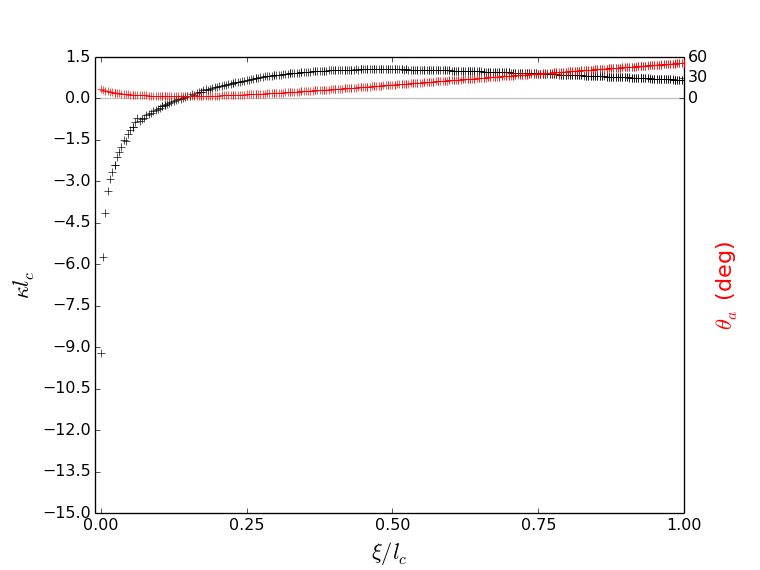}\\
(c)&(d)\\
\includegraphics[width=2.25in]{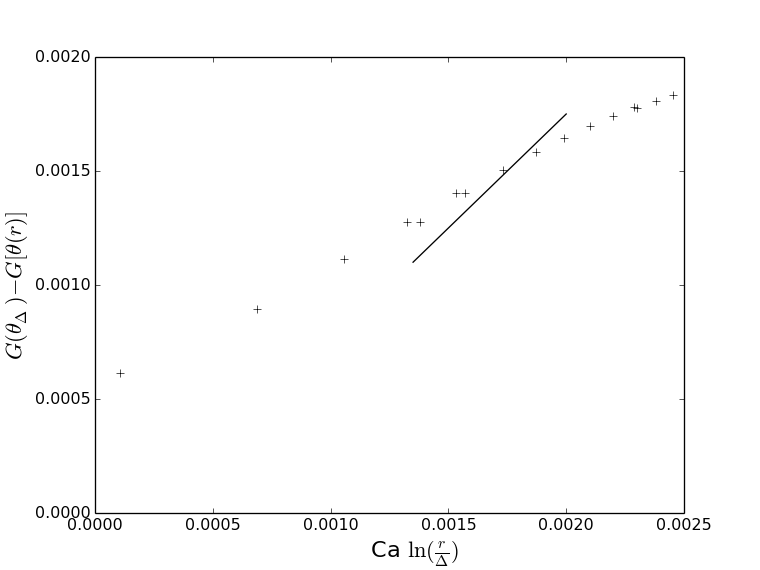} &
\includegraphics[width=2.25in]{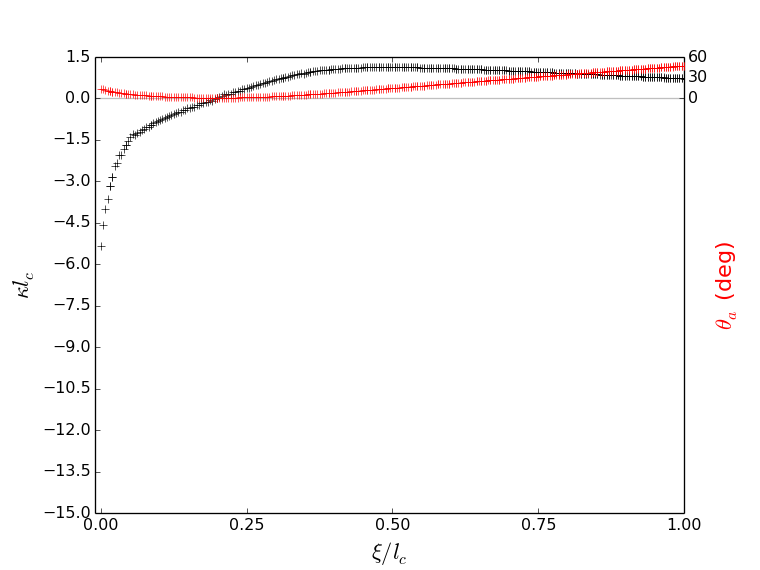}\\
(e)&(f)\\
\end{tabular}
\end{center}
\caption[]{
Left panel: Comparison of the computed (black symbols) $G(\theta_e=\theta_\Delta) - G[\theta(r)]$
versus $\Ca \ln (r/\Delta)$ with the prediction of Eq.~(\ref{fit-phi}) with the best
$\phi$ value of (a) $0.64$ (Setup A), (c) $0.85$ (Setup B), and (e) $0.77$ (Setup C) (black solid line).
The fit is performed three to four grid points away from the contact line 
to minimize the inaccuracies  amplified in our numerical method at the grid scale near the contact line.
Right panel: Nondimensional curvature (black symbols) 
and the slope that the interface makes with the substrate
(red symbols), as a function of the nondimensional vertical distance 
of the interface from the contact line, $\zeta/l_c$.
$\theta_{\rm \Delta}=15^\circ$ and $\Delta/l_c = 0.0035$ (Setup A) and $0.004$ (Setup B and C), 
at  $\mbox{Ca}=0.0009$, $\tau= 1.52$ (Setup A), $\mbox{Ca}=0.0009$, $\tau= 1.2$ (Setup B),
and  $\mbox{Ca}=0.00095$, $\tau= 1.52$ (Setup C).}
\label{fig:15}
\end{figure}

\begin{figure}[]
\begin{center}
\begin{tabular}{cc}
\includegraphics[width=2.25in]{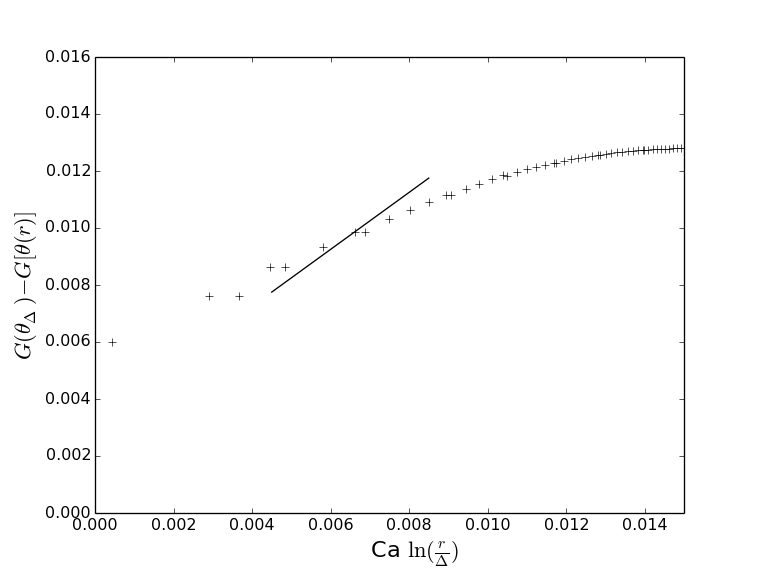} &
\includegraphics[width=2.25in]{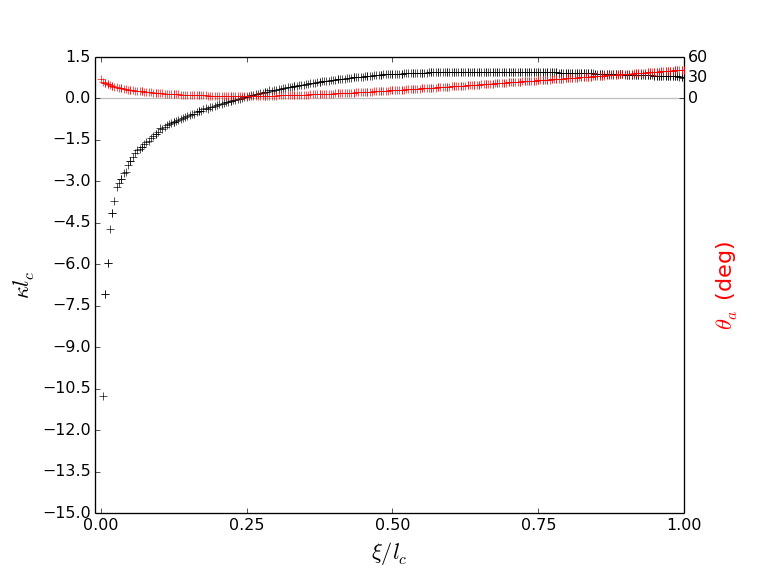}\\
(a)&(b)\\
\includegraphics[width=2.25in]{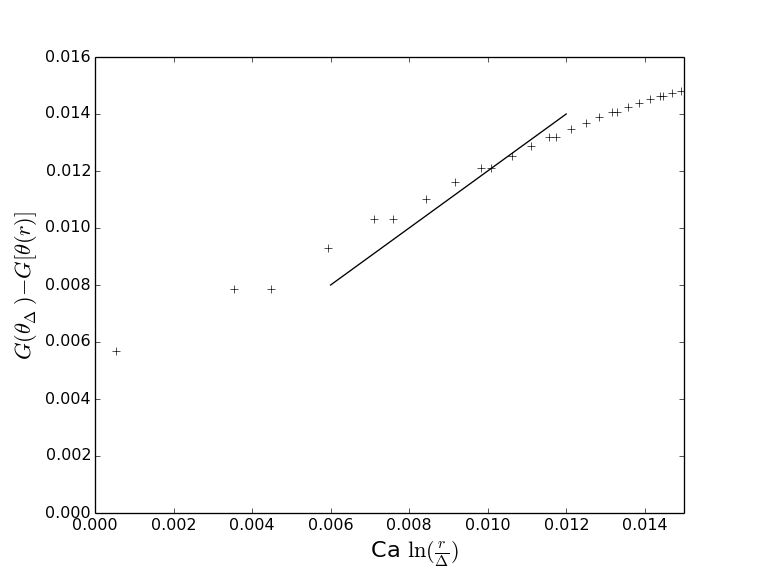} &
\includegraphics[width=2.25in]{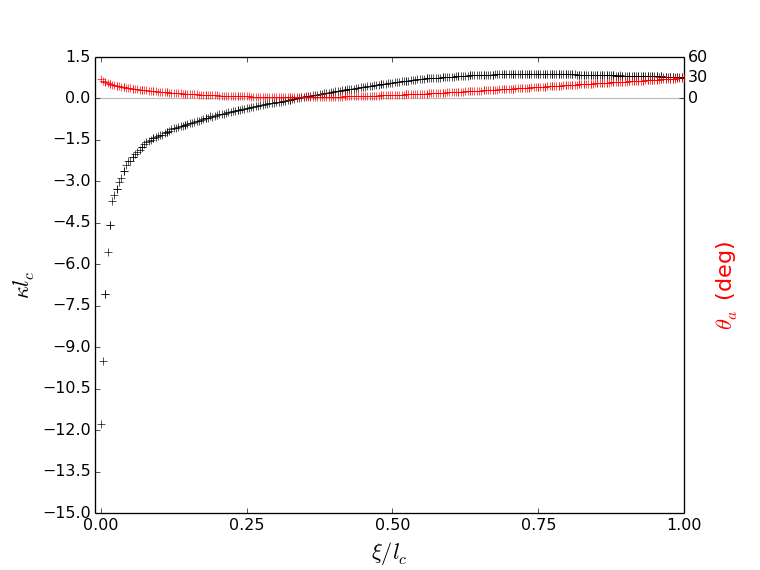}\\
(c)&(d)\\
\end{tabular}
\end{center}
\caption[]{
Left panel: Comparison of the computed (black symbols) $G(\theta_e=\theta_\Delta) - G[\theta(r)]$
versus $\Ca \ln (r/\Delta)$ with the prediction of Eq.~(\ref{fit-phi}) with the best
$\phi$ value of (a) $2.15$ (Setup B) and (c) $1.51$ (Setup C) (black solid line).
The fit is performed three to four grid points away from the contact line 
to minimize the inaccuracies  amplified in our numerical method at the grid scale near the contact line.
Right panel: Nondimensional curvature (black symbols) 
and the slope that the interface makes with the substrate
(red symbols), as a function of the nondimensional vertical distance 
of the interface from the contact line, $\zeta/l_c$.
$\theta_{\rm \Delta}=30^\circ$ and $\Delta/l_c = 0.004$ (Setup B and C), 
at  $\mbox{Ca}=0.004$, $\tau= 5.37$ (Setup B) and  $\mbox{Ca}=0.0049$, $\tau= 5.95$ (Setup C).}
\label{fig:30L8}
\end{figure}

\begin{figure}[]
\begin{center}
\begin{tabular}{cc}
\includegraphics[width=2.25in]{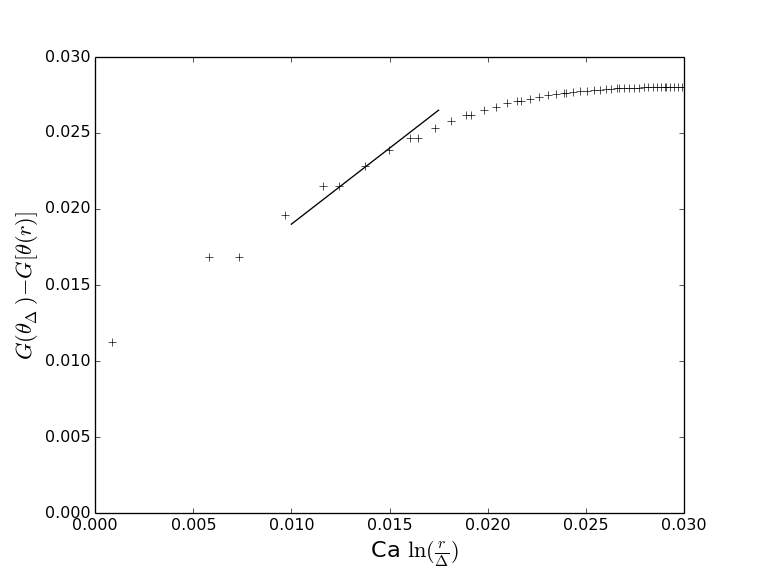} &
\includegraphics[width=2.25in]{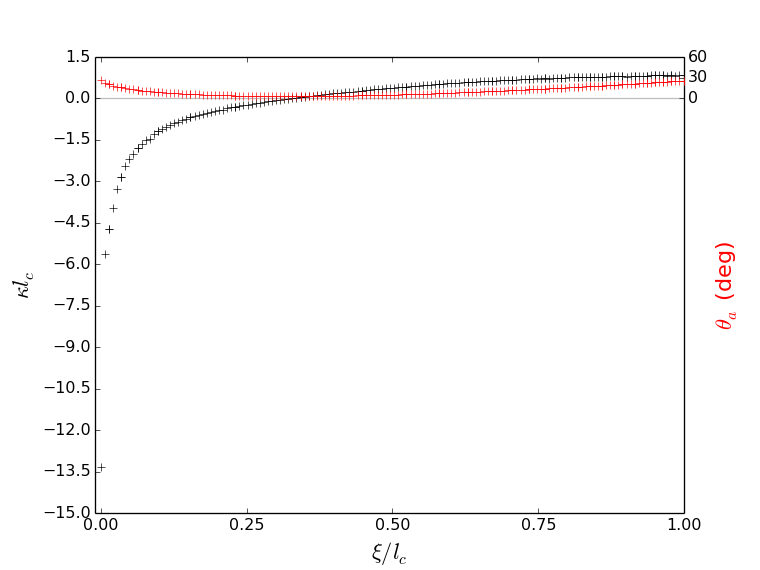}\\
(a)&(b)\\
\includegraphics[width=2.25in]{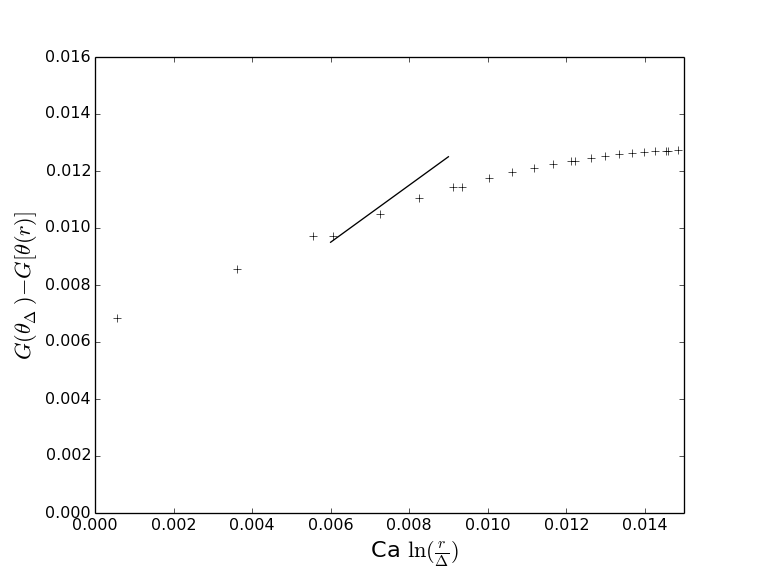} &
\includegraphics[width=2.25in]{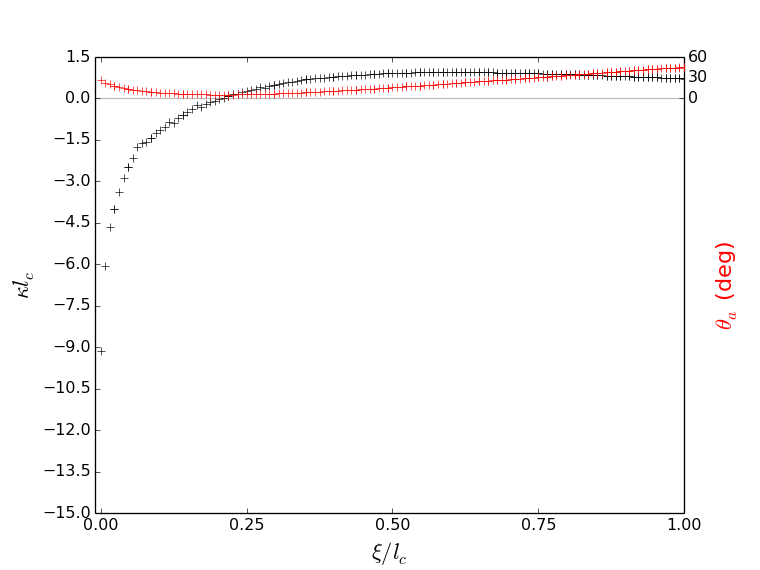}\\
(c)&(d)\\
\includegraphics[width=2.25in]{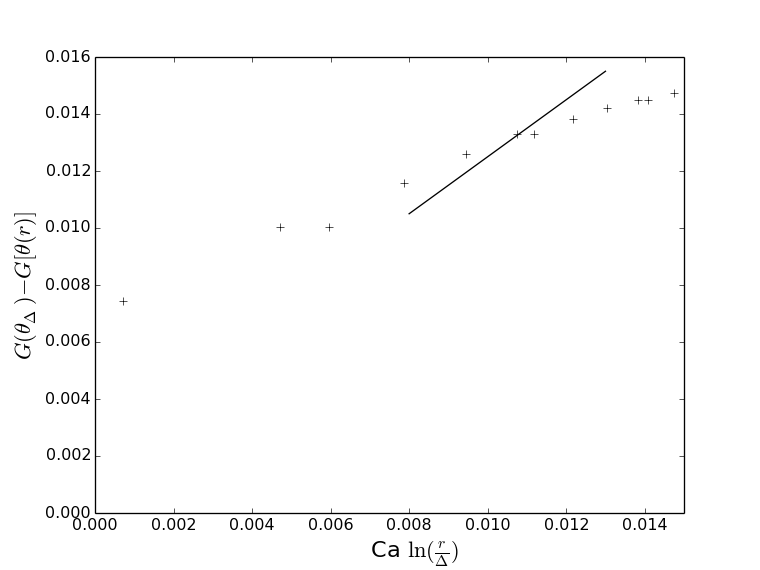} &
\includegraphics[width=2.25in]{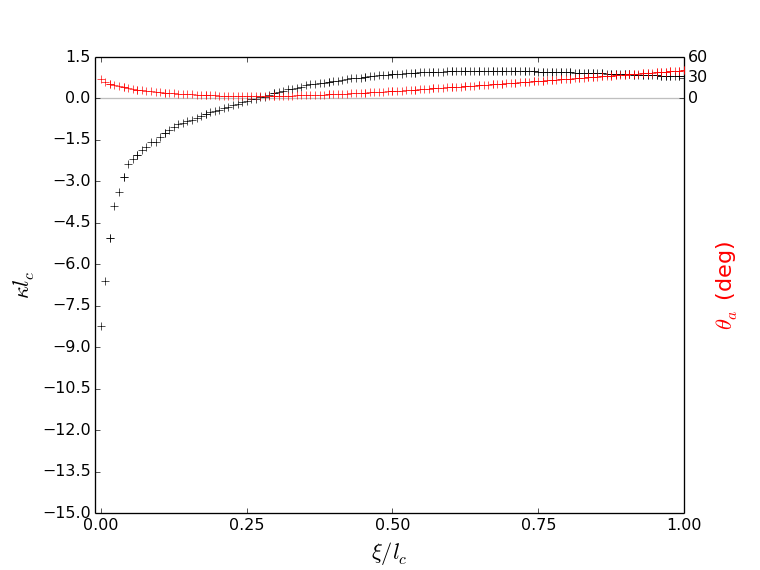}\\
(e)&(f)\\
\end{tabular}
\end{center}
\caption[]{
Left panel: Comparison of the computed (black symbols) $G(\theta_e=\theta_\Delta) - G[\theta(r)]$
versus $\Ca \ln (r/\Delta)$ with the prediction of Eq.~(\ref{fit-phi}) with the best
$\phi$ value of (a) $2.71$ (Setup A), (c) $2.01$ (Setup B), and (e) $1.47$ (Setup C) (black solid line).
The fit is performed three to four grid points away from the contact line 
to minimize the inaccuracies  amplified in our numerical method at the grid scale near the contact line.
Right panel: Nondimensional curvature (black symbols) 
and the slope that the interface makes with the substrate
(red symbols), as a function of the nondimensional vertical distance 
of the interface from the contact line, $\zeta/l_c$.
$\theta_{\rm \Delta}=30^\circ$ and $\Delta/l_c = 0.007$ (Setup A) and $0.008$ (Setup B and C), 
at  $\mbox{Ca}=0.004$, $\tau= 1.52$ (Setup A), $\mbox{Ca}=0.005$, $\tau= 3.18$ (Setup B),
and  $\mbox{Ca}=0.0065$, $\tau= 3.04$ (Setup C).}
\label{fig:30L7}
\end{figure}

\begin{figure}[]
\begin{center}
\begin{tabular}{cc}
\includegraphics[width=2.25in]{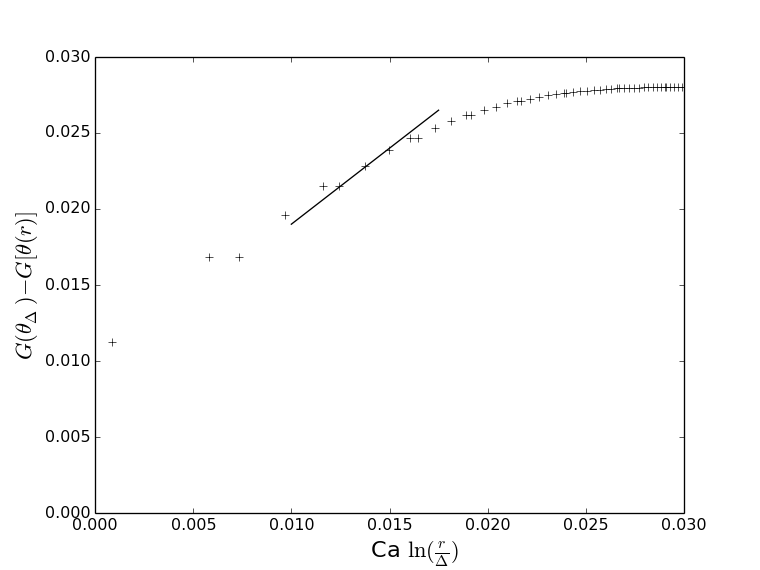} &
\includegraphics[width=2.25in]{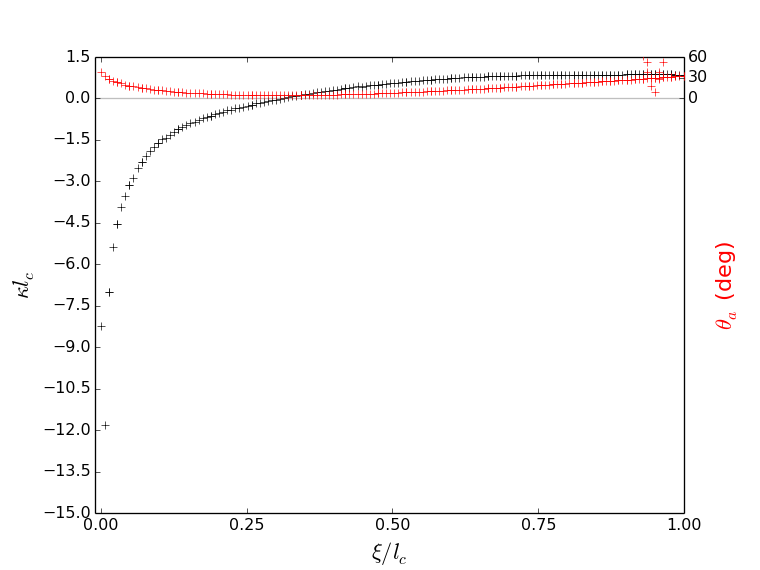}\\
(a)&(b)\\
\includegraphics[width=2.25in]{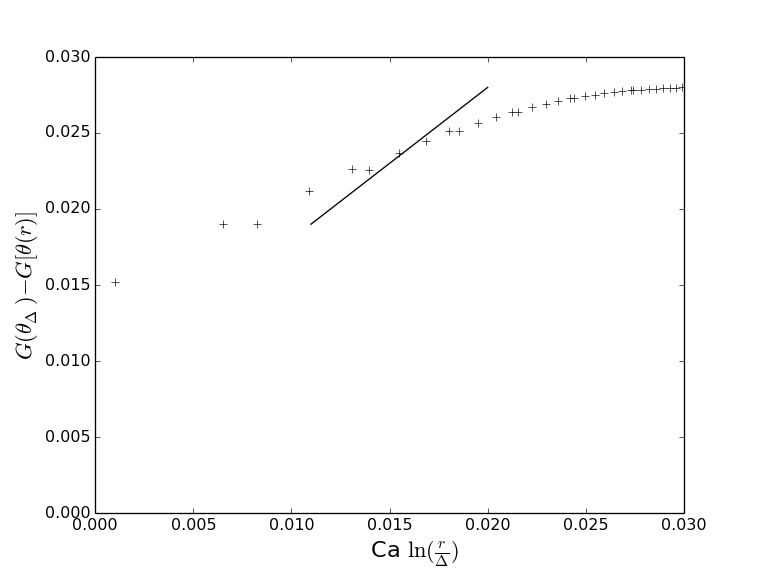} &
\includegraphics[width=2.25in]{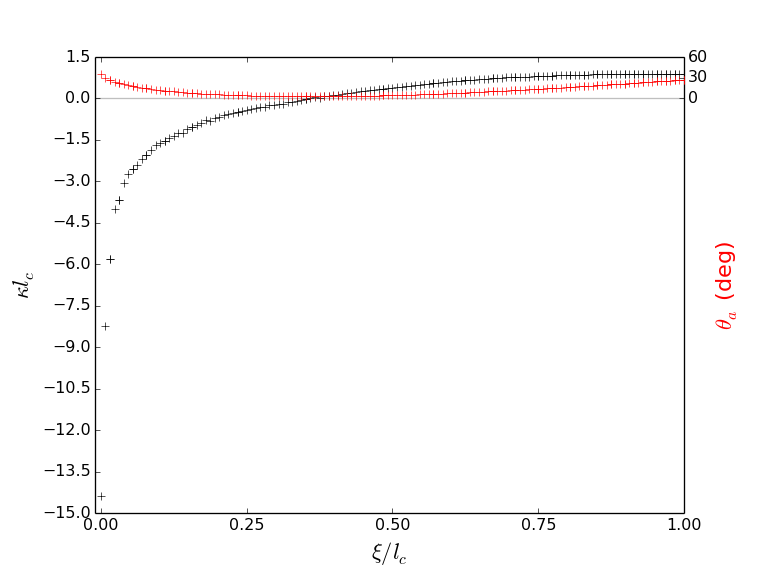}\\
(c)&(d)\\
\includegraphics[width=2.25in]{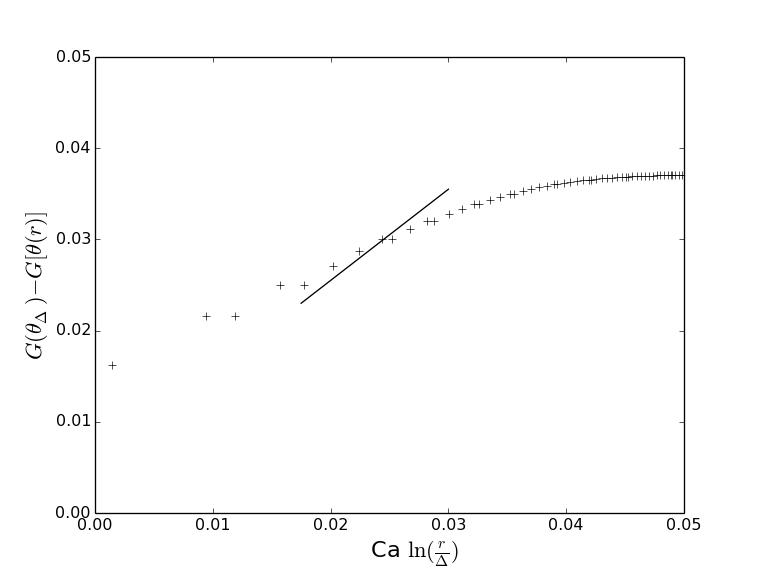} &
\includegraphics[width=2.25in]{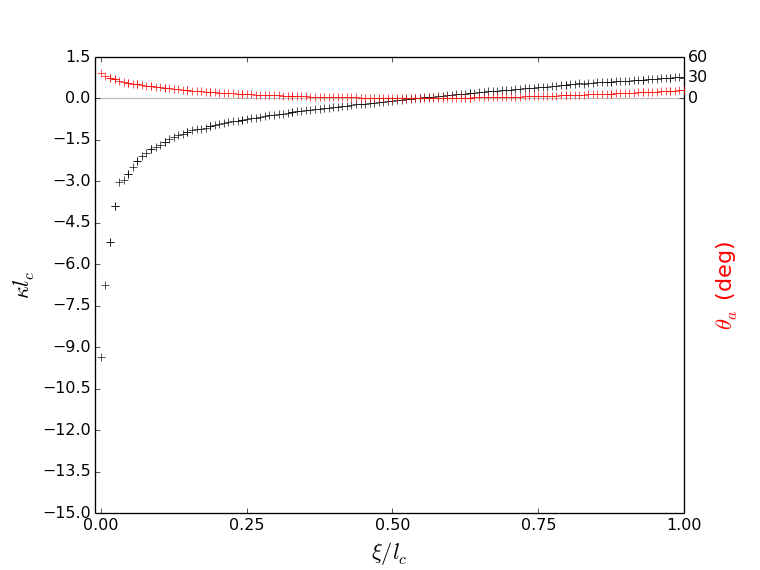}\\
(e)&(f)\\
\end{tabular}
\end{center}
\caption[]{
Left panel: Comparison of the computed (black symbols) $G(\theta_e=\theta_\Delta) - G[\theta(r)]$
versus $\Ca \ln (r/\Delta)$ with the prediction of Eq.~(\ref{fit-phi}) with the best
$\phi$ value of (a) $3.08$ (Setup A), (c) $2.42$ (Setup B), and (e) $1.52$ (Setup C) (black solid line).
The fit is performed three to four grid points away from the contact line 
to minimize the inaccuracies  amplified in our numerical method at the grid scale near the contact line.
Right panel: Nondimensional curvature (black symbols) 
and the slope that the interface makes with the substrate
(red symbols), as a function of the nondimensional vertical distance 
of the interface from the contact line, $\zeta/l_c$.
$\theta_{\rm \Delta}=40^\circ$ and $\Delta/l_c = 0.007$ (Setup A) and $0.008$ (Setup B and C), 
at  $\mbox{Ca}=0.008$, $\tau= 5.55$ (Setup A), $\mbox{Ca}=0.009$, $\tau= 5.52$ (Setup B),
and  $\mbox{Ca}=0.013$, $\tau= 6.27$ (Setup C).}
\label{fig:40}
\end{figure}

\begin{figure}[]
\begin{center}
\begin{tabular}{cc}
\includegraphics[width=2.25in]{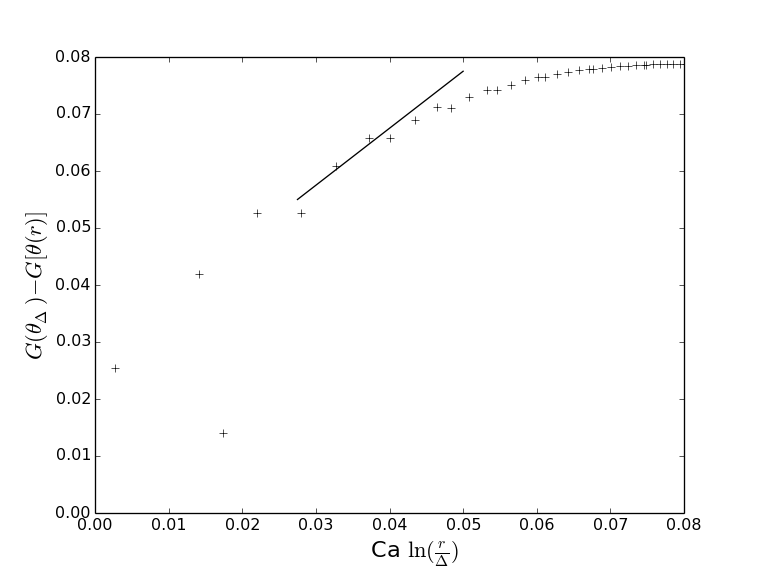} &
\includegraphics[width=2.25in]{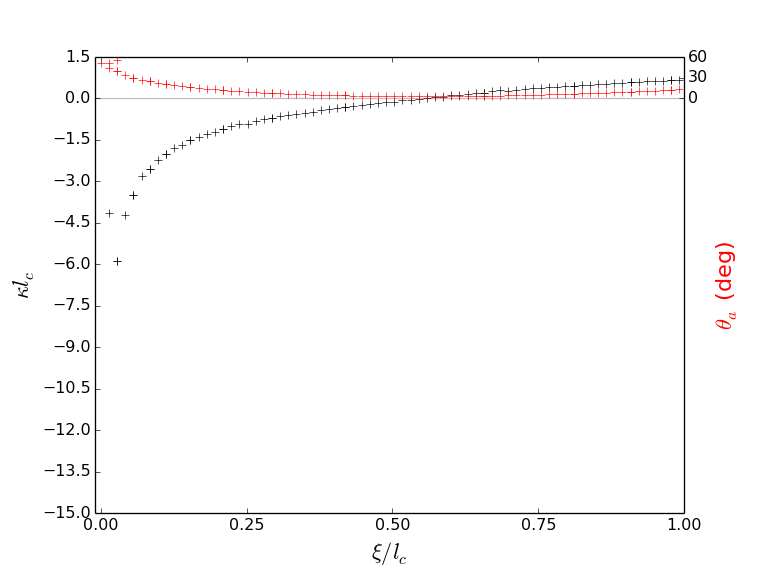}\\
(a)&(b)\\
\includegraphics[width=2.25in]{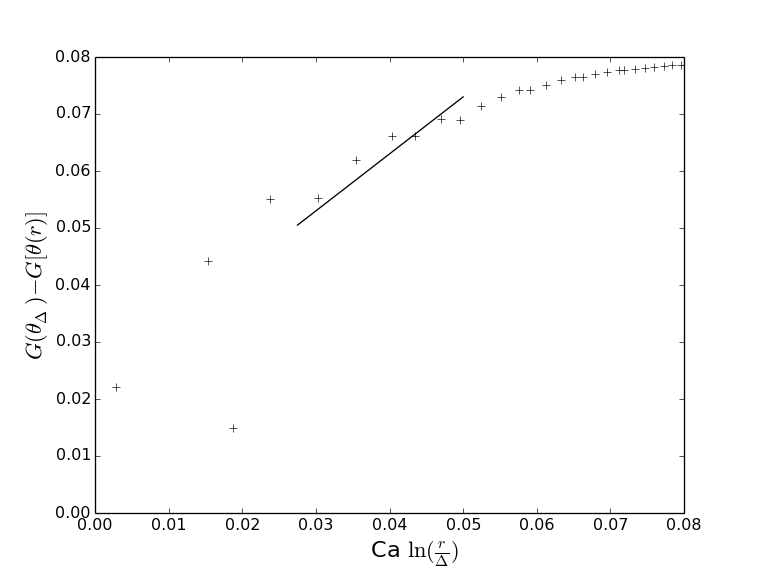} &
\includegraphics[width=2.25in]{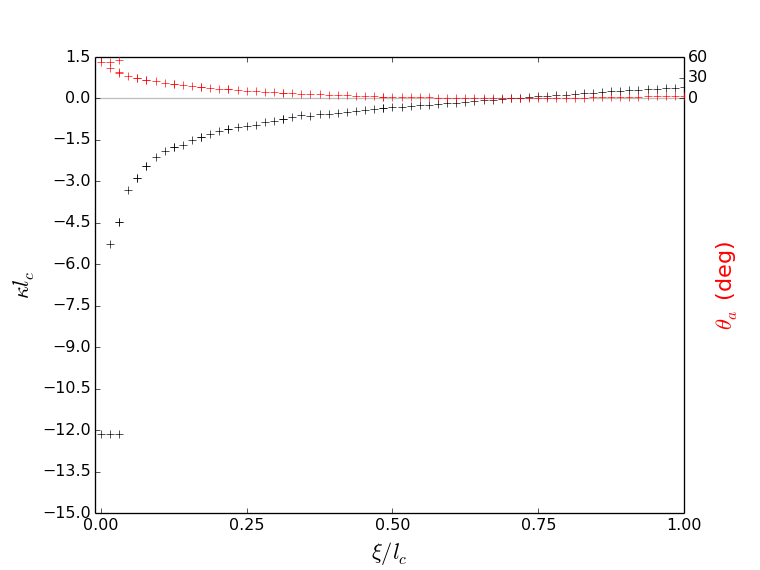}\\
(c)&(d)\\
\includegraphics[width=2.25in]{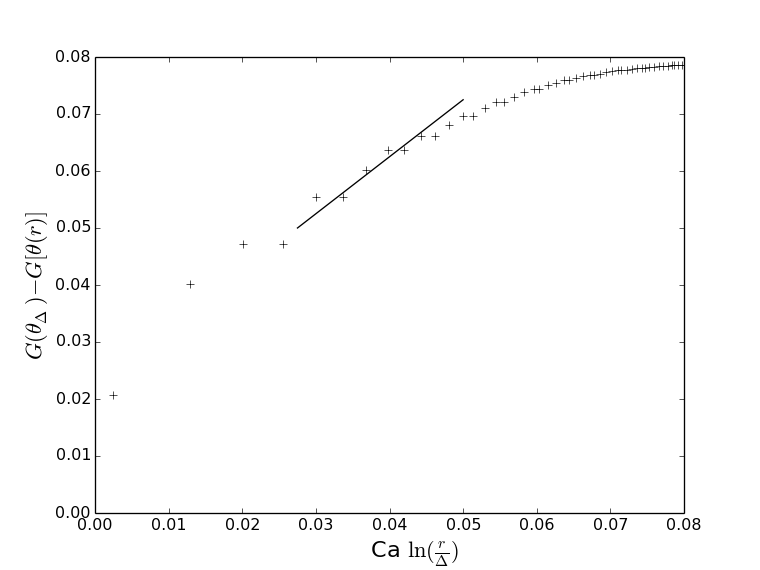} &
\includegraphics[width=2.25in]{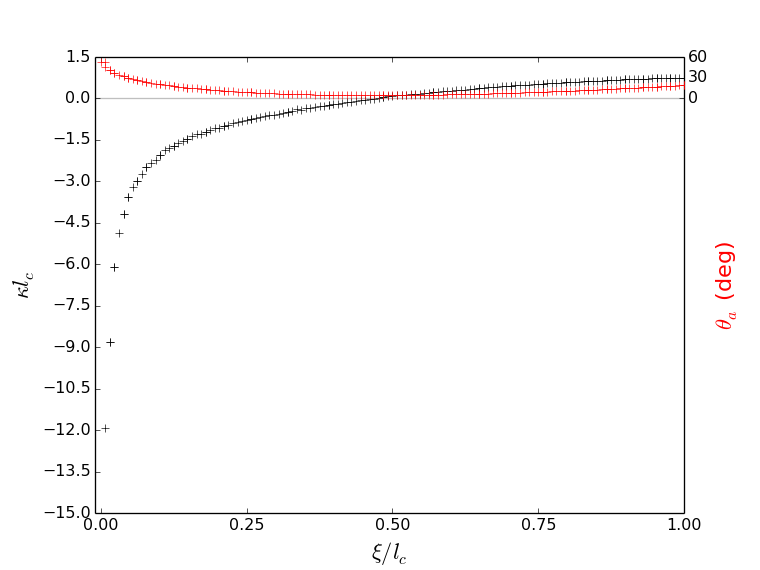}\\
(e)&(f)\\
\end{tabular}
\end{center}
\caption[]{
Left panel: Comparison of the computed (black symbols) $G(\theta_e=\theta_\Delta) - G[\theta(r)]$
versus $\Ca \ln (r/\Delta)$ with the prediction of Eq.~(\ref{fit-phi}) with the best
$\phi$ value of (a) $3.14$ (Setup A), (c) $2.42$ (Setup B), (e) $2.78$ (Setup B)(black solid line).
The fit is performed three to four grid points away from the contact line 
to minimize the inaccuracies  amplified in our numerical method at the grid scale near the contact line.
Right panel: Nondimensional curvature (black symbols) 
and the slope that the interface makes with the substrate
(red symbols), as a function of the nondimensional vertical distance 
of the interface from the contact line, $\zeta/l_c$.
$\theta_{\rm \Delta}=60^\circ$ and $\Delta/l_c = 0.014$ (a-b) (Setup A), $0.016$ (c-d) (Setup B), and $0.008$ (e-f) (Setup B)
at  $\mbox{Ca}=0.024$, $\tau= 8.85$ (a-b) (Setup A), $\mbox{Ca}=0.024$, $\tau= 10.48$ (c-d) (Setup B),
and  $\mbox{Ca}=0.022$, $\tau= 11.12$ (e-f) (Setup B).}
\label{fig:60}
\end{figure}

\begin{figure}[]
\begin{center}
\begin{tabular}{cc}
\includegraphics[width=2.25in]{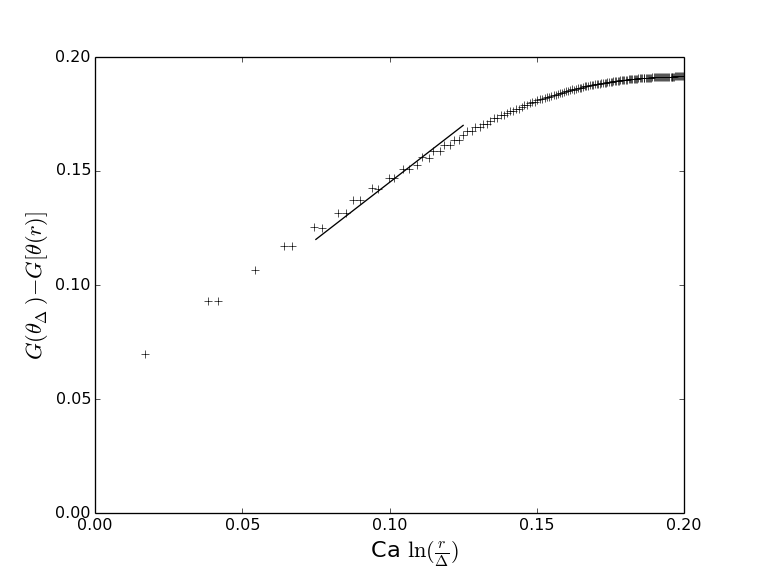} &
\includegraphics[width=2.25in]{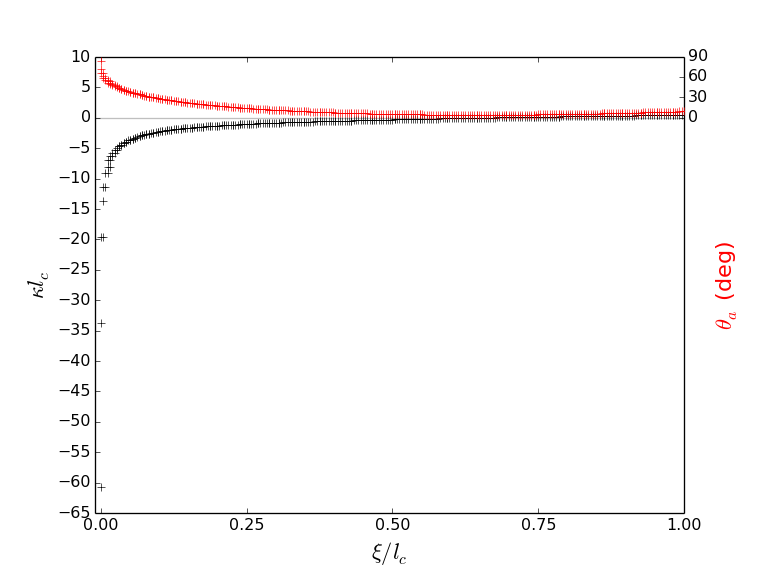}\\
(a)&(b)\\
\includegraphics[width=2.25in]{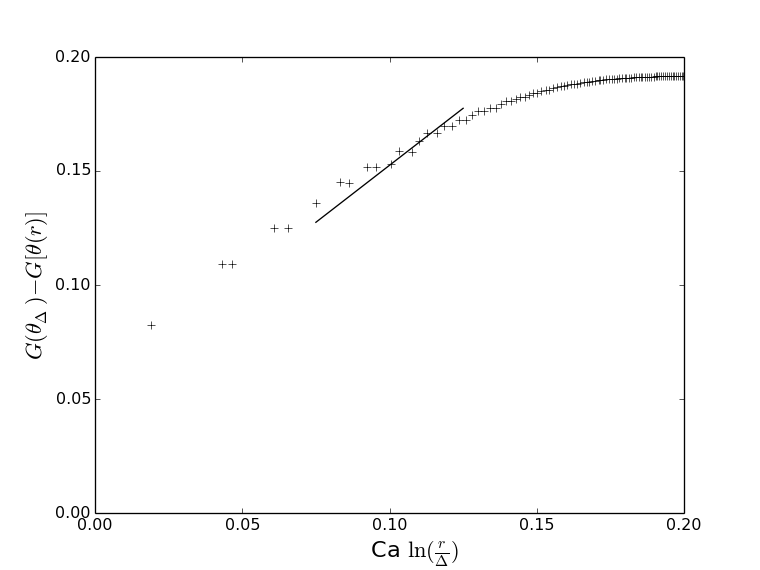} &
\includegraphics[width=2.25in]{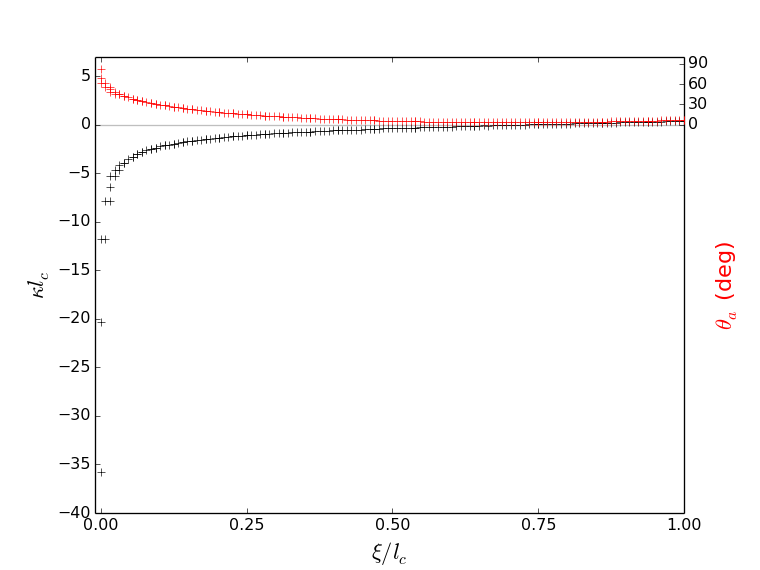}\\
(c)&(d)\\
\includegraphics[width=2.25in]{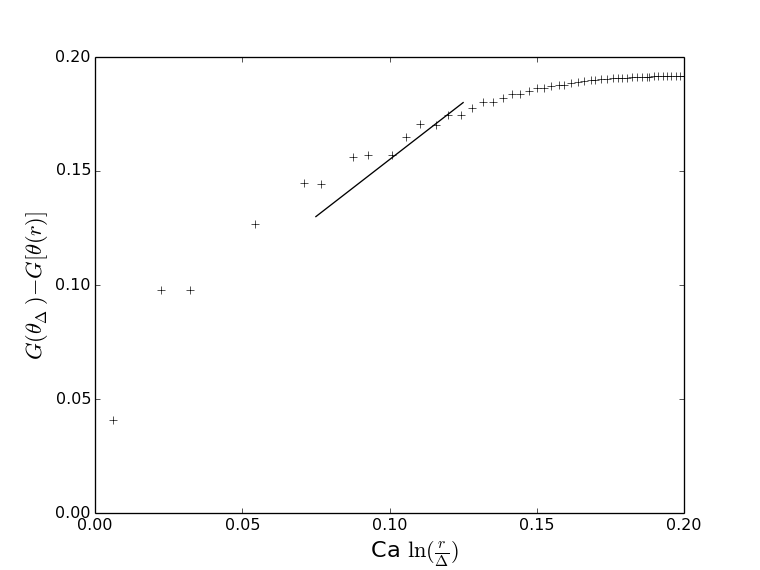} &
\includegraphics[width=2.25in]{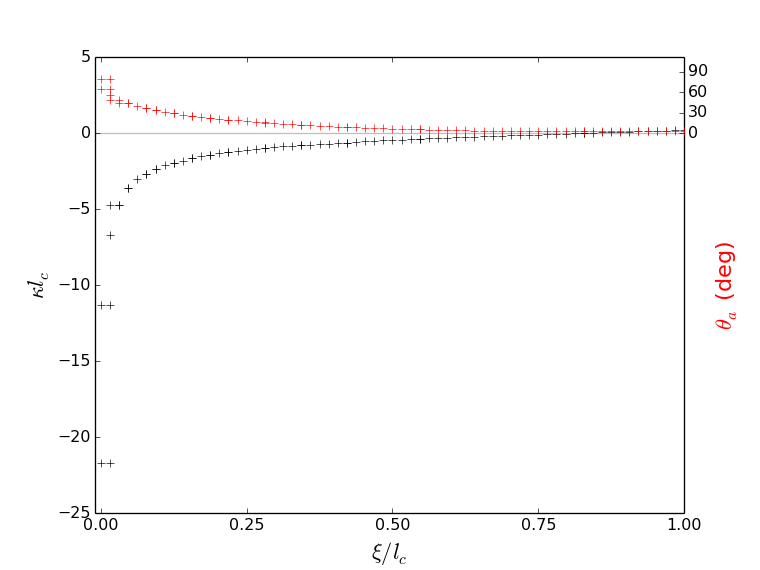}\\
(e)&(f)\\
\end{tabular}
\end{center}
\caption[]{
Left panel: Comparison of the computed (black symbols) $G(\theta_e=\theta_\Delta) - G[\theta(r)]$
versus $\Ca \ln (r/\Delta)$ with the prediction of Eq.~(\ref{fit-phi}) with the best
$\phi$ value of (a) $2.91$, (c) $3.05$, (e) $2.72$ (black solid line).
The fit is performed three to four grid points away from the contact line 
to minimize the inaccuracies  amplified in our numerical method at the grid scale near the contact line.
Right panel: Nondimensional curvature (black symbols) 
and the slope that the interface makes with the substrate
(red symbols), as a function of the nondimensional vertical distance 
of the interface from the contact line, $\zeta/l_c$.  Setup B,
$\theta_{\rm \Delta}=90^\circ$ and $\Delta/l_c = 0.004$ (a-b), $0.008$ (c-d), and $0.016$ (e-f) 
at  $\mbox{Ca}=0.042$, $\tau= 17.41$ (a-b), $\mbox{Ca}=0.047$, $\tau= 22.76$ (c-d),
and  $\mbox{Ca}=0.055$, $\tau= 22.28$ (e-f).}
\label{fig:90}
\end{figure}

\begin{figure}[]
\begin{center}
\begin{tabular}{cc}
\includegraphics[width=2.25in]{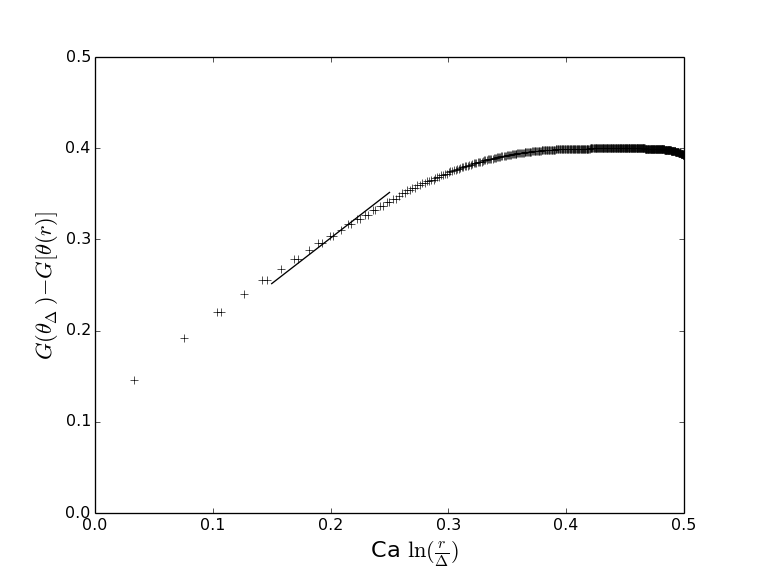} &
\includegraphics[width=2.25in]{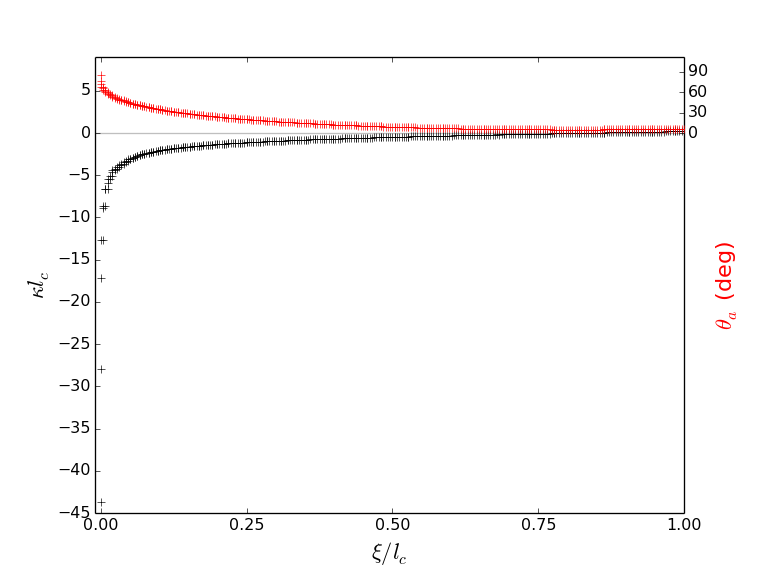}\\
(a)&(b)\\
\includegraphics[width=2.25in]{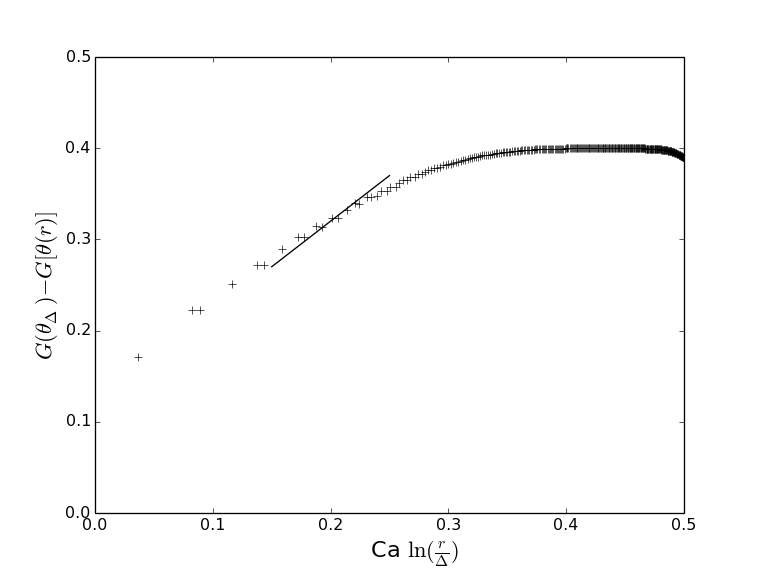} &
\includegraphics[width=2.25in]{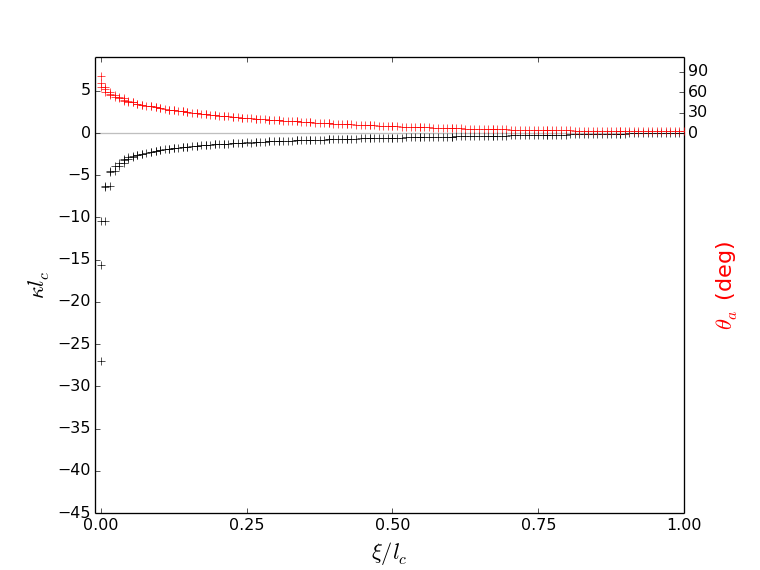}\\
(c)&(d)\\
\includegraphics[width=2.25in]{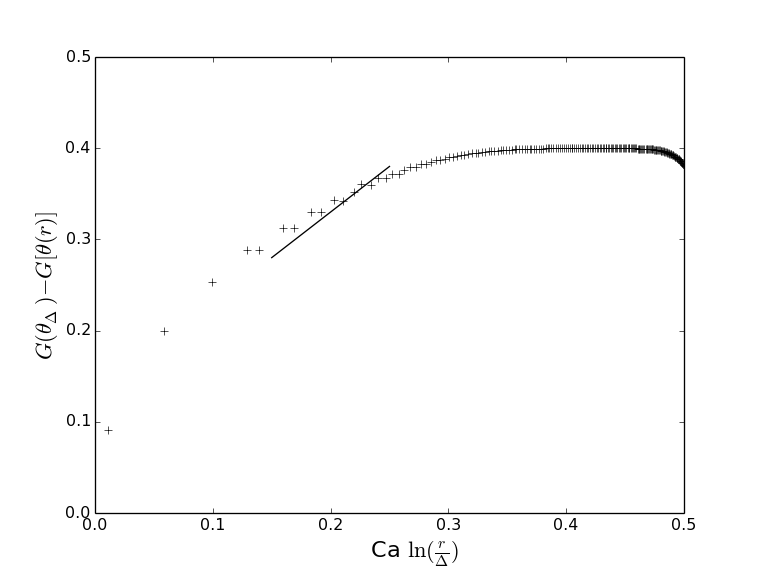} &
\includegraphics[width=2.25in]{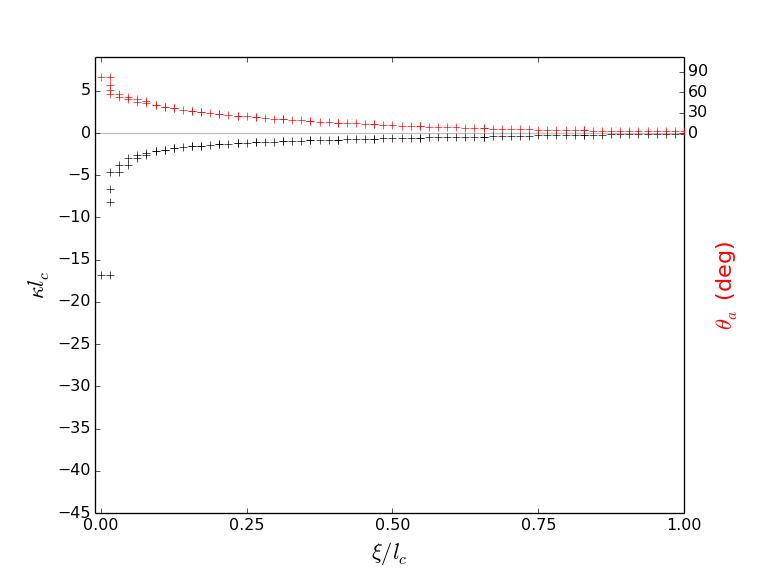}\\
(e)&(f)\\
\end{tabular}
\end{center}
\caption[]{
Left panel: Comparison of the computed (black symbols) $G(\theta_e=\theta_\Delta) - G[\theta(r)]$
versus $\Ca \ln (r/\Delta)$ with the prediction of Eq.~(\ref{fit-phi}) with the best
$\phi$ value of (a) $3.42$, (c) $3.79$, (e) $3.67$ (black solid line).
The fit is performed three to four grid points away from the contact line 
to minimize the inaccuracies  amplified in our numerical method at the grid scale near the contact line.
Right panel: Nondimensional curvature (black symbols) 
and the slope that the interface makes with the substrate
(red symbols), as a function of the nondimensional vertical distance 
of the interface from the contact line, $\zeta/l_c$.  Setup C,
$\theta_{\rm \Delta}=90^\circ$ and $\Delta/l_c = 0.004$ (a-b), $0.008$ (c-d), and $0.016$ (e-f) 
at  $\mbox{Ca}=0.082$, $\tau= 15.79$ (a-b), $\mbox{Ca}=0.09$, $\tau= 22.5$ (c-d),
and  $\mbox{Ca}=0.1$, $\tau= 39.52$ (e-f).}
\label{fig:90VR50}
\end{figure}

\begin{figure}[]
\begin{center}
\begin{tabular}{cc}
\includegraphics[width=2.25in]{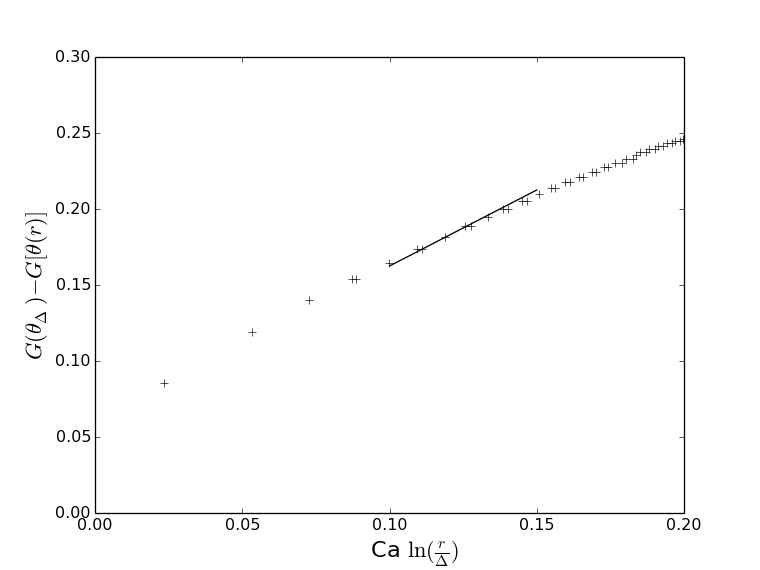} &
\includegraphics[width=2.25in]{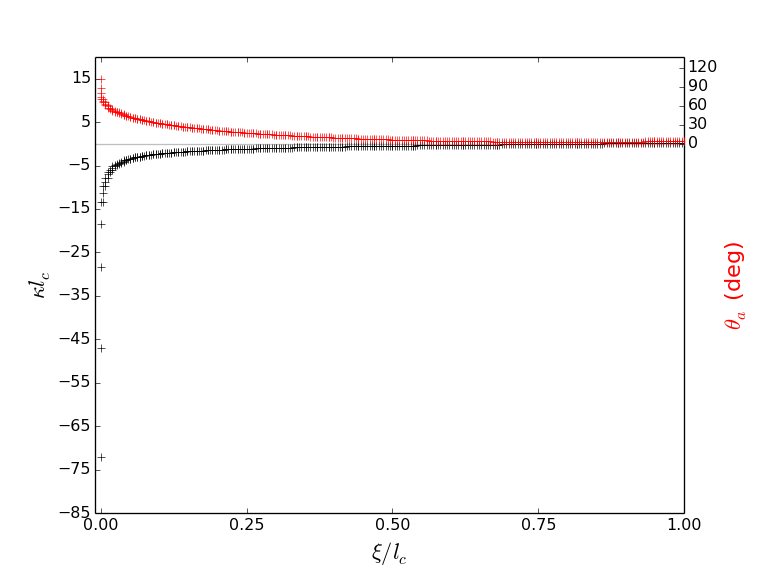}\\
(a)&(b)\\
\includegraphics[width=2.25in]{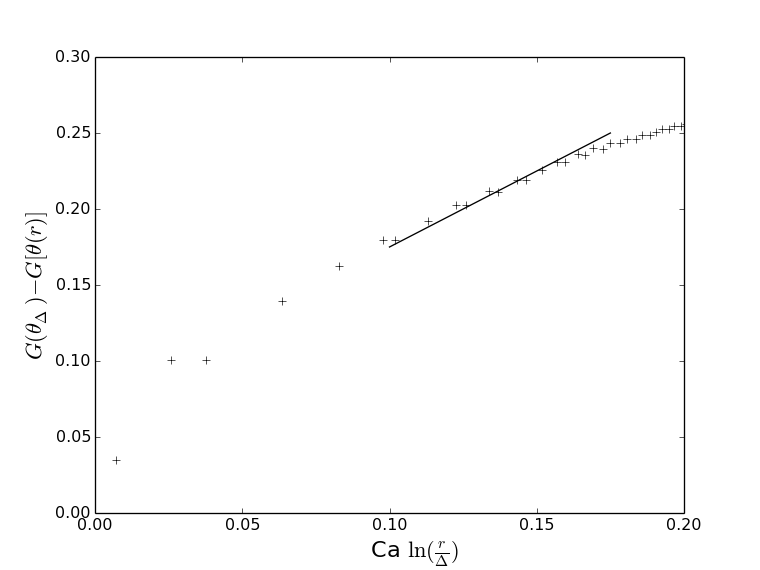} &
\includegraphics[width=2.25in]{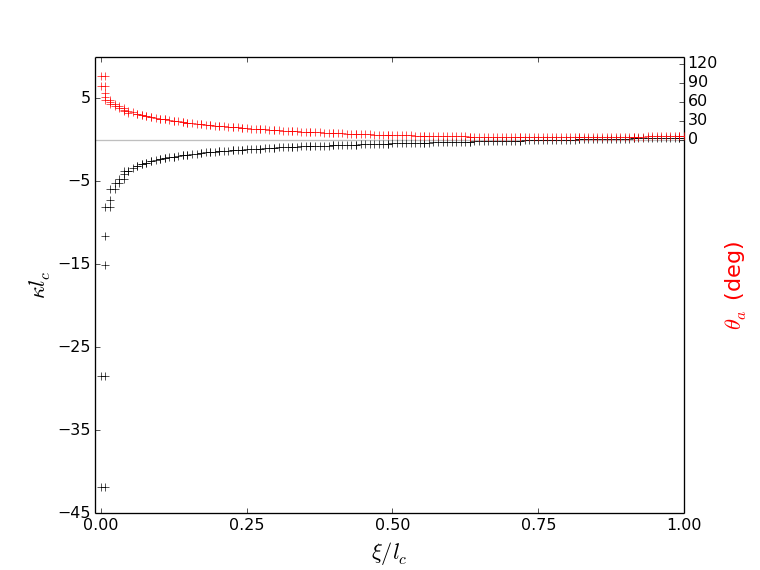}\\
(c)&(d)\\
\includegraphics[width=2.25in]{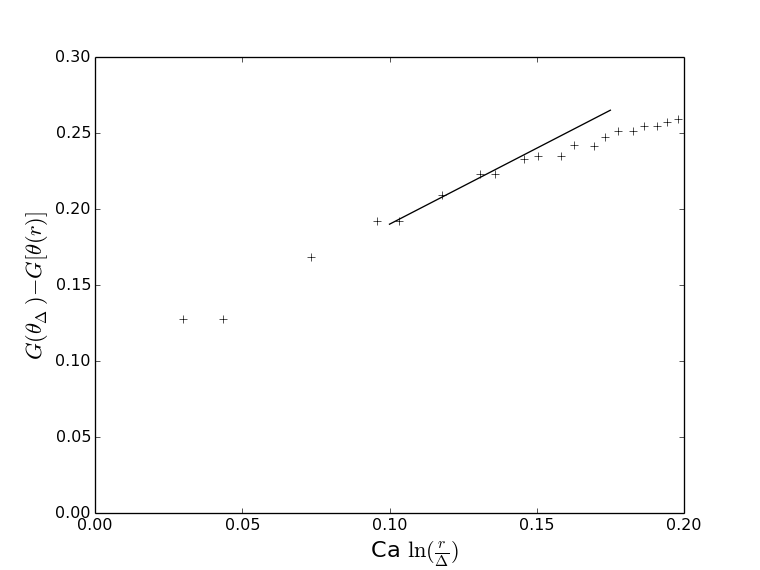} &
\includegraphics[width=2.25in]{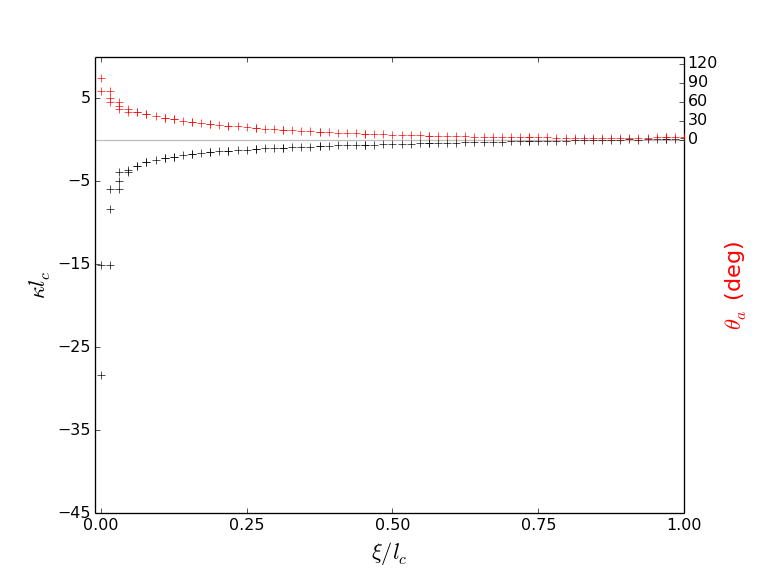}\\
(e)&(f)\\
\end{tabular}
\end{center}
\caption[]{
Left panel: Comparison of the computed (black symbols) $G(\theta_e=\theta_\Delta) - G[\theta(r)]$
versus $\Ca \ln (r/\Delta)$ with the prediction of Eq.~(\ref{fit-phi}) with the best
$\phi$ value of (a) $2.94$, (c) $3.22$, (e) $3.37$ (black solid line).
The fit is performed three to four grid points away from the contact line 
to minimize the inaccuracies  amplified in our numerical method at the grid scale near the contact line.
Right panel: Nondimensional curvature (black symbols) 
and the slope that the interface makes with the substrate
(red symbols), as a function of the nondimensional vertical distance 
of the interface from the contact line, $\zeta/l_c$.  Setup B,
$\theta_{\rm \Delta}=110^\circ$ and $\Delta/l_c = 0.004$ (a-b), $0.008$ (c-d), and $0.016$ (e-f) 
at  $\mbox{Ca}=0.058$, $\tau= 21.67$ (a-b), $\mbox{Ca}=0.064$, $\tau= 32.88$ (c-d),
and  $\mbox{Ca}=0.074$, $\tau= 29.92$ (e-f).}
\label{fig:110}
\end{figure}

\begin{figure}[]
\begin{center}
\begin{tabular}{cc}
\includegraphics[width=2.25in]{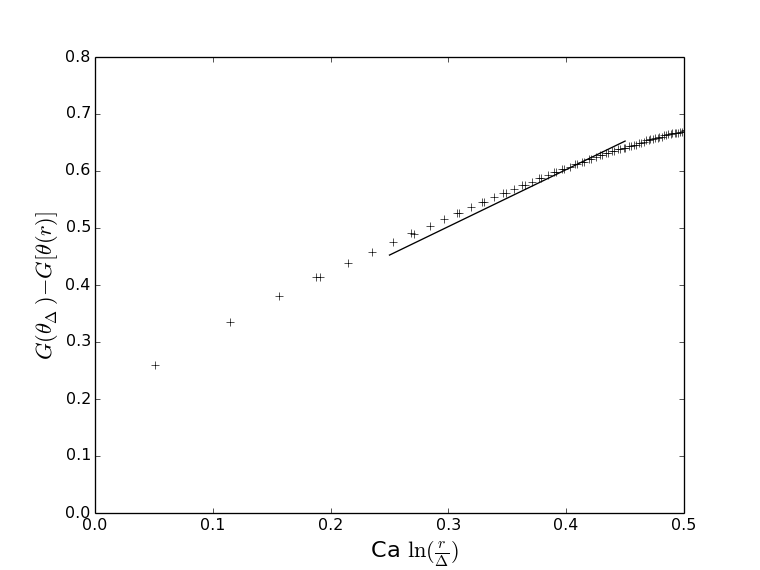} &
\includegraphics[width=2.25in]{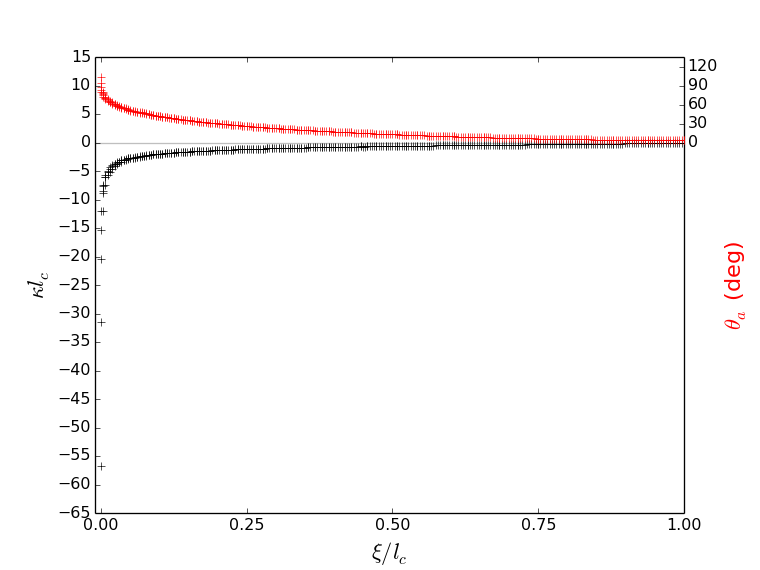}\\
(a)&(b)\\
\includegraphics[width=2.25in]{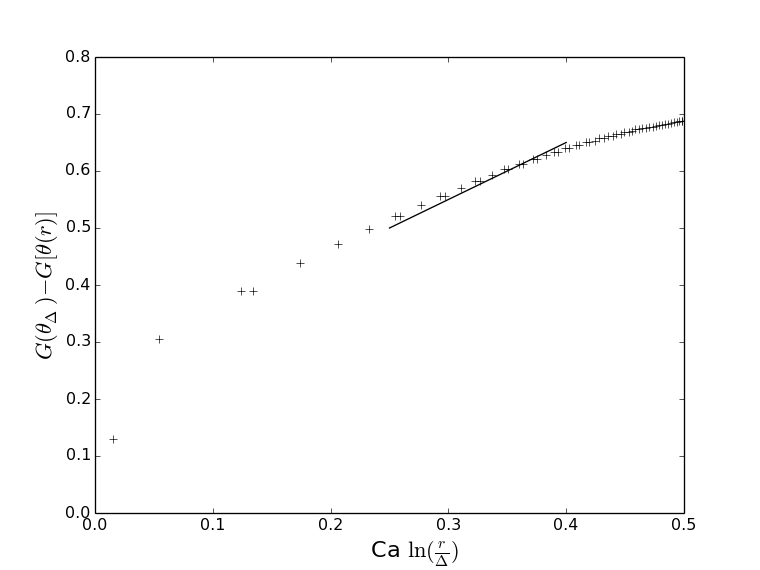} &
\includegraphics[width=2.25in]{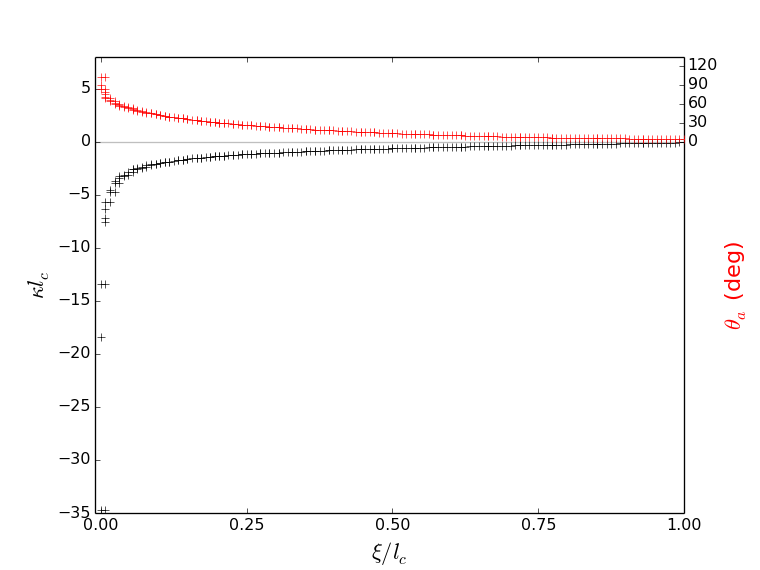}\\
(c)&(d)\\
\includegraphics[width=2.25in]{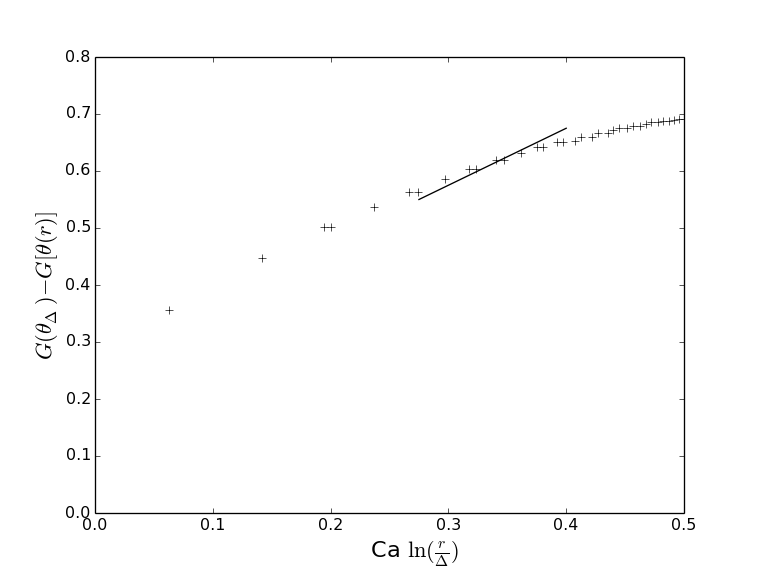} &
\includegraphics[width=2.25in]{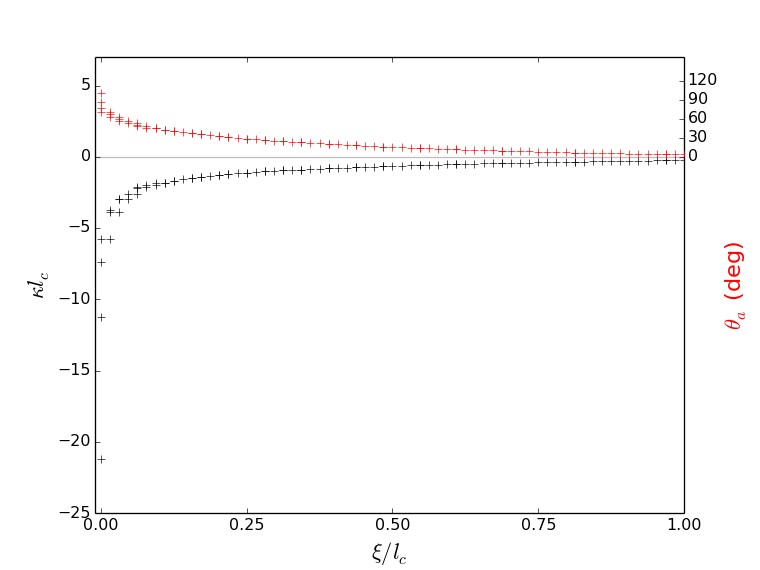}\\
(e)&(f)\\
\end{tabular}
\end{center}
\caption[]{
Left panel: Comparison of the computed (black symbols) $G(\theta_e=\theta_\Delta) - G[\theta(r)]$
versus $\Ca \ln (r/\Delta)$ with the prediction of Eq.~(\ref{fit-phi}) with the best
$\phi$ value of (a) $5.05$, (c) $6.37$, (e) $6.08$ (black solid line).
The fit is performed three to four grid points away from the contact line 
to minimize the inaccuracies  amplified in our numerical method at the grid scale near the contact line.
Right panel: Nondimensional curvature (black symbols) 
and the slope that the interface makes with the substrate
(red symbols), as a function of the nondimensional vertical distance 
of the interface from the contact line, $\zeta/l_c$.  Setup C,
$\theta_{\rm \Delta}=110^\circ$ and $\Delta/l_c = 0.004$ (a-b), $0.008$ (c-d), and $0.016$ (e-f) 
at  $\mbox{Ca}=0.125$, $\tau= 30.05$ (a-b), $\mbox{Ca}=0.135$, $\tau= 47.76$ (c-d),
and  $\mbox{Ca}=0.155$, $\tau= 35.43$ (e-f).}
\label{fig:110VR50}
\end{figure}

Figs.~\ref{fig:15} to \ref{fig:110VR50} depict  
the curvature plots (black symbols) along with the angle (red symbols)
$\theta_a=\theta(\zeta/l_c)$ that the interface makes with the substrate
as a function of $\zeta/l_c$. We find that the maximum curvature is obtained
at the contact line and that this maximum curvature increases with decreasing the mesh
size. As shown, there is a substantial bending of the interface for $\theta_{\rm \Delta}=110^\circ$;
for example, for $\Delta=0.004$,  curvature is tending to zero around $-0.75 l_c$.
However, the maximum curvature does not reach the asymptotic value of $\sqrt 2 l_c^{-1}$ when 
the asymptotic parameter $\eps = \Ca / G(\theta_e)$ is not small enough. As shown,
for small angles, where the smallness of $\eps$ is guaranteed,  the maximum curvature 
approaches $\sqrt 2 l_c^{-1}$.
Figs.~\ref{fig:15} to Figs.~\ref{fig:110VR50} also show that the angle is indeed near zero at inflection 
point when the curvature vanishes as assumed in Sec.~\ref{sec-gen-theo}.

We also compare the computed 
interface slope close to the contact line region with the 
prediction of the Cox's theory by
plotting relation (\ref{fit-phi}) on Figs.~\ref{fig:15} to \ref{fig:110VR50}.
We plot the computed (black symbols) $G(\theta_e=\theta_\Delta) - G[\theta(r)]$
versus $\Ca \ln (r/\Delta)$,
using the computed interface slope, $\theta(r)$, 
at the distance $r=\sqrt{x_{i}^2+y_{i}^2}$,
where $x_{i}$ and $y_{i}$ are the $x$ and $y$ coordinates of the cell center
of an interfacial cell relative to the contact line coordinates. 
Figs.~\ref{fig:15} to \ref{fig:110VR50} also show the
comparison of the computed (black symbols) $G(\theta_e=\theta_\Delta) - G[\theta(r)]$
versus $\Ca \ln (r/\Delta)$ with the prediction of Eq.~(\ref{fit-phi}) using the best
fitted value of $\phi$ (solid lines).
However, we also find three to four aberrant grid points at the smallest values of $r$ corresponding probably to 
the limited accuracy of our numerical method at the grid scale near the contact line. 
Moreover, if the relation  (\ref{fit-phi}) from the Cox-Voinov theory were exact, 
the slope of $G(\theta_e=\theta_\Delta) - G[\theta(r)]$
versus $\Ca \ln (r/\Delta)$ should be unity. In fact, there is a deviation from unity which is
largest at small angles ($15^\circ$ and $30^\circ$) and smallest at large angles.
This is particularly significant since at small angles and small $q$ the theory has a simple derivation from lubrication theory and the assumption of a parabolic flow in the thin liquid wedge. 
A possible explanation is that this theoretical flow is not well approximated 
by the numerical method. Indeed, for a small angle, there is a large region where there are
very few grid points across the thin liquid wedge, with the boundary conditions on the solid and the free surface being imperfectly approximated by the finite volume method and the interpolations used for viscous stresses. This type of fit thus appears to be a good test of the accuracy of the method in the vicinity of the contact line.

The fitting also yields a best value for $\phi$ in the range where Cox's theory is expected
to be asymptotically valid $\Delta \ll r \ll r_{\rm max}$. The expected upper limit 
$r_{\rm max} =  \Ca^{1/3}l_c$ will be derived below. 
We consider the results for $0.004\le\Delta/l_c\le0.016$ and show
that the fitted values of $\phi$ are similar and thus approximately independent of $\Delta$ when 
$\Ca$ is maintained at criticality. We extend the discussion on these results next.

\begin{figure}[]
\begin{center}
\begin{tabular}{cc}
\includegraphics[width=2.15in]{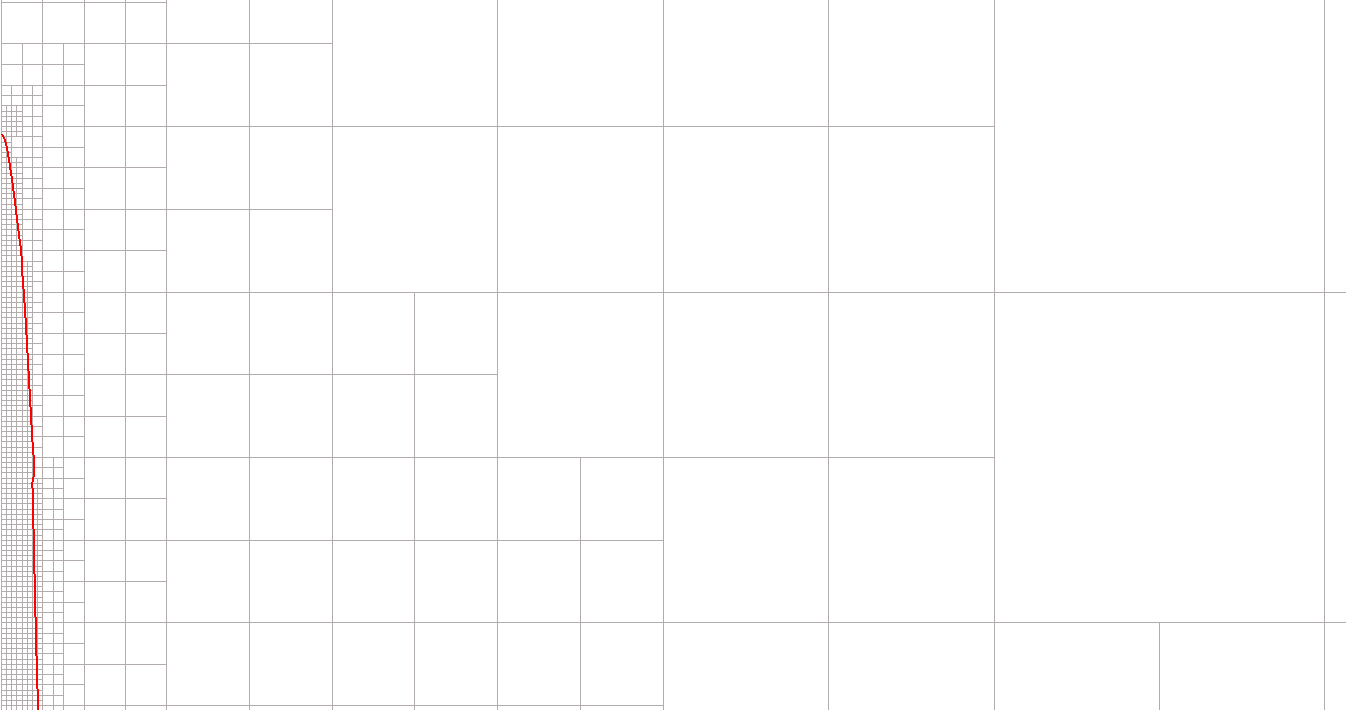} &
\includegraphics[width=2.15in]{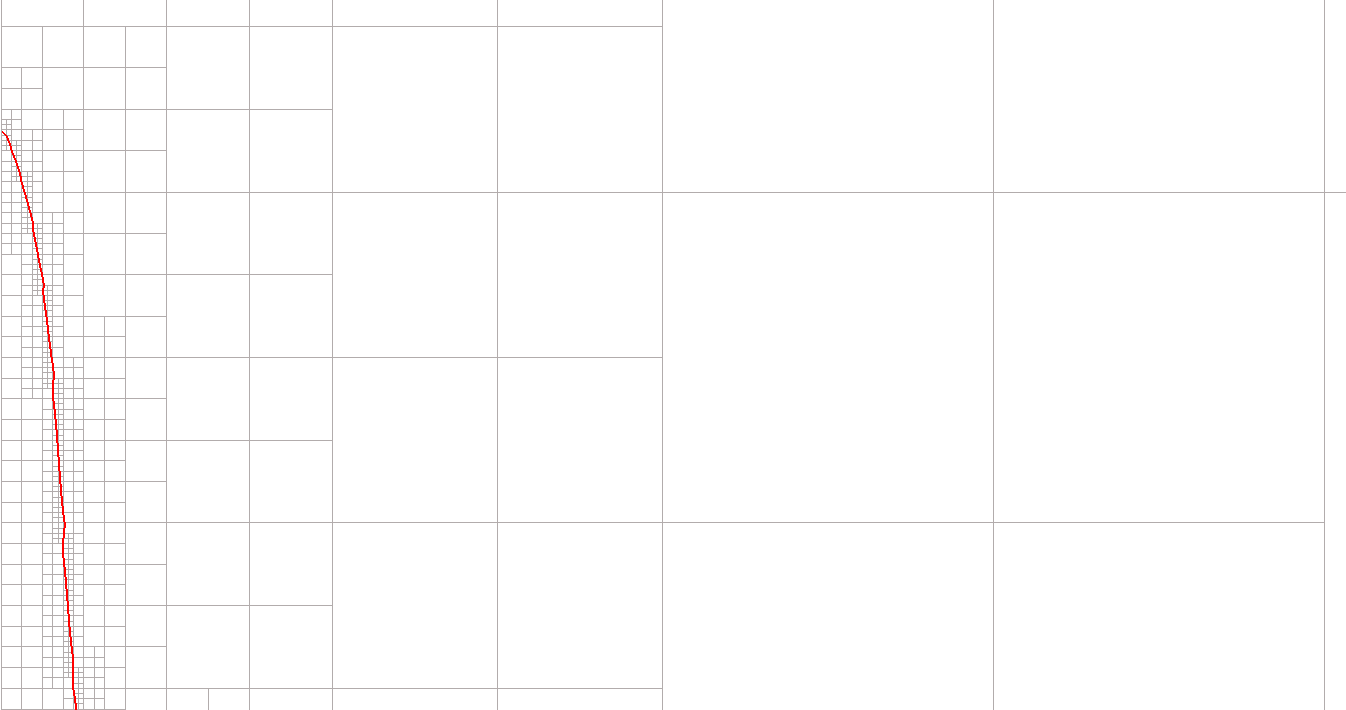}\\
(a)&(b)\\
\includegraphics[width=2.15in]{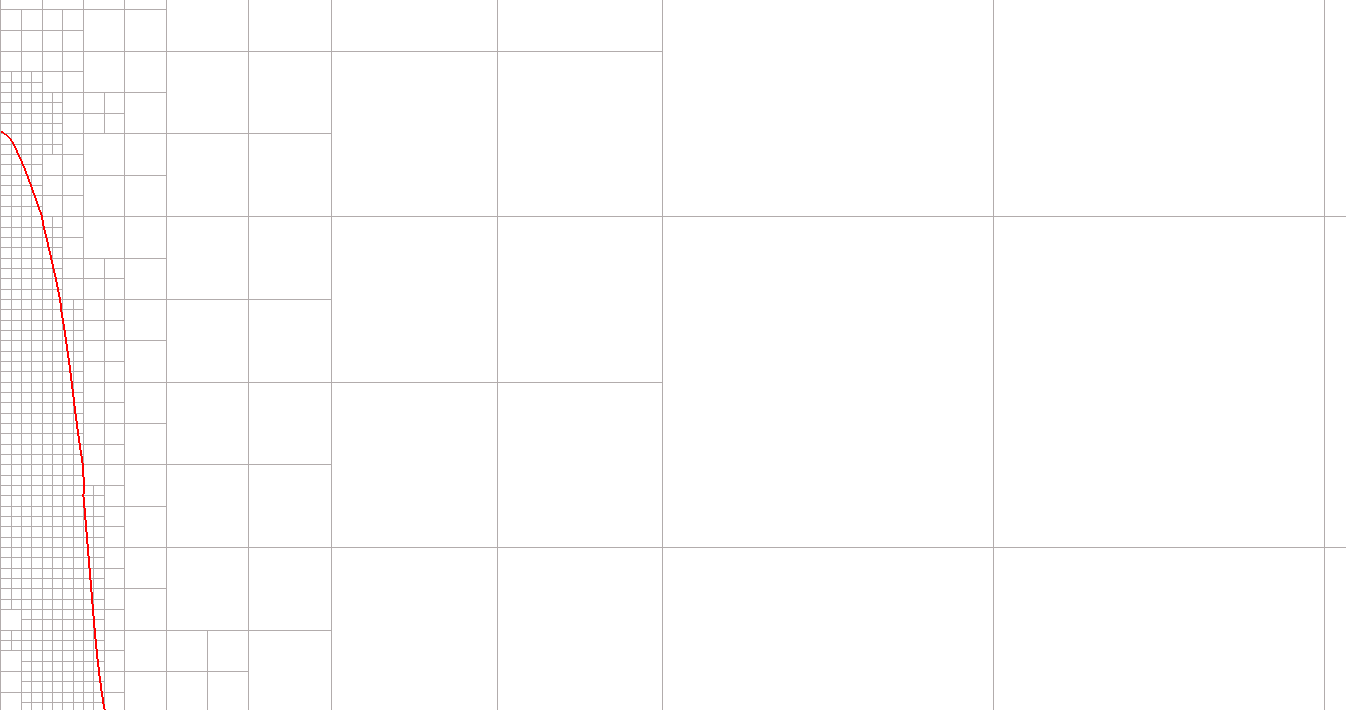} &
\includegraphics[width=2.15in]{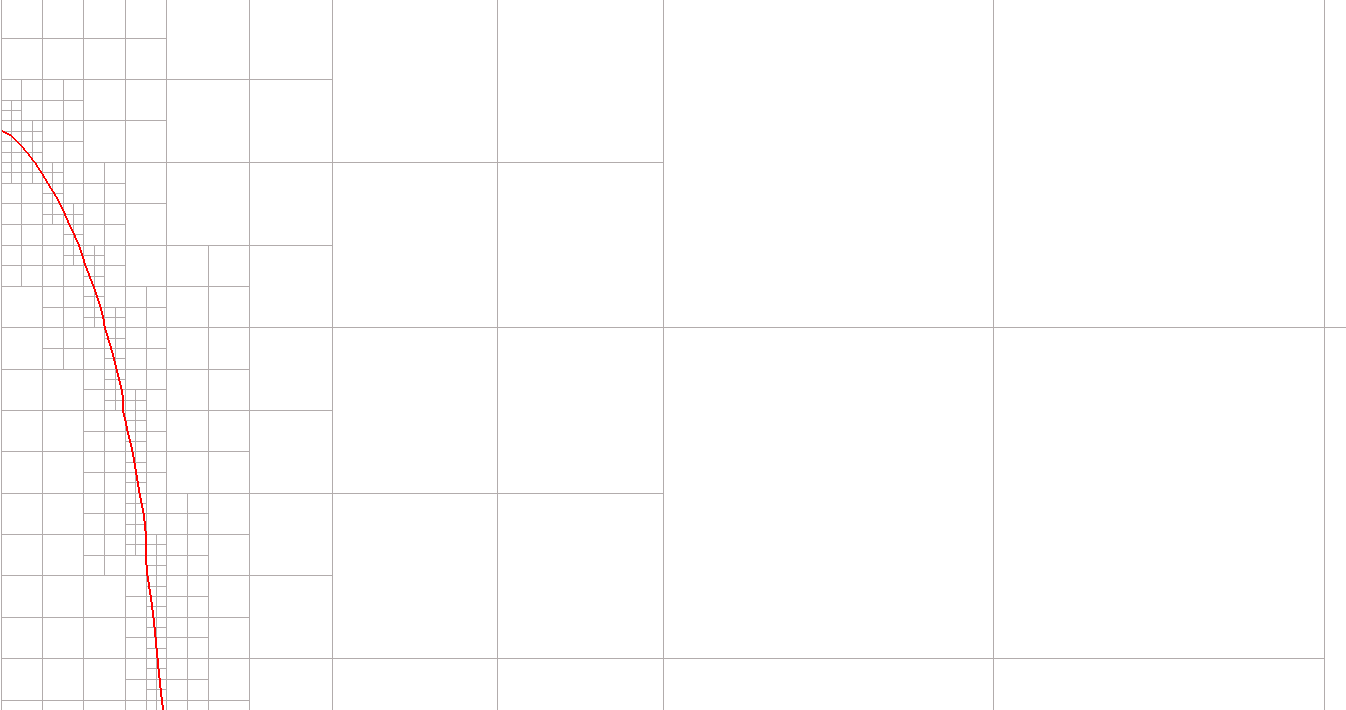}\\
(c)&(d)\\
\includegraphics[width=2.15in]{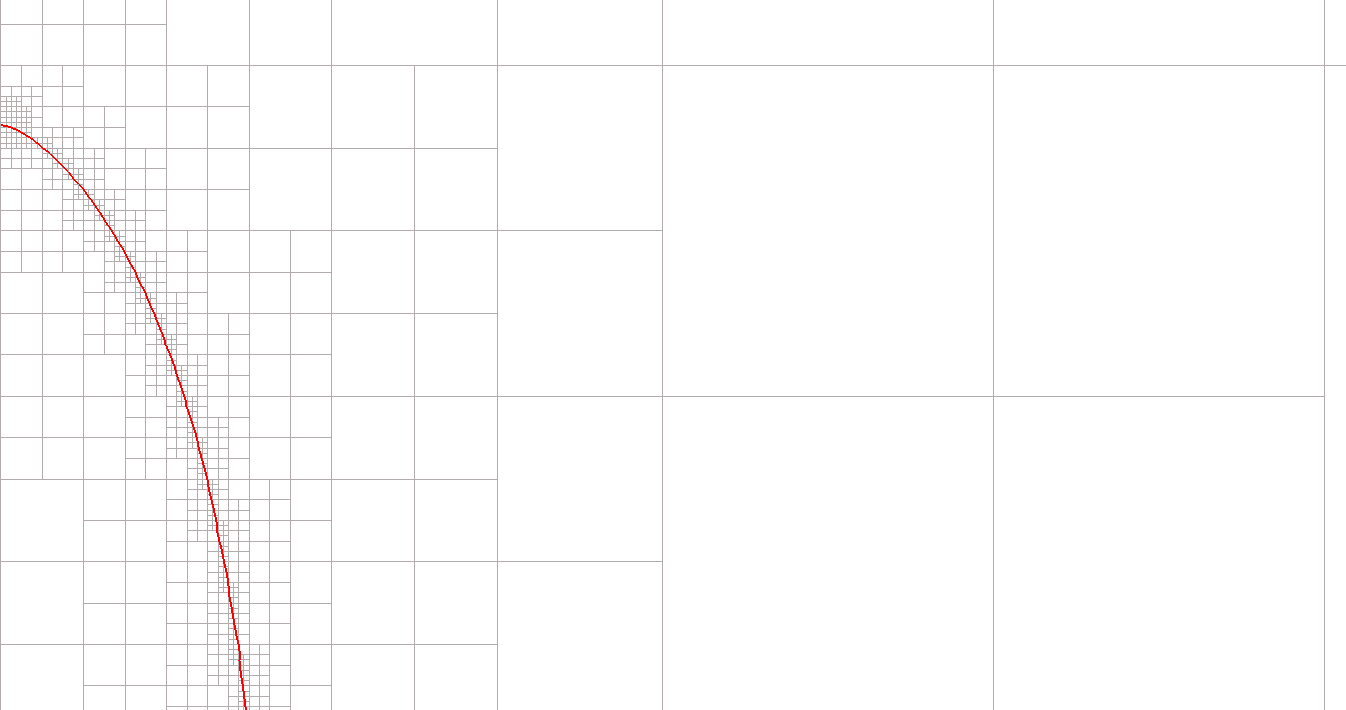} &
\includegraphics[width=2.15in]{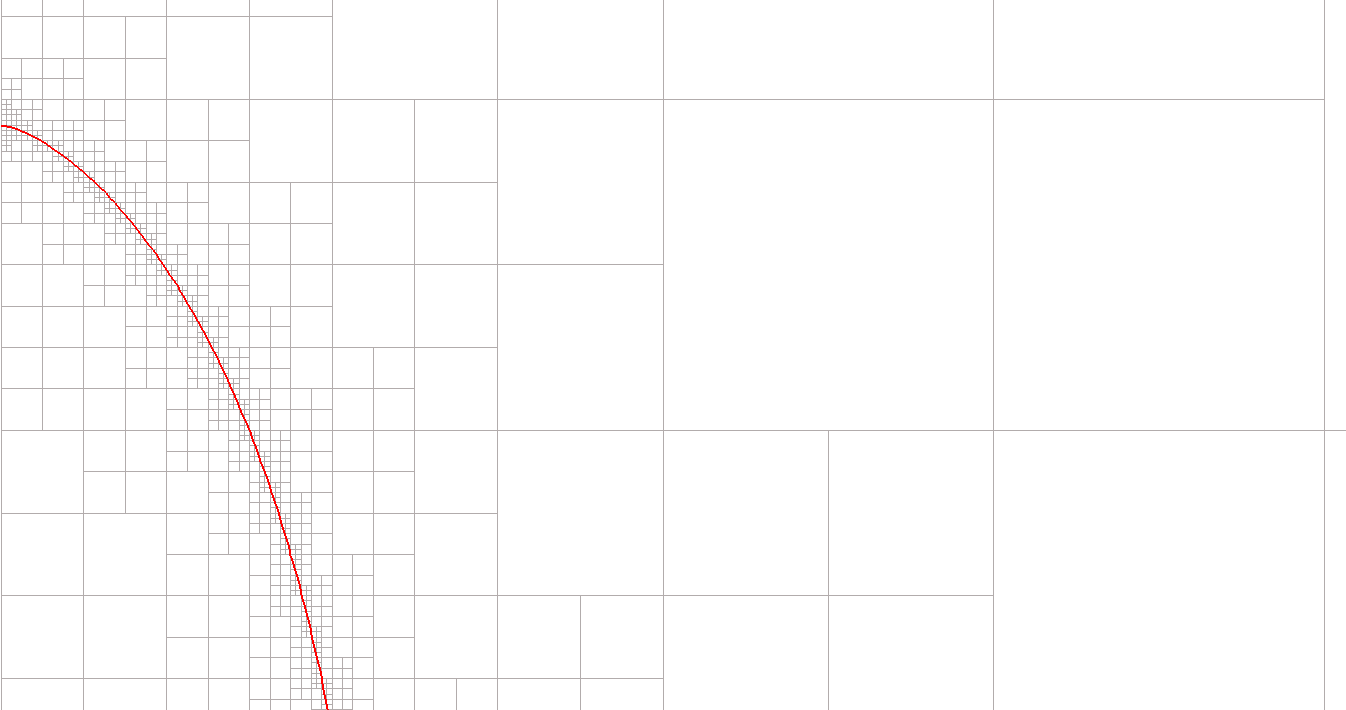}\\
(e)&(f)\\
\includegraphics[width=2.15in]{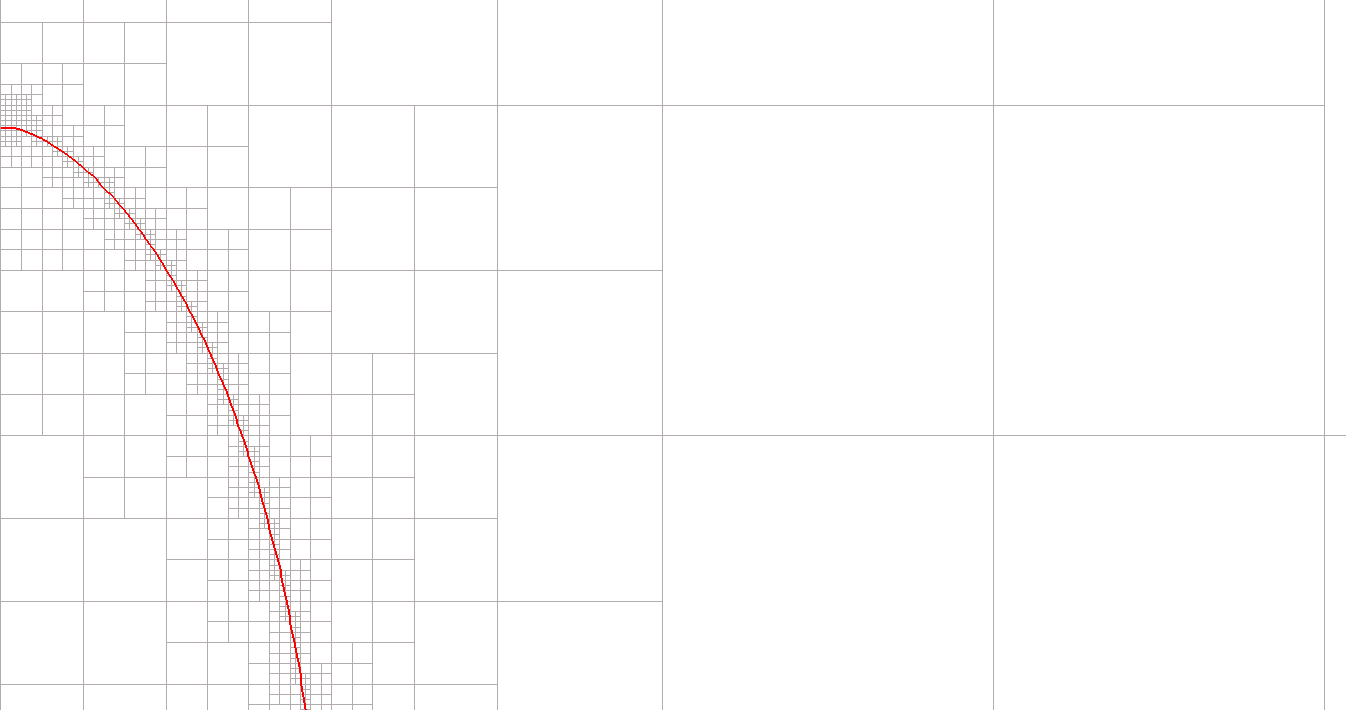} &
\includegraphics[width=2.15in]{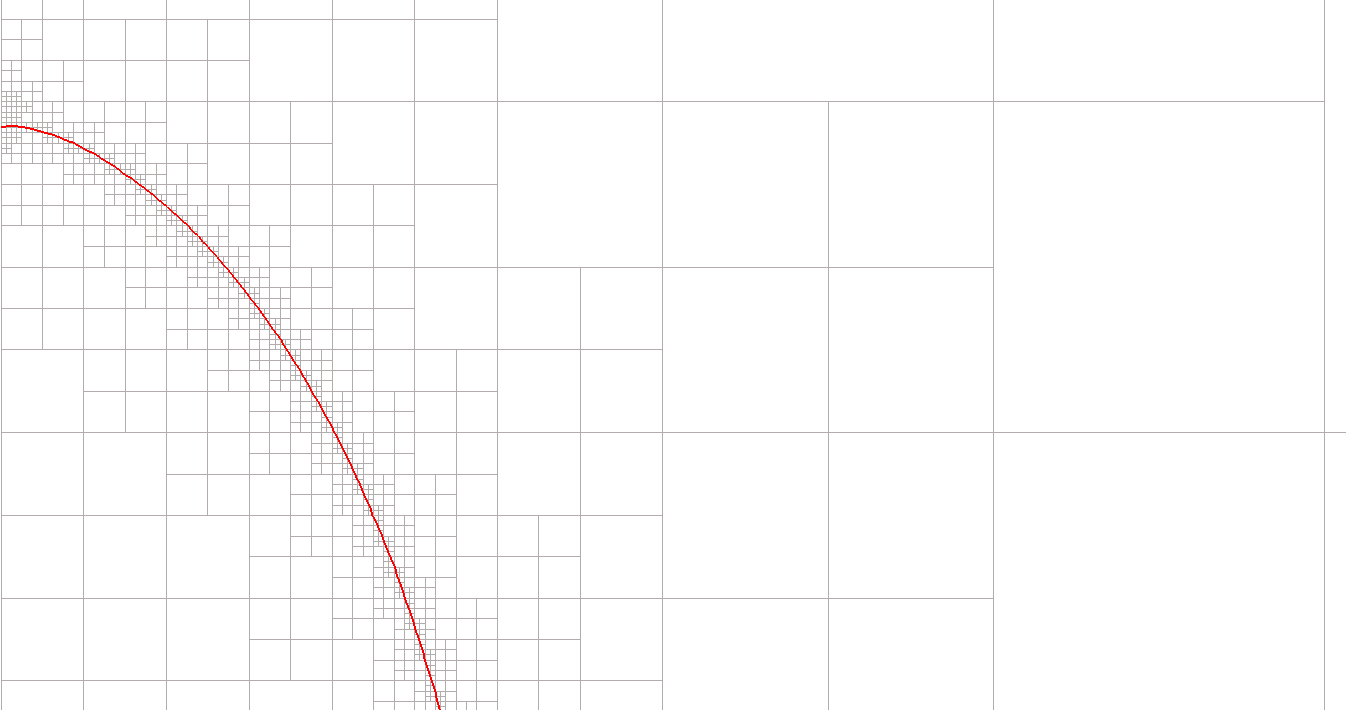}\\
(g)&(h)\\
&\includegraphics[width=2.15in]{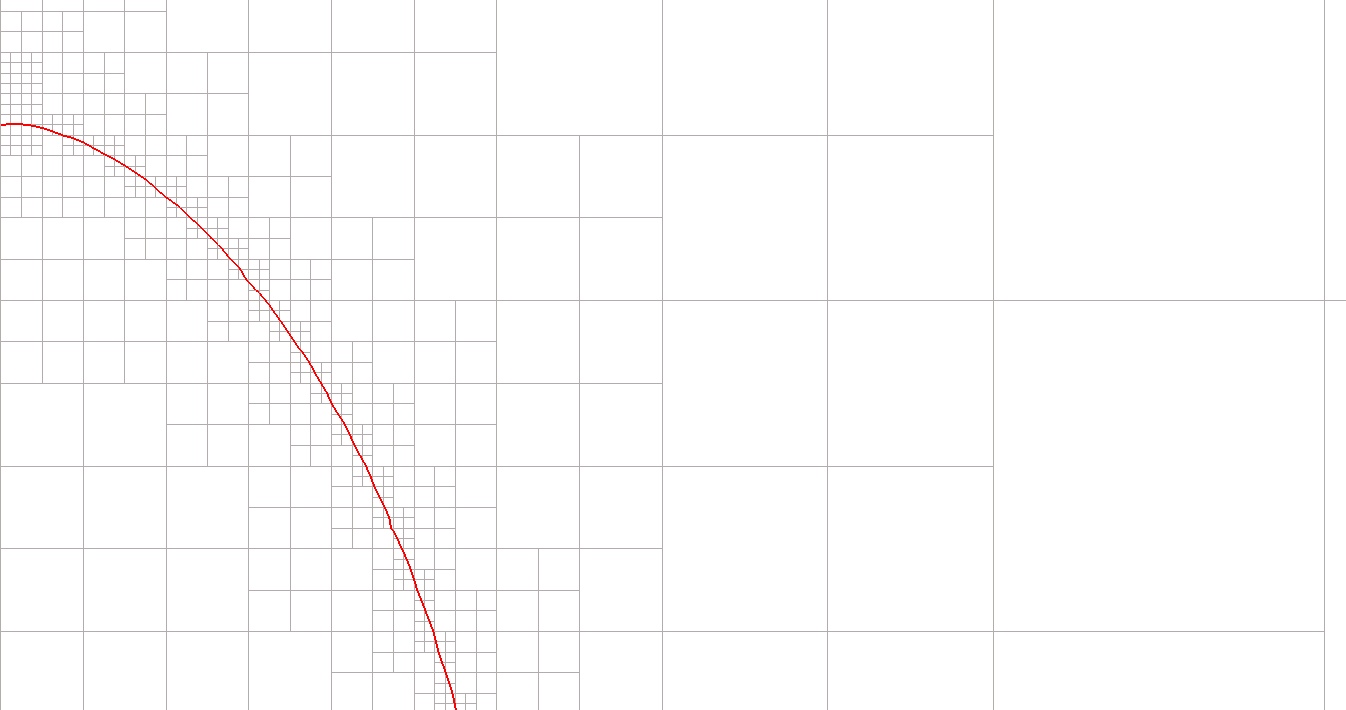} \\
&(i)
\end{tabular}
\end{center}
\caption[]{
Interface profiles and the background adaptive mesh refinement for
(a) Setup A,  $\theta_{\rm \Delta}=15^\circ$, $\Delta/l_c=0.003$, $\mbox{Ca}=0.0009$, $\tau= 1.52$,
(b) Setup C,  $\theta_{\rm \Delta}=30^\circ$, $\Delta/l_c=0.004$, $\mbox{Ca}=0.004$, $\tau= 5.37$, 
(c) Setup A,  $\theta_{\rm \Delta}=40^\circ$, $\Delta/l_c=0.007$, $\mbox{Ca}=0.008$, $\tau= 5.55$,
(d) Setup B,  $\theta_{\rm \Delta}=60^\circ$, $\Delta/l_c=0.016$, $\mbox{Ca}=0.022$, $\tau= 11.12$, 
(e) Setup B,  $\theta_{\rm \Delta}=90^\circ$, $\Delta/l_c=0.004$, $\mbox{Ca}=0.042$, $\tau= 17.41$,
(f) Setup C,  $\theta_{\rm \Delta}=90^\circ$, $\Delta/l_c=0.004$,  $\mbox{Ca}=0.082$, $\tau= 15.79$, 
(g)Setup B,  $\theta_{\rm \Delta}=110^\circ$, $\Delta/l_c=0.004$, $\mbox{Ca}=0.058$, $\tau= 21.67$,
(h) Setup C,  $\theta_{\rm \Delta}=110^\circ$, $\Delta/l_c=0.004$, $\mbox{Ca}=0.125$, $\tau= 30.05$, and
 (i) Setup C,  $\theta_{\rm \Delta}=110^\circ$, $\Delta/l_c=0.008$, $\mbox{Ca}=0.135$, $\tau= 47.76$.}
\label{fig:interfaces15-110}
\end{figure}

Figs.~\ref{fig:interfaces15-110} illustrate the interface profiles corresponding to
examples in Figs.~\ref{fig:15} to \ref{fig:110VR50}. The results show the interface shapes,
close to the contact line where $r \ll r_{\rm max}$, for various cases, demonstrating how the interface
profiles vary at the transition. The background mesh depicts the adaptive mesh refinement
that is utilized for the simulations. The results show the monotonically increasing film thickness
that connects to the flat region of the interface, corresponding to the zero apparent interface angle, 
that is in turn connected to the static liquid reservoir meniscus (only the film near the contact line is shown here).
For $\theta_{\rm \Delta}=110^\circ$, there is a significant bending of the interface, more markedly close to
the contact line. This bending of the interface close to the contact line is better captured at higher
mesh resolutions. As shown, for $\theta_{\rm \Delta}\le30^\circ$, even the smallest grid size appears to be just
sufficient, illustrating the challenging computations at smaller angles.   

\subsection{General case}
\label{sec-large-slope}
In the arbitrary angle case or when $q$ does not vanish, one cannot use lubrication theory as done in
\cite{Eggers2004b}. 
However we can proceed by assuming a vanishing apparent contact angle at the transition. 
For a vertical plate, this means that at the inflection point the slope vanishes
($\eta' = 0$ when $\eta''=0$). 
We have indeed confirmed this observation by our numerical results. As shown in  
Figs.~\ref{fig:15} to \ref{fig:110VR50} (b,d,f), for the range of contact angles and mesh sizes considered 
here, $15^\circ\le\theta_{\rm \Delta}\le110^\circ$ and $0.004\le\Delta/l_c\le0.016$,
the vanishing slope of the interface coincides with the inflection point at the transition,
confirming the criteria for the dewetting transition.


Thus there is a region IV of small slope overlapping with region II 
and III in which the lubrication approximation is
valid. However because the slope is not small everywhere in region II the lubrication approximation does not
apply there and asymptotic matching is less obvious than
in the literature. One possibility would be to use a generalized differential equation also valid 
for large slopes as done in \cite{Snoeijer06}. A mapping from the equation valid for
large slopes in \cite{Snoeijer06} to the lubrication approximation
for small slopes has been performed in \cite{Qin:18at}, allowing to reuse the result in 
\cite{Eggers2004b}; however, for several reasons explained below, we prefer to perform directly a matching 
from Cox's theory valid from region II to region IV.

In what follows, we first perform a heuristic derivation of the central relation of our theory, that gives the 
critical value of the capillary number $\Ca_{cr}$ implicitly as function of the other parameters of the problem.
We then compare these results to those of lubrication theory obtained for small angles, and to our numerics for
small angles. 

\begin{figure}[]
\begin{center}
\includegraphics[width=4in]{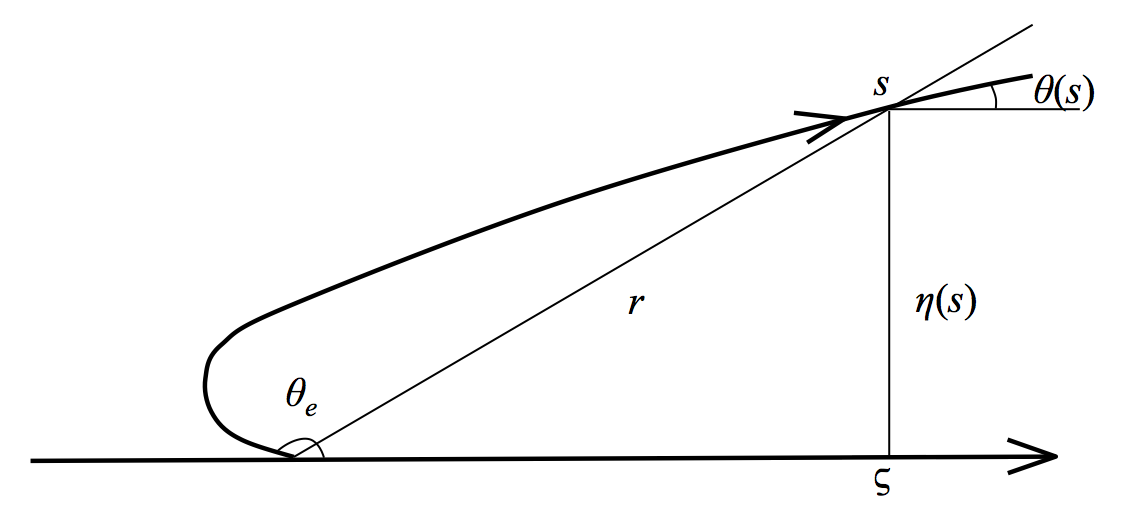}
\end{center}
\caption{The geometry of the wedge. The case of a large $\theta_e$ angle at the microscopic scale is 
illustrated. The various geometrical quantities 
$s,r,\eta,\zeta,\theta$ and $\theta_e$ are discussed in the text. The figure emphasizes the important fact that our theory is also valid for $\theta_e > \pi/2$ but that despite this fact we have $r/s$ ~ $\zeta/s ~ 1$ at large $r$.}
\label{curvilinear-geometry}
\end{figure}

The heuristic description  starts by determining the scale of $\Ca$ and the thickness of the liquid in region III near the inflexion point.
Using the Cox-Voinov analysis, the slope is given by Eq.~(\ref{Voinovform}) 
and we write the condition that it vanishes in region IV. Let the inflection point occur
at $\zeta \sim \zeta_I$ with $\zeta_I \gg r_m$. The bending occurs over a large distance $\eta_I$ and the
``final phase'' of the bending where the slope and the curvature are small occupies almost all
of region II. Thus region II ``looks like'' a very thin wedge seen at scale $\eta_I$ (paralell to the wall) 
and $\zeta_I$ (perpendicular to the wall). The meaning of all the geometrical quantities
$s,r,\eta,\zeta,\theta$ and $\theta_e$ is summarized on Fig.~\ref{curvilinear-geometry}. 
As a  result of the thinness,
we can approximate $r \sim \zeta$ in that region. 
Writing the small slope assumption $\theta = 0$ at $r=\zeta_I$ in  Eq.~(\ref{Voinovform}) gives
\be
\Ca_{cr} = \frac{G(\theta_e)}{ \ln (\zeta_I/r_m)},
\nd
and $\Ca_{cr}$ is small if $1/\ln (\zeta_I/r_m)$ is small. The small parameter in the expansion
is 
\be
\epsilon = 1/\ln(\zeta_I/r_m) = {\Ca_{\rm cr}}/{G(\theta_e)}.
\nd
This determines the scale of $\Ca$ and we now turn to determining 
the thickness $\eta_I$  of the liquid in region III near the inflection point. 
This is not as trivial as it may seem since in addition to the $\eta_I \sim \Ca^{2/3} l_c$
solutions near the inflection point that match the meniscus, there are also solutions with an 
asymptotically flat interface,
$\eta \rightarrow \eta_\infty$ as $\zeta \rightarrow \infty$. These solutions have $\eta_\infty \sim \Ca^{1/2} l_c$ and 
involve a balance of gravity and viscosity instead of a balance of surface tension and viscosity. 
A solution may be found trivially by letting $\eta=$ Constant 
in the generalized lubrication equation or in the large slope equation in \cite{Snoeijer06}. 
It is thus a fixed point solution. The following serves to show that one matches 
to the correct inflection point solution, not to the fixed point solution. 
Recalling that $\Ca$ is small, it is possible to deduce that the curvature is still small
in most of region II thus $ds/dr \sim 1$ where $s$ is the curvilinear abscissa along the interface. 
Then
\be
\eta_I \sim \int_{r_m}^{\zeta_I} \sin \left\{ G^{-1} \left[ G(\theta_e) - \epsilon G(\theta_e) \ln\left( \frac{r}{r_m} \right) \right] \right\} dr,
\nd
where we have used the fact that  $r_I \sim \zeta_I$, $d\eta/ds = \sin \theta(s)$ and $\Ca \simeq \Ca_{\rm cr}$. 
After some work it can be shown that this integral behaves as 
\be
\eta_I \sim \eps^{1/3}  G(\theta_e)^{1/3} \exp \left(\frac 1 \epsilon \right) r_m,
\nd
while by definition
\be
\zeta_I \sim  \exp \left(\frac 1 \epsilon \right)  r_m.
\nd
We thus have $\eta_I \sim \Ca^{1/3} \zeta_I$ as expected for the inflection point
in the thin film dewetting transition.
In region IV, the length scales are thus $\eta_I$ across the film and $\zeta_I$ parallel to the film. 
The curvature thus scales as 
\be
\eta^{\prime\prime} \sim \eta_I/\zeta_I^2 \sim \eps^{1/3} G(\theta_e)^{1/3} \exp \left(- \frac 1 \epsilon \right)  r_m^{-1}.
\nd
To match region IV and region III, one should have the curvature $\eta^{\prime\prime}$  of same
order as the curvature  $l_c^{-1}$ of the static meniscus. Writing $\eta^{\prime\prime} \sim l_c^{-1}$ one obtains
\be
\eps^{1/3}  G(\theta_e)^{1/3}   \exp \left(- \frac 1 \epsilon \right) \frac{l_c}{r_m} \sim 1. \label{condexp}
\nd
Identifying $\epsilon$ as above
\be
 \frac{C(q) \Ca_{cr}^{1/3} l_c}{r_m} \exp \left[ - \frac{G(\theta_e)}{\Ca_{cr}} \right] = 1, 
\nd
where we have introduced a constant $C(q) = \Or(1)$.

In the free surface case when $q=0$, we may match region III with region IV in a manner similar to the one in
\cite{Eggers2004b}, as shown in \ref{app2}, to obtain 
\be
C(0) = \frac{3^{1/3} 2^{-1/3}}{\pi \red{\rm e} {\rm Ai}^2(s_{\rm max}) \kappa_\infty l_c}.
\nd
Or equivalently (see \ref{cacr1})
\be
 \frac{3^{1/3} 2^{-1/3} \Ca_{cr}^{1/3}}{\pi  \red{\rm e} {\rm Ai}^2(s_{\rm max}) r_m \kappa_\infty} \exp \left[ - \frac{G(\theta_e)}{\Ca_{cr}} \right] = 1. \label{matchcq}
\nd
It is possible that $C(q)$ does not differ very much from $C(0)$ if in region IV
the outer fluid stress is negligible. Indeed 
in the thin layer the stress scales as $\mu_1 V_s/\eta(y)$ while the stress in the outer fluid 
scales as the much smaller $\mu_2 V_s / y$, realizing an approximation of the free surface condition.

  We note that expression (\ref{matchcq}) is equivalent at small
  angles $\theta_e$ to the one obtained in \cite{Eggers2004b}. It is also
  equivalent to the expression obtained in \cite{Qin:18at} at all angles   apart 
  for the fact that we make no
  hypothesis on the behavior of $r_m$ at large angles, unlike
  \cite{Qin:18at} and \cite{Snoeijer06}. Moreover, as we wrote above,
  we prefer not to obtain our expression through a generalized
  lubrication equation, but rather through asymptotic matching. We
  believe that the effect of some models of the microscopic physics,
  such as the slip length model, in region I, cannot be found for large
  angles by asymptotic analysis but would require a full solution of
  the Stokes equation analogous to that of Huh and Scriven \cite{HuhScriv}, but for
  different boundary conditions. 

We also note that at the transition the condition (\ref{condexp}) yields
$\eta_I \sim \Ca^{2/3} l_c$, and that we used essentially the same matching argument as 
Derjaguin \cite{Derjaguin1943} and Landau and Levich \cite{LandauLevich1942}. 
This can also be compared to the prediction of
Voinov \cite{Voinov00} which is similar but with a constant $C(0)$ differing from that obtained
by asymptotic matching. 

Considering now the numerical case, we substitute the numerical expression (\ref{phi-num}) for the microscopic
length scale $r_m$ in the critical matching expression (\ref{matchcq}) to obtain the central relation of our theory
\be
 \frac{C(q) \phi(\theta_\Delta,q) \Ca_{cr}^{1/3} l_c}{\Delta} 
\exp \left[ - \frac {G(\theta_\Delta)}{\Ca_{cr}} \right] = 1, \label{cacr_cq}
\nd
where we identified $\theta_\Delta$ with $\theta_e$. 

It is interesting to investigate explicit expressions for $\Ca_{\rm cr}$ obtained by an asymptotic analysis of 
Eq.~(\ref{cacr_cq}). To perform this analysis, we let $\delta =  1/\ln({l_c/\Delta})$. This small parameter
is of the same order as $\eps$ above but unlike $\eps$ is explicitly defined using the parameters of the problem. 
To give an order of magnitude for $\delta$ and $\epsilon$, if we use, as in the above simulation, a 
grid size $\Delta /l_c = 1/2^8$, we obtain $\delta \approx 0.18$. 
Thus our small parameters is small but not exceedingly so.
It is useful to introduce the  auxiliary parameter ${\hat \mu}$ defined as
${\hat \mu} = - 1/\ln[ C(q) \phi(\theta_\Delta,q) G(\theta_\Delta,q)]$. 
The parameter ${\hat \mu}$ will be  small for transcendentally small $\theta_\Delta$. 
Otherwise if $\hat \mu$ is considered order 1, 
solutions of  Eq.~(\ref{cacr_cq}) expand as 
\be
\Ca_{\rm cr} = \delta G(\theta_\Delta,q) \left[ 1 - \delta \ln \delta - \delta {\hat \mu}^{-1} + \Or(\delta^2 \ln \delta)
+ \Or(\delta^2 {\hat \mu}^{-2}) \right].  \label{asympcr}
\nd
At small angles, with the expected scaling $\phi(\theta_\Delta) \sim \theta_\Delta$, one gets
$\hat \mu \sim - 1/(4 \ln \theta_\Delta)$. Moreover the higher order terms in Eq.~(\ref{asympcr}) are small
if ${\hat \mu} \gg \delta$ which is
verified even for the smallest $\theta_\Delta$ and typical values of $l_c/\Delta$ in our calculations. 
For Setup B, with $\theta_\Delta = 110^\circ, \Delta/l_c = 1/2^8, q=1$, and using the best fit estimate
$\phi \simeq 3$, the first order
of expansion of Eq.~(\ref{asympcr})  gives  $\Ca_{\rm cr} \simeq G \delta \simeq 0.0558$, using all the terms in 
 expansion (\ref{asympcr})  gives  $\Ca_{\rm cr} \simeq 0.0626$, while a full iterative solution of
Eq.~(\ref{cacr_cq}) yields $\Ca_{\rm cr} \simeq 0.0637$.
For Setup C, in the case  $\theta_\Delta = 110^\circ, \Delta/l_c = 1/2^8, q=0.02$, and using the best fit estimate
$\phi \simeq 3.5$, the first order gives  $\Ca_{\rm cr} \simeq G \delta \simeq 0.127$, using all the terms in 
 expansion (\ref{asympcr})  gives  $\Ca_{\rm cr} \simeq 0.169$, while a full iterative solution of
Eq.~(\ref{cacr_cq}) yields $\Ca_{\rm cr} \simeq 0.132$.  Thus in a wide range of parameters, the approximation
\be
\Ca_{\rm cr} = \frac{G(\theta_\Delta,q)}{\ln({l_c/\Delta})},
\nd
is correct within 15\% accuracy.

\subsection{Comparison with numerics}

In our simulations, we measure $\Ca_{cr}$ and specify $\Delta$
and $\theta_{\Delta}$. For the two special cases described above, we compare the value of
$\Ca_{\rm cr}$ from full simulations to the solutions of the central relation,  Eq.~(\ref{cacr_cq}).
The simulations values are in remarkable agreement with the above solutions. 
For setup B, with $\theta_\Delta = 110^\circ, \Delta/l_c = 1/2^8, q=1$, we find numerically
$Ca_{cr} = 0.0555$ compared to the theoretical value $0.0637$. 
For setup C, in the case $\theta_\Delta = 110^\circ, \Delta/l_c = 1/2^8, q=0.02$,
we find numerically
$Ca_{cr} = 0.12125$ compared to the theoretical value $0.132$. 

Alternately, to perform the comparison, Eq.~(\ref{cacr_cq}) can be inverted to provide an estimate of $\tilde \phi = C(q)/C(0) \phi$ given the 
numerical estimate of $\Ca_{cr}$. The estimate is
\be
\tilde \phi(\theta_\Delta,q) =
\frac{\pi\red{\mathrm e} {\rm Ai}^2(s_{\rm max})}{3^{1/3} 2^{-5/6} }
\frac {\Delta}{\Ca_{cr}^{1/3} l_c}
\exp \left[ \frac {G(\theta_\Delta,q)}{\Ca_{cr}} \right]. \label{phicomp}
\nd
If our theory is valid, the right hand side should not depend on $\Delta$. Moreover 
if lubrication theory can be used in region IV, as is possibly the case
even for $q>0$, the above expression should yield the gauge function $\phi$ defined in relation
with Cox's analysis and estimated numerically above.
 
\begin{figure}[]
\begin{center}
\begin{tabular}{c}
\includegraphics[width=4in]{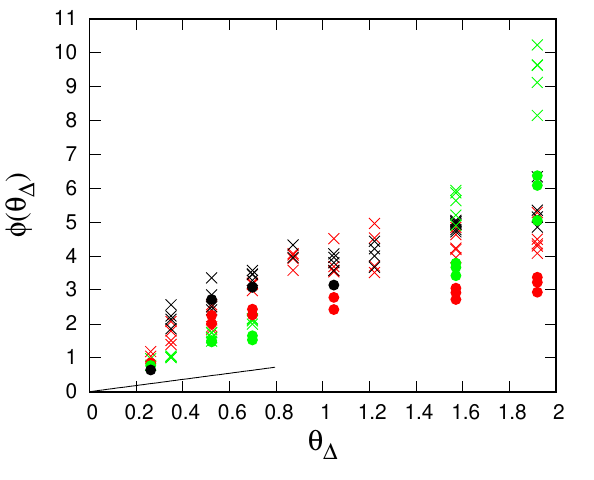}
\end{tabular}
\end{center}
\caption
{The gauge function $\phi$ plotted using expression (\ref{phicomp}) above, 
for Setups A ({\color{black}$\times$}), B ({\color{red}$\times$}), and C ({\color{green}$\times$}),
compared to the computed values from the best fit of $\phi$ for 
Setups A ({\color{black}$\bullet$}), B ({\color{red}$\bullet$}), and C ({\color{green}$\bullet$}).
The solid line is the prediction from lubrication theory  $\phi = {\mathrm e} \theta_\Delta/3$.
For a given angle $\theta_\Delta$, the various values of $\phi$ correspond to the 
various values of the grid size $\Delta$ for each Setup. }
\label{phifig}
\end{figure}

In Fig.~\ref{phifig}, we plot the values of the RHS of  Eq.~(\ref{phicomp}) 
for Setups A, B and C, along with the computed values from the best fit of $\phi$ presented in Sec.~\ref{sec:comp}
(see Figs.~\ref{fig:15} to \ref{fig:110VR50}).
In the near-free-surface Setup C ($q=1/50$),  a very clean estimate of the gauge function $\phi$
is obtained at small angles and we find 
$\phi(\theta_\Delta) \simeq  \theta_\Delta$. Numerical results from the direct comparison
with Eq.~(\ref{fit-phi}) are also in reasonable agreement with the predictions at small angles. 
Eq.~(\ref{effslip}) therefore suggests an effective slip length $\lambda \simeq \Delta$.
Thus our numerical model may be viewed as having an
effective slip of the order of a grid cell. 

\begin{figure}[]
\begin{center}
\begin{tabular}{cc}
\includegraphics[width=2.5in]{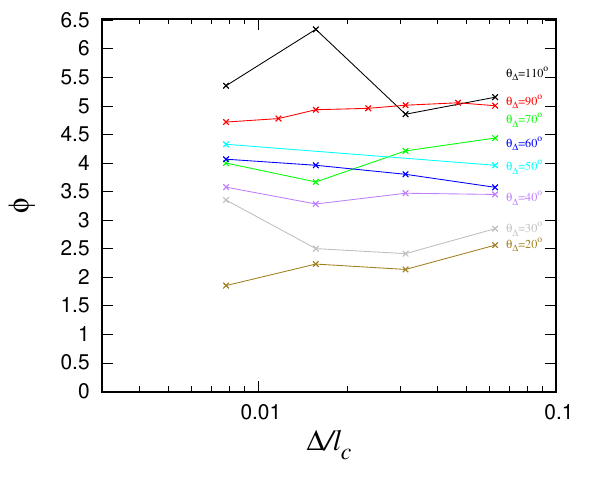} &
\includegraphics[width=2.5in]{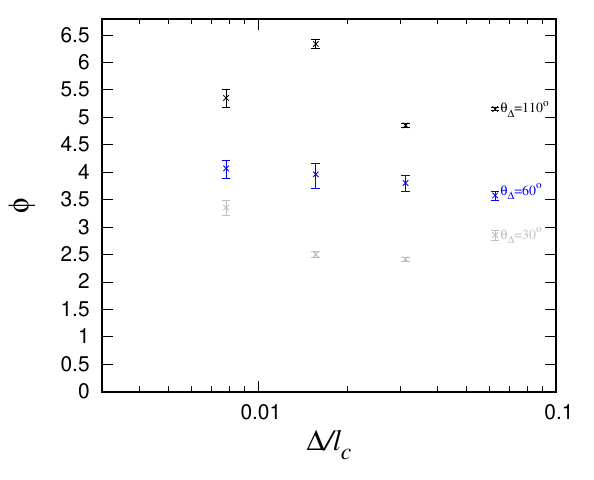}\\
(a)&(b)\\
\includegraphics[width=2.5in]{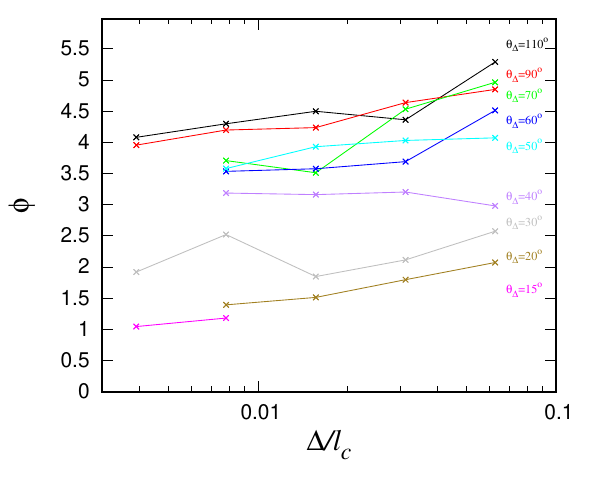}&
\includegraphics[width=2.5in]{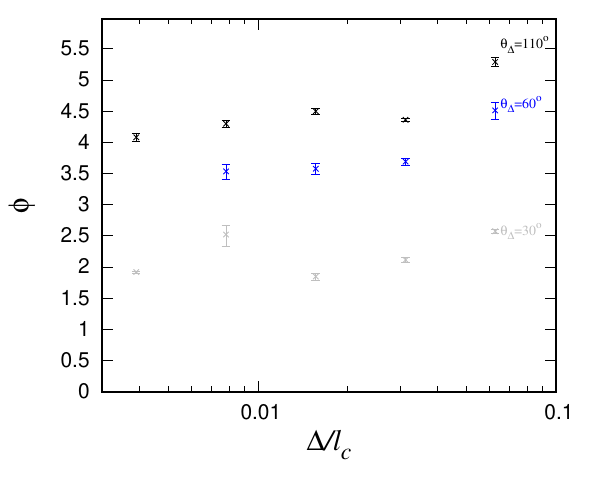}\\
(c)&(d)\\
\includegraphics[width=2.5in]{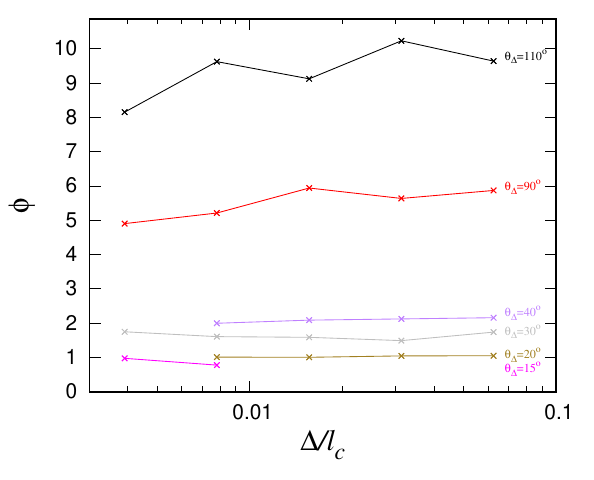}&
\includegraphics[width=2.5in]{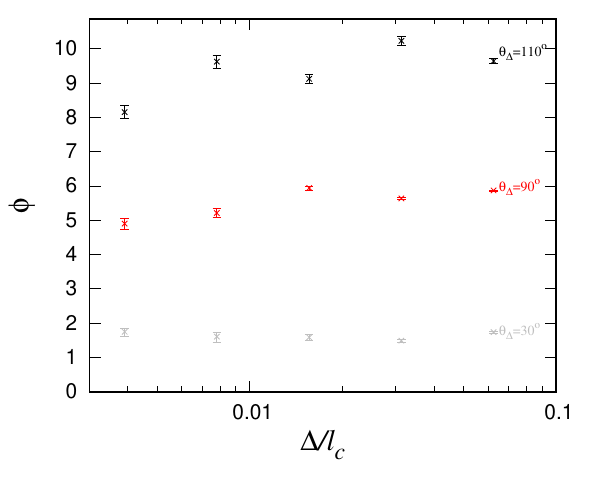}\\
(e)&(f)\\
\end{tabular}
\end{center}
\caption[]{
$\phi(\theta_\Delta)$ as a function of $\Delta/l_c$ for
(a) Setup A,  (c) Setup B,  and (d) Setup C,  
and exemplifying the errors
in the computation of $\phi$ for
(b) Setup A,  (d) Setup B,  and (f) Setup C.}
\label{fig:phi_error}
\end{figure}

In Setups A and B, the plot is more scattered indicating deviations from our theory, which is not surprising
given the not-so-small value of the asymptotic parameter. However, a rough proportionality 
to the angle $\theta_\Delta$ is observed at small angles. Moreover, for all the Setups,
large contact angle data do not follow the linear proportionality to the  angle $\theta_\Delta$, with 
the viscosity ratio having a more prominent effect on this deviation at large angles. To shed more light
on the behavior of $\phi$ as a function of the mesh size, as well as exemplifying the errors
in the computation of $\phi$, in Fig.~\ref{fig:phi_error}, we plot  $\phi(\theta_\Delta)$ versus
$\Delta/l_c$. The results show some small variations; this is however expected because the error
in estimating $\phi$ also depends on $\Delta$, while decreasing the mesh size does not reduce the
transition capillary by much.

\newcommand\solidrule[1][2mm]{\rule[0.5ex]{#1}{.9pt}}

\section{Dynamic inner contact angle} 
\label{sec:dynamic}

A possible reinterpretation of Cox's theory is to consider that the
inner region angle is not the equilibrium angle but an angle $\theta_{in}$ depending
on the contact line velocity so that
\be
\theta_{in} = f_\theta (\Ca,q), \label{dynamic_innner}
\nd
where dependence on the fluids and solid material properties is implicit.
Then Eq.~(\ref{eq:cox_eq_2_m}) must be rewritten as 
\be
G[\theta(r)] = G(\theta_{in}) - \Ca \ln (r/\lambda)  - \Ca \frac{ Q_i }{f(\theta_{in},q)}
+ o(\Ca), \label{eq:cox_eq_in}
\nd
and we can proceed as before. We note however that there are two possibilities.

1) The microscopic angle is discontinuous, that is there are two angles,
the advancing angle $\theta_a$ and the receding angle $\theta_r$, so that
$\theta_r < \theta_e < \theta_a$ and that for small $\Ca$,
$\theta_{in} \sim \theta_a$ for an advancing contact line and 
$\theta_{in} \sim \theta_r$ for a receding contact line. The conclusion is that
the asymptotic analysis is unchanged with $\theta_e$ replaced by either $\theta_r$ or $\theta_a$.

2)  The microscopic angle is continuous so that at first order
\be
\theta_{in} \sim \theta_e + \Ca f^{\prime}_{\Ca}, \label{linear}
\nd
 with  $f^{\prime}_{\Ca}$ a constant. 
In this case, and the small $\Ca$ assumption, Eq.~(\ref{eq:cox_eq_in}) can be rewritten
at first order in $\Ca$ in the form of Eq.~(\ref{eq:cox_eq_2_m})
\be
G[\theta(r)] = G(\theta_{e}) - \Ca \ln (r/\lambda)  - \Ca \frac{ Q^\prime_i }{f(\theta_{e},q)}
+ o(\Ca) \label{eq:cox_eq_in_2}
\nd
the change in microscopic angle with $\Ca$ being absorbed into 
the new integration constant $Q^\prime_i$. 
We note that the relation (\ref{linear}) is in agreement with the 
Molecular Kinetics Theory (MKT) of Blake and Haynes \cite{Blake1969},
which states that the contact line is displaced by small random molecular jumps at the surface of the solid
due to thermal fluctuations. 
The average size of these small jumps is determined by the intermolecular interactions between the liquid and solid.
The frequency of these random molecular jumps and the average distance of each jump
can either be determined empirically or by direct comparison to molecular dynamics simulations.
This model postulates that there is an out-of-balance surface tension force, of non-hydrodynamic origin,
as a result of the contact line moving on a solid surface. 
The model then relates  the contact line speed 
to this driving force, resulting in a dynamic contact angle.

The idea of advancing and receding angles is related to the idea of physical 
roughness of the surface. This may connect to the idea of a discontinuity in the global numerical solution 
itself.  We have conducted 
tests of the numerical model for advancing menisci for $\Ca$ down to $10^{-6}$, 
although the dewetting transition itself was studied only for $\Ca > 10^{-4}$. 
For very small $\Ca$, the contact line evolves in an irregular manner, with 
intermittent spikes in velocity akin to the motion on a rough surface. 
These spikes have a time periodicity of $T=\Delta/V_s$ as if the interface was pinned
with spatial periodicity equal to the grid size $\Delta$. However, no such irregularity
is observed for $\Ca = 10^{-4}$. Thus we may infer that at least in the range of parameters 
used in this study $10^{-3} \lesssim \Ca \lesssim 0.1$, the numerical solution is a continuous
function of the contact line position. 

{
\section{Consequences for dynamic contact line computations in general}
\label{sec:neo-AZB}

How to conduct realistic simulations of flows including a moving contact line in practice is
a difficult problem that has elicited significant research efforts (See \cite{Legendre2015}
for a systematic comparison of some models.) Here we only discuss how our 
Cox type analysis of numerical models connects with other work on mesh independent
simulations. In the computation of a number of problems involving dynamic contact lines,
such as droplet spreading, drop impact or drop sliding on inclined plates, the result typically
varies with grid resolution because of hydrodynamic effects. 
One already-suggested approach \cite{Legendre2015} 
is to specify numerically the angle 
\be
G(\theta_\Delta) = G(\theta_{in}) - \Ca \ln (\Delta/r_m) \label{maglio_model}
\nd
for some $\theta_{in}$ that may be function of the capillary number as in Eq.~(\ref{dynamic_innner}),
and for some microscopic length $r_m$. 
We remind the reader that  $\theta_{\Delta}$ is the numerically imposed contact angle 
while $\theta_{in}$ is an experimentally measured or determined quantity whose meaning is the contact angle at
the microscopic scale $r_m$. Since $r_m$ and $\Delta$ are not necessary equal and 
not even of the same order of magnitude, we distinguish these two quantities.
The loose rationale for expression~(\ref{maglio_model}) is that if $\theta_\Delta$ is the interface
angle at distance $\Delta$, then the Cox theory, i.e.~Eq.~(\ref{Voinovform}), reduces to
Eq.~(\ref{maglio_model}). 
A more systematic approach amounts to note that if Eq.~(\ref{maglio_model}) is used
then in region II of our asymptotic analysis,
we have from Eqs.~(\ref{maglio_model}) and (\ref{phi-num})  
\be
G(\theta) = G(\theta_{in}) - \Ca \ln \left[ \frac{r \phi(\theta_\Delta)}{r_m}\right]. \label{indep}
\nd
The above equation predicts in region II an angle independent of the grid size. If the simulation is sufficiently well resolved and grid independent in the other regions of the domain, then via asymptotic matching of the outer scales with region II, the simulation should also be mesh-independent globally. 
This most systematic approach differs from the naive interpretation that $\theta_{\Delta}$ is the angle 
at scale $\Delta$ by a factor $\ln \phi(\theta_\Delta) = -f(\theta_\Delta,q)/Q_i$. 

The approach in the previous publication by two of the authors \cite{Afkhami_jcp09} is identical to the above one except for an approximation where 
\be
G(\theta) - G(\theta_{in}) \simeq [\cos \theta - \cos \theta_{in}] / 5.63
\nd
valid only in the range defined by Sheng and Zhou \cite{Sheng92}, 
that is  $q=1$ and $\vert\cos{\theta}\vert<0.6$. The approach embodied in Eq.~(\ref{maglio_model})
is more general, and can be used to define a procedure to perform 
mesh-independent simulations of realistic problems. 
(However, predictive simulations  of moving contact lines are not possible if a physical model at the microscopic scale, beyond the Navier-Stokes equations, is not available.)

In such a procedure, one of the main issues is whether contact line physics are obtained by experiment
or by a reduction to a microscopic theory such as molecular dynamics or phase field. 

To develop a strategy for realistic simulations where experiments would be used, 
the first step would be to perform a series of simulations at varying $\Delta$ and
for the conditions of the experiments. Obviously, simulations and experiments
are not performed near the critical $\Ca$ but below it. 
 Comparing the angles observed in the region where the theory is still valid
(region II) would cross-validate the numerics, 
the Cox's theory, and the experimental reality. This would in turn fix the parameters $\theta_{in}$ and 
$r_m$ that could be used in simulations. 
This approach is of course made difficult by the fact that there
is no evidence so far that a single pair $\theta_{in}, r_m$ 
could predict experiments over a range of 
different flow configurations. 
Actually the authors of \cite{YueFeng2011} state that ``In the literature we have 
not found a single pair of experiments in different geometries using precisely the same materials.''. 
This highlights
the difficulty of a predictive simulation approach based on experiments. Moreover, it should be
noted that the authors of references \cite{blake1999experimental} and \cite{Wilson:2006ks} 
claim precisely the opposite, that no inner angle 
$\theta_{in}$, even a dynamical one depending on the capillary number, 
can predict the whole range of experiments they have performed or simulated.  

Another approach could be based on mesoscale or nanoscale physics, 
using molecular dynamics or phase field
simulations to predict interface shapes, and again attempting to fit the predicted shapes to 
those predicted by Navier-Stokes simulations performed using  Eq.~(\ref{maglio_model}) and
a pair $\theta_{in}, r_m$. 
The models would have to be simulated up to scales much larger than the
nanoscale or the mesoscale so the Cox asymptotics become observable. 
This could be difficult since in particular the phase field models involve an intrinsic length scale $l_d$ (in the notations of ref.~\cite{YueFeng2011}) that may be large enough to preclude the appearance
of a ``Cox region'' such as region II inserted between the mesoscale model scale $l_d$ and the larger scales of the simulation
such as the capillary length $l_c$. 

All the above considerations must be subject to the proviso that the Cox region II is really observable. 
There are several circumstances when that cannot be the case. 
For both withdrawing or advancing contact angles, above a critical $\Ca$ number, a transition occurs 
to a liquid film or air film solutions, marking the disappearance of region II. 
These transitions fix an upper limit on $\Ca$ for moving contact lines to simply exist. A lower limit
for our approach to be applicable is obtained at small capillary numbers. Indeed,
the Cox solution is valid only if region II exists, which may be very narrow for small $\Ca$.
Indeed region II transitions directly to region III at small $\Ca$ for viscous forces of the same order
 than the hydrostatic pressure gradient
$\mu U / \ell^2 \sim \rho g$ which is equivalent to $\ell ~\sim l_c \Ca^{-1/2}$ where the capillary length appears. 
For $\Ca=10^{-6}$ and the classic capillary length $l_c = 2.1 10^{-3}m$  we find
$\eta_I \sim 2.1 \mu$m. This distance may become so small that simulations at this scale are not practical
and it is preferable to set the  grid size at larger scales at which the contact angle
is simply the apparent contact angle or the angle of the meniscus solution. 
}

\section{Conclusions}\label{sec:con}


We focus on the problem of
forced dewetting transition of a partially
wetting plate withdrawn from a reservoir using direct numerical simulations.

We compare numerical solutions in the vicinity of the contact line
with the theory of Cox and Voinov and point out the existence of a gauge function 
$\phi$ that corresponds to one of the integration constants in these theories. We find 
that Cox-Voinov theory is not well verified at small angles pointing to necessary improvements
in the numerical treatment of the dynamic contact line. These improvements may be related to the
treatment of boundary conditions or to the treatment of viscosity in mixed cells. They should be the focus
of future studies on this topic. 

Using the gauge function $\phi$, we provide a numerically-validated approximate
correlation for the critical capillary number, $\Ca_{cr}$, at which
the dewetting transition occurs, as a function of the mesh size, $\Delta$, and the 
numerically-imposed equilibrium contact angle, $\theta_\Delta$.

Using the asymptotic hydrodynamic theory of the vicinity of the contact line
and matching it to the static theory of menisci, we generalize the
hydrodynamic theory of the dewetting transition. We use it to 
derive an equation for the effect of $\Delta$ and  $\theta_\Delta$
on the larger-scale
regions of the simulation. 
The critical capillary number is then predicted
by an implicit equation for $\Ca_{cr}$. 
This equation contains the unknown gauge function $\phi$ that characterizes the contact 
line dynamics and is akin to a coefficient that determines the amount of 
slip. This gauge function is specific to the numerical model used.
Our numerical simulations allow to quantify this ``numerical slip'' and
confirm that it varies linearly with the grid size $\Delta$. 
{
Of particular interest is the work of Snoeijer in \cite{Snoeijer06}
which generalizes the lubrication theory to large angles.
The numerical verification of the lubrication equation developed in \cite{Snoeijer06}
is topic of our future work.}

We suggest a manner in which simulations can be made convergent upon grid
refinement, despite the singularity at the contact line. 
This involves adjusting the numerical contact angle as a function
of the grid size. This adaptation of the contact angle involves
the numerical gauge function $\phi$ and improves in several
ways over the model proposed in \cite{Afkhami_jcp09}. Indeed, it is valid for 
arbitrary angles and viscosity ratios. However, {the microscopic parameters and even 
more generally the microscopic physics are 
 not known, except for the case treated in this paper of a postulated Navier-slip model}. Thus 
the applicability of the grid-independent model may be limited.

The perspectives of this work are a systematic determination of the gauge
function $\phi$ for a range of numerical or physical contact line models used in practice and the 
verification of the procedure for grid-independent computations in a number of
flows. Such grid independent simulations should be performed in conjunction with 
experiments on contact line dynamics. 
 
\section*{Acknowledgements}

This research was supported by Electricit\'e de France (EdF), Contract No. 8610-5910129228 
(``Advances in Subgrid Models for Subcooled Flow Boiling in Pressurized Water Reactors'') 
and NSF grant Nos.~DMS-1320037
and CBET-1604351. S.~A.~gratefully acknowledges the support from CNRS for 
his visit to Institut Jean Le Rond d'Alembert
during the preparation of this paper. 
S.~Z.~gratefully acknowledges the support from CNRS and MOST (Taiwan) for 
supporting his visit to NTU during the preparation and writing of this paper. 

 The authors thank CINES and its team for the grant of computer time and technical assistance on
the OCCIGEN supercomputer in the framework of the GENCI allocation x20162b7325. 
The authors also thank D.~Fuster, G.~Galliero,  C.~Josserand, D.~Legendre, I.~Lunati, 
A.~Malan, H. Tchelepi, M. Abu-Al-Saud 
and A.-B.~Wang  for fruitful discussions. S.~Z.~gratefully acknowledges helpful discussions with
Y.~Pomeau on contact line modeling, and thanks Jens Eggers for his helpful correspondence about the derivation of equation (\ref{phicomp}).

\appendix

\section{Matching procedure in the small angle, free-surface case}
\label{app2}

In the small $\theta_e$ case, lubrication theory can be used following ref.~\cite{Eggers2004b}. 
We already expressed the first step of that theory when we obtained the slope in the vicinity 
of the contact line. It was seen that lubrication theory can be used when 
$\theta_e \ll Ca^{1/3}$. The range  of validity of the
theory is estimated in ref.~\cite{Eggers:2005uq} and for the {\em advancing}
case to be $\Ca/\theta_e^3 < 0.05$. 
This is equivalent for small angles
to $\Ca/G(\theta_e) \sim \delta < 0.5$ which is not too small but larger than the value in 
the numerical case above. 

In region IV, the lubrication equation is well known to be the Airy equation 
\cite{duffy1997third}
\be
\eta^{'''} = 3 \Ca/\eta^2, \label{lub0}
\nd
and we assume the scaling
\be
\eta(\zeta) = 3^{1/3} \Ca^{2/3} H(\zeta \Ca^{-1/3}),
\nd
so that the thickness of the film is $\Ca^{2/3}$ much smaller than its horizontal extent
$\Ca^{1/3}$. 
Eq.~(\ref{lub0}) can be solved using Airy functions.
The analysis given in \cite{duffy1997third} is reproduced in \ref{Airy}. 
One finds that for $\zeta \ll \Ca^{1/3} l_c$
\be
\eta'(\zeta) \sim  \{{9\Ca}\ln[\pi/(2^{2/3}\beta^2 \red{\mathrm e} \zeta) ]\}^{1/3} + \red{\Or \left[ \vert \ln(\beta^2 \zeta) \vert^{-5/3} \right]}\label{hpmain}
\nd 
where $\beta$ is a parameter characterizing the Airy function solution of Eq.~(\ref{lub0}).
Matching Eq.~(\ref{hpmain}) with the Cox's solution Eq.~(\ref{coxhp3}) gives
\be
\beta^2 \sim \frac {\pi}{2^{2/3} \red {\mathrm e}r_m} \exp \left[ - \frac {\theta_e^3}{9\Ca} \right]. \label{b2}
\nd
The matching is performed for $\zeta$ small in region IV variables but large in region II variables,
$r_m \ll \zeta \ll \Ca^{1/3}l_c$. This is consistent with an upper bound $r_{\rm max} =  \Ca^{1/3}l_c$ on the 
validity of the Cox's solution Eq.~(\ref{Voinovform}).
At the upper end of region IV,  for $\Ca^{1/3}l_c  \ll \zeta \ll l_c$, one can find the curvature 
as shown in \ref{Airy}
\be
\kappa_\infty^{\rm IV}  \sim  (3 \Ca)^{1/3} 
\left[ \frac {2^{1/6} \beta}{\pi {\rm Ai}(s_{1})}\right]^2.  \label{kinfpp}
\nd
Eq.~(\ref{kinfpp}) is obtained from the Airy function
solution, which is parameterized by both  $\beta$ and $s_1$. 
The determination of $s_1$ is more subtle \cite{Eggers2004b,Chan12}. It is seen that
expression  (\ref{kinfpp})  predicts a range of possible curvatures depending on the 
value of $s_1$. 
The smallest possible curvature obtains when the Airy function 
assumes its global maximum ${\rm Ai}(s_{\rm max}) \simeq 0.53$ for
$s_1 = s_{\rm max} \simeq -1.0$. 
Thus for a given $\Ca$ and $\beta$, a minimum curvature is given by
\be
\kappa_{\rm IV,min}  = (3 \Ca)^{1/3} 
\left[\frac {2^{1/6} \beta}{\pi {\rm Ai}(s_{\rm max})}\right]^2 . \label{kinf2}
\nd
{Indeed if  the curvature determined at the lower limit of region III
is larger, that is  $\kappa^{\rm III} > \kappa_{\rm IV,min}$,
it is always possible to match with a solution parameterized by some $s_1$
such that ${\rm Ai}(s_1) < {\rm Ai}(s_{\rm max})$ and 
${\rm Ai}(s_{\rm max})$ is small enough. However, if 
curvature $\kappa^{\rm III} < \kappa_{\rm IV,min}$, the matching is impossible. 
Thus the critical $\Ca$ is given by  $\kappa^{\rm III,max} = \kappa_{\rm IV,min}$}
\be
 (3 \Ca_{cr})^{1/3} 
\left[ \frac {2^{1/6} \beta}{\pi {\rm Ai}(s_{\rm max})}\right]^2 = \kappa^{\rm III,max}. \label{kinf3}
\nd
We now eliminate $\beta$ between Eq.~(\ref{b2}) and Eq.~(\ref{kinf3}) and use the notation
$\kappa^{\rm III,max} = \kappa_\infty$,
\be
\frac{3^{1/3} 2^{-1/3}}{\pi {\rm Ai}^2(s_{\rm max})} 
\frac {\Ca_{cr}^{1/3}}{ \red{\mathrm e} r_m}  
\exp \left( - \frac {\theta_e^3}{9\Ca_{cr}} \right)  = \kappa_\infty.  \label{cacr1}
\nd
In the slip-length model case, with $r_m$ given by Eq.~(\ref{phi-lub}) and $\phi$ given by Eq.~(\ref{eh2}), we obtain
\red{\be
\frac{ \theta_e}{ 18^{1/3} \pi {\rm Ai}^2(s_{\rm max})} 
\frac {\Ca_{cr}^{1/3}}{\lambda}  
\exp \left( - \frac {\theta_e^3}{9\Ca_{cr}} \right)  = \kappa_\infty \label{cacr1e}
\nd}
This is in agreement with Eggers's result in \cite{Eggers2004b}. 
Eggers considers the case of a plate inclined with a small slope $\theta_p$ above the horizontal,
while we consider a vertical plate. This does not change the nature of the asymptotics and as already
pointed out in  ref.~\cite{Eggers2004b}, the theory transposes as well to the case of a vertical
plate (although the full numerical solution is then more difficult to obtain). 
In the small
plate angle case $\theta_p$ is equal to the curvature $\kappa_\infty$ of the meniscus solution, while in our case
the curvature is given by  Eq.~(\ref{kappaIII}). 
%
%
There is agreement between Eq.~(\ref{cacr1e}) and Eq.~(9) of  \cite{Eggers2004b} 
if we take note of the use of dimensionless
variables in  \cite{Eggers2004b}, while we use dimensional variables, and substitute
$\kappa_\infty $ for $\theta$ (using ref.~\cite{Eggers2004b} notation for $\theta_p$). 

\section{Analysis of the Airy equation}
\label{Airy}

We first outline the Airy function solution of Eq.~(\ref{lub0})
given in \cite{duffy1997third}.
With the transformation $\eta = (3\mbox{Ca})^{1/3} H$, Eq.~\ref{lub0} becomes
\be
H^{'''}(\zeta) = \frac{1}{H^2(\zeta)}.
\ee
This equation can be turned into Airy's equation for a new variable $z(s)$,
that is $z^{''}=sz$, upon the substitutions
$$H(\zeta) = z^{-2}(s),  \label{yint}
$$ 
and 
$$
\frac{d\zeta}{ds}=-2^{1/3}z^{-2}(s), \label{xint}
$$ 
whose solution is
\be
z(s)=\alpha \mbox{Ai}(s) + \beta \mbox{Bi}(s),
\ee
which implies
\be
H =\frac{1}{[\alpha \mbox{Ai}(s) + \beta \mbox{Bi}(s)]^2},
\label{eq:inner}
\ee
where $\alpha$ and $\beta$ are two arbitrary constants,
while $\zeta(s)$ is an antiderivative of $-2^{1/3}z^{-2}(s)$. 
After integrating explicitly for $\zeta(s)$,
Duffy and Wilson \cite{duffy1997third} conclude that for $\beta>0$ 
and any given $\alpha$, there is a single branch that behaves as desired
for the contact line problem, that it 
grows monotonically from $H=0$ for $\zeta=0$ and for $s\to \infty$ and 
has $H\to\infty$ for $s\to s_1$ and  $\zeta\to\infty$ 
where $s_1$ is a root of 
\be
\alpha \mbox{Ai}(s_1) + \beta \mbox{Bi}(s_1)=0.
\label{eq:root}
\ee 
That single branch cannot go to infinity inside $s\in[s_1,\infty)$,
which implies that $s_1$ is the largest among the countably infinite set of roots of Eq.~(\ref{eq:root}). 
Thus the solution can be characterized either 1)  by the arbitrary pair $\alpha,\beta$  ($\beta>0$) 
or 2)  by $\beta$  ($\beta>0$)
and $s_1$ chosen arbitrarily  with $\alpha$ given by 
$\alpha = -  \beta \mbox{Bi}(s_1)/ \mbox{Ai}(s_1)$
provided there is no larger root of Eq.~(\ref{eq:root}). In what follows, we use the second
characterization.
It can be verified that 
\be
\zeta=2^{1/3}\pi\frac{\mbox{Ai}(s)}{\beta [\alpha \mbox{Ai}(s) + \beta \mbox{Bi}(s)]}, \label{xexp}
\ee
is an antiderivative of  $-2^{1/3}z^{-2}(s)$ and that it satisfies $\zeta=0$ for $s\to\infty$. Thus the
solution given parametrically by Eqs.~(\ref{eq:inner}) and (\ref{xexp})
starts at $H=0$ for $\zeta=0$ and grows monotonically with $H\to\infty$ for $\zeta \to \infty$.
Expanding the solution for $s\gg 1$, Duffy and Wilson  \cite{duffy1997third} find that
{\be
H(\zeta) \sim \zeta [-3  ( \ln \zeta + c )]^{1/3}, \label{hp1}
\nd }
where $c=  \ln[ \pi/(2^{2/3}\beta^2)]$. 
Differentiating, and returning to the original variable $\eta$,
leads to 
\be
\eta'(\zeta) \sim  \{{9\Ca}\ln[\pi/(2^{2/3}\beta^2 \red{\mathrm e} \zeta) ]\}^{1/3} + \red{\Or \left[ \vert \ln(\beta^2 \zeta) \vert^{-5/3} \right]}\label{hp}
\nd 
For $\zeta\gg1$, Eq.~(\ref{eq:inner}) reduces to
\be
H(\zeta)=\frac{1}{2}\left[\frac{2^{1/6}\beta}{\pi\mbox{Ai}(s_1)}\right]^2\zeta^2
- \frac{2^{2/3}\mbox{Ai}^{'}(s_1)}{\mbox{Ai}(s_1)}\zeta +O(1).
\label{eq:inner3}
\ee
Differentiating twice, one obtains the second derivative. As discussed in Sec.~\ref{sec:theory},
it is possible to equate the curvature with the second derivative 
since 1) ${\rm Ai}'(s_1) =0$, and 2)
the curvature is of the order of $l_c^{-1}$ and the matching is
performed at distances small in the region III variables thus over distances 
$\zeta \ll l_c$. Over such distances, the curvature of the parabola (Eq.~(\ref{eq:inner3}))
is close to the curvature at its apex. 
Reverting to the original variable $\eta$, we 
obtain the curvature $\kappa_\infty \simeq \eta''(\zeta)$ in the 
asymptotic range $\Ca^{-1/3}l_c  \ll \zeta \ll l_c$ as
\be
\kappa_\infty \sim (3 \Ca)^{1/3} \left[ \frac {2^{1/6} \beta}{\pi {\rm Ai}(s_{1})}\right]^2.  \label{kinf}
\nd

\section*{References}

\bibliographystyle{elsarticle-num}

\end{document}